\documentclass[fleqn,usenatbib]{mnras}

\usepackage[utf8]{inputenc}
\usepackage[T1]{fontenc}
\usepackage{ae,aecompl}
\usepackage{graphicx}	
\usepackage{amsmath}	
\usepackage{amssymb}	
\usepackage{siunitx}
\usepackage{multicol}
\usepackage{listings}
\usepackage{subcaption}
\usepackage{longtable}
\usepackage[justification=justified, singlelinecheck=false]{caption}
\usepackage{silence}
\usepackage[table]{xcolor}
\usepackage{newtxtext,newtxmath}

\graphicspath{{images/}}

\newcommand{\gaia}{\textit{Gaia}}
\newcommand{\gdrtwo}{\textit{Gaia}~DR2}
\newcommand{\gdrthree}{\textit{Gaia}~DR3}

\newcommand{\teffespucd}{\texttt{teff\_espucd}}
\newcommand{\gaiag}{\ensuremath{G}}
\newcommand{\gaiabp}{\ensuremath{G_{\mathrm{BP}}}}
\newcommand{\gaiarp}{\ensuremath{G_{\mathrm{RP}}}}

\newcommand{\vtotal}{\ensuremath{V_{\text{total}}}}

\newcommand{\rd}{\textsuperscript{rd}}
\newcommand{\thdate}{\textsuperscript{th}}
\newcommand{\teff}{\ensuremath{T_{\mathrm{eff}}}}
\newcommand{\teffexpect}{\ensuremath{\widehat{T_{\mathrm{eff}}}}}
\newcommand{\angstrom}{\ensuremath{\textup{\AA}}}
\newcommand{\logg}{\ensuremath{\log g}}
\newcommand{\banyan}{BANYAN~$\Sigma$}
\newcommand{\kms}{\ensuremath{\mathrm{km\,s}^{-1}}}
\newcommand{\masyr}{\ensuremath{\mathrm{mas\,yr}^{-1}}}

\title[GTC/OSIRIS spectra of late-M and L dwarfs]{The \gaia\ Ultracool Dwarf Sample -- IV.
GTC/OSIRIS optical spectra of \gaia\ late-M and L dwarfs}
\author[W.\,J. Cooper et~al.]{W.\,J.~Cooper,$^{1,2}$\thanks{E-mail: w.cooper@herts.ac.uk}
H.\,R.\,A.~Jones,$^{1}$ 
R.\,L.~Smart,$^{2}$  
S.\,L.~Folkes,$^{1}$ 
J.\,A.~Caballero,$^{3}$ 
\newauthor
F.~Marocco,$^{4}$ 
M.\,C.~G\'alvez~Ortiz,$^{3}$ 
A.\,J.~Burgasser,$^{5}$ 
J.\,D.~Kirkpatrick,$^{4}$ 
L.\,M.~Sarro,$^{6}$ 
\newauthor
B.~Burningham,$^{1}$ 
A.\,Cabrera-Lavers,$^{7}$ 
P.\,E.~Tremblay,$^{8}$ 
C.~Reyl\'e,$^{9}$ 
N.~Lodieu,$^{10,11}$ 
\newauthor
Z.\,H.~Zhang$^{12,13}$ 
N.\,J.~Cook,$^{14}$ 
J.\,F.~Faherty,$^{15}$ 
D.~Garc\'ia-\'Alvarez,$^{7,10}$ 
\newauthor
D.~Montes,$^{16}$ 
D.\,J.~Pinfield,$^{1}$ 
A.\,S.~Rajpurohit,$^{17}$ 
J.~Shi$^{18,19}$ 
\\
$^{1}$Centre for Astrophysics Research, University of Hertfordshire, Hatfield, Hertfordshire, AL10 9AB, UK\\
$^{2}$Istituto Nazionale di Astrofisica, Osservatorio Astrofisico di Torino, Strada Osservatorio 20, I-10025 Pino Torinese, IT\\
$^{3}$Centro de Astrobiolog\'ia (CAB), CSIC-INTA, Camino Bajo del Castillo s/n, Campus ESAC, E-28692 Villanueva de la Ca\~nada, Madrid, ES\\
$^{4}$IPAC, Mail Code 100-22, Caltech, 1200 E. California Boulevard, Pasadena, CA 91125, US\\
$^{5}$Center for Astrophysics and Space Science, University of California San Diego, La Jolla, CA 92093, US\\
$^{6}$Departamento de Inteligencia Artificial, ETSI Informática, UNED, Juan del Rosal, E-16 28040 Madrid, ES\\
$^{7}$GRANTECAN, Cuesta de San Jos\'e s/n, E-38712, Bre\~na Baja, La Palma, ES \\
$^{8}$Department of Physics, University of Warwick, Coventry CV4 7AL, UK\\
$^{9}$Institut UTINAM, CNRS UMR6213, Université de Bourgogne Franche-Comt\'e, OSU THETA Franche-Comt\'e-Bourgogne, \\
\hspace{0.1cm} Observatoire de Besan\c{c}on, BP 1615, 25010, Besan\c{c}on Cedex, FR\\
$^{10}$Instituto de Astrof{\'{\i}}sica de Canarias, E-38205 La Laguna, Tenerife, ES\\
$^{11}$Universidad de La Laguna, Departamento de Astrof{\'{\i}}sica, E-38206 La Laguna, Tenerife, ES\\
$^{12}$School of Astronomy and Space Science, Nanjing University, 163 Xianlin Avenue, Nanjing 210023, CN \\
$^{13}$Key Laboratory of Modern Astronomy and Astrophysics, Nanjing University, Ministry of Education, Nanjing 210023, CN \\
$^{14}$Institute for Research on Exoplanets, Université de Montréal, Département de Physique, C.P. 6128 Succ. Centre-ville, Montréal, QC H3C 3J7, CA\\
$^{15}$Department of Astrophysics, American Museum of Natural History, Central Park West at 79th Street, NY 10024, US\\
$^{16}$Departamento de F{\'i}sica de la Tierra y Astrof{\'i}sica \& IPARCOS-UCM (Instituto de F\'{i}sica de Part\'{i}culas y del Cosmos de la UCM),\\
\hspace{0.1cm} Facultad de Ciencias F{\'i}sicas, Universidad Complutense de Madrid, E-28040 Madrid, ES\\
$^{17}$Astronomy and Astrophysics Division, Physical Research Laboratory, Navrangapura, Ahmedabad, 380009, IN\\
$^{18}$College of Astronomy and Space Sciences, University of Chinese Academy of Sciences, Beijing 100049, CN\\
$^{19}$Key Laboratory of Optical Astronomy, National Astronomical Observatories,
       Chinese Academy of Sciences, Beijing 100012, CN
}

\date{Accepted 2024 September 5. Received 2024 September 5; in original form 2023 July 7}

\pubyear{2024} 

\begin{document}
\label{firstpage}
\pagerange{\pageref{firstpage}--\pageref{lastpage}}
\maketitle

\begin{abstract}
    As part of our comprehensive, ongoing characterisation of the low-mass end of the main sequence
    in the Solar neighbourhood,
    we used the OSIRIS instrument at the 10.4\,m Gran Telescopio Canarias
    to acquire low- and mid-resolution (R${\approx}$300 and R${\approx}$2500) optical spectroscopy of 53 late-M and L ultracool dwarfs.
    Most of these objects are known but poorly investigated and lacking complete kinematics.
    We measured spectral indices, determined spectral types (six of which are new) and
    inferred effective temperature and surface gravity from BT-Settl synthetic spectra fits for all objects.
    We were able to measure radial velocities via line centre fitting and cross correlation for 46 objects,
    29 of which lacked previous radial velocity measurements.
    Using these radial velocities in combination with the latest \gdrthree\ data,
    we also calculated Galactocentric space velocities.
    From their kinematics, we identified two candidates outside of
    the thin disc and four in young stellar kinematic groups.
    Two further ultracool dwarfs are apparently young field objects:
    2MASSW J1246467$+$402715 (L4$\beta$), which has a potential, weak lithium absorption line, and G~196--3B (L3$\beta$),
    which was already known as young due to its well-studied primary companion.
\end{abstract}


\section{Introduction} \label{sec:gtcintro}
Ultracool dwarfs (UCDs) are objects with effective temperatures
$\teff \lessapprox 2700$\,K~\citep[spectral type $\gtrsim$ M7\,V,][]{1999ApJ...519..802K}
continuing on from the low-mass tail of the main sequence, that consist of spectral types late-M, L, T and Y dwarfs.
These UCDs consist of a combination of low-mass stars and brown dwarfs.
Brown dwarfs are sub-stellar objects incapable of hydrogen fusion and are defined by mass,
between the deuterium minimum mass burning limit,
${\sim}13$\,Jupiter masses~\citep{1996ApJ...460..993S, 2000ApJ...542L.119C} and the hydrogen minimum mass burning limit,
${\sim}72$\,Jupiter masses~\citep{1997A&A...327.1039C, 1997A&A...327.1054B}.
The majority of known UCDs are within the Solar neighbourhood~\citep[e.g.\ ][]{GUCDS2, 2021ApJS..253....7K, 2023A&A...669A.139S}
with typically dim apparent optical magnitudes (\gaia\ $\gaiag \gtrapprox 17$\,mag).
The closest stars to the Sun have been catalogued throughout the history of astronomy.
For example, the Catalogue of Nearby Stars (CNS) from~\citet{1957MiABA...8....1G}
has been updated with every all-sky photometric and
astrometric survey, including the most recent release using~\gdrthree\ data~\citep[CNS5,][]{2023A&A...670A..19G}.
This Solar neighbourhood has been further described in the `The Solar Neighborhood' series by the
Research Consortium on Nearby Stars (\href{http://www.astro.gsu.edu/RECONS/}{RECONS}
\footnote{\url{http://www.astro.gsu.edu/RECONS/}}) team with publications
from~\citet{1994AJ....108.1437H} to~\citet{2022AJ....163..178V}.
Specifically, M dwarfs within 30\,pc were covered in another series of articles 
from~\citet{1999A&A...344..897D} to~\citet{2005A&A...441..653C}.
Volume limited samples such as the recent~\citet[100\,pc,][]{2021A&A...649A...6G},~\citet[20\,pc,][]{2021ApJS..253....7K}
and~\citet[10\,pc,][]{2021A&A...650A.201R} works
provide important constraints on the initial mass
function~\citep{1955ApJ...121..161S, 1986FCPh...11....1S, 2001MNRAS.322..231K, 2003PASP..115..763C},
which underpins all aspects of astrophysics from stars to galaxies to cosmology.

Spectral features of low mass stars, M, L and T dwarfs, and their definitions were initially described
by~\citet{1998MNRAS.301.1031T},~\citet{1999ApJ...519..802K},~\citet{1999AJ....118.2466M},~\citet{2002ApJ...564..421B},
~\citet{2002ApJ...564..466G} and~\citet{2005ARA&A..43..195K}.
The bulk of the flux emitted by L dwarfs lies in the near infrared (NIR) and continues strongly towards the
mid-infrared spectral regions for later spectral type UCDs.
However, several features of youth, e.g.\ a weak sodium doublet, $\lambda \lambda$8183,8195\,\AA~\citep{1997ApJ...479..902S},
are apparent in mid- to high-resolution optical spectra.
Additionally, in the optical regime features such as the $\lambda$9850--10200\,\AA~FeH
Wing-Ford band~\citep{1997ApJ...484..499S} can be seen,
which can be indicative of low or high metallicity.
Optical spectra have an advantage in that there are fewer and weaker telluric absorption bands than
in ground-based infrared spectra,
where water and oxygen bands can dominate~\citep{2007AandA...473..245R, 2015AandA...576A..77S}.
However, only the closest and brightest UCDs can be observed with optical spectroscopy due to the low relative flux;
further and fainter UCDs require large aperture telescopes and long exposure times.

UCDs have typically been selected from photometric criteria using optical and near- to mid-infrared imaging surveys,
supported by proper motion analysis.
Examples of optical surveys include SuperCOSMOS~\citep{2001MNRAS.326.1279H}, \gaia~\citep{2016A&A...595A...1G},
Pan-STARRS~\citep[PS1,][]{2016arXiv161205560C} and the SDSS~\citep{2000AJ....120.1579Y, 2009ApJS..182..543A},
in which UCDs appear red.
Notable infrared surveys and catalogues include 2MASS~\citep{2003yCat.2246....0C, 2006AJ....131.1163S},
DENIS~\citep{1997Msngr..87...27E},
VISTA's VVV/VIRAC/VHS~\citep{2010NewA...15..433M, 2018MNRAS.474.1826S, 2021yCat.2367....0M} and
UKIDSS~\citep{2007MNRAS.379.1599L}.
Further infrared is the WISE~\citep{2010AJ....140.1868W} survey, which was expanded upon in the
unWISE/catWISE~\citep{2019ApJS..240...30S, 2021ApJS..253....8M, 2023AJ....165...36M} catalogues.
These NIR surveys are complemented by additional surveys constraining UCDs in open clusters such as the 
Pleiades~\citep{1995MNRAS.272..630S, 2000MNRAS.313..347P, 2012yCat..74221495L},
or elsewhere~\citep{2000MNRAS.314..858L, 2000Sci...290..103Z, 2013MNRAS.433..457B}.

The photometry of UCDs is important because the change in colour across the optical and NIR
regime~\citep{2002ApJ...564..452L} correlates with physical and atmospheric properties.
These changing processes, such as dust, condensate cloud formation and subsequent clearing as an atmosphere cools,
are well covered in the literature~\citep[e.g.\ ][]{2002ApJ...568..335M, 2002AJ....124.1170D, 2008ApJ...689.1327S}.
Understanding a changing atmosphere for different ages with a range of masses has allowed the
computing of `cooling tracks'~\citep{1997ApJ...491..856B, 2015A&A...577A..42B}.
Accounting for theoretical atmospheric physics has been used in model grids such as
BT-Settl~\citep{2011ASPC..448...91A}, or Sonora~\citep{2021ApJ...920...85M, 2021ApJ...923..269K},
and when interpreting the results of retrieval
techniques~\citep[e.g.\ ][]{2017MNRAS.470.1177B, 2022ApJ...940..164C}.
Being able to constrain the mass and/or age has underpinned modern observational UCD astronomy,
but is challenging due to the mass/age degeneracy~\citep{1997ApJ...491..856B}.
For example, benchmark systems~\citep[e.g.\ ][]{2006MNRAS.368.1281P, 2009ApJ...692..729D} allow us to constrain the age
of a brown dwarf with the coeval main sequence primary.
The metallicity and surface gravity of an object of a given spectral type are the major variables
affecting the photometric colour~\citep{2009ApJ...702..154S},
see references to `blue' and `red' L dwarfs~\citep[e.g.\ ][]{2009AJ....137....1F, 2010AJ....139.1808S}.
Any works that infer spectral type, surface gravity and effective temperature must take into account the
atmospheric physics, as these directly correlate with observable features.

\gaia\ is a European Space Agency mission, launched in 2013 to make high-precision measurements of positions,
parallaxes, and proper motions of well over a billion sources and photometry in 
three different photometric filters (\gaiabp, \gaiag, \gaiarp).
The third \gaia\ data release~\citep[EDR3 and DR3 --][respectively]{GDR3, 2023A&A...674A...1G}
containing astrometric and photometric measurements, was in December 2021,
with the remaining measurements and inferred parameters, including spectra, in June 2022\footnote{
The astrometry and photometry in \gdrthree\ used in this work is identical to that within \gaia\ EDR3
whilst the astrophysical parameters are purely from \gdrthree;
hence, both data releases are cited here.
}.

Obtaining the full 6D (right ascension, declination, proper motions, parallax, radial velocity:
$\alpha, \delta, \mu_{\alpha}\cos{\delta}, \mu_{\delta}, \varpi, v_r$) positional and kinematic information
is fundamental to fully characterise the populations of UCDs within a volume limited sample~\citep[e.g.\ ][]{2021AJ....161...42B}.
Precise measurements of radial velocities (RVs) are obtained from high signal-to-noise observations taken
with high resolution spectrographs with resolving powers of R${\sim}$100\,000,
leading to uncertainties ${\sim}$1--5\,m\,s$^{-1}$.
This has only been achievable for the nearest, brightest
UCDs~\citep[e.g.\ ][]{2019A&A...627A..49Z}.
\citet{2010ApJ...723..684B} achieved $\delta v_r \approx$ 50--200\,m\,s$^{-1}$ with the Keck Near-Infrared Spectrometer (NIRSPEC),
which had a resolution of R${\approx}$25\,000.
The `Brown Dwarf Kinematics Project' has gathered further UCD RVs~\citep{2015ApJS..220...18B, 2021ApJS..257...45H}
with both the NIRSPEC and the Magellan Echellette (MagE, R${\sim}4100$, $\delta v_r \approx$ 2--3\,\kms) spectrographs.
By comparison, the lower-resolution spectroscopy such as those discussed in this work (R${\approx}$2500)
is only capable of theoretical minimum uncertainties of ${\gtrsim}$5\,\kms;
this is still useful when constraining the kinematics of the Solar neighbourhood.
Parallaxes and proper motions of UCDs were historically gathered from ground based time-domain
campaigns~\citep[e.g.\ PARSEC:][]{2011AJ....141...54A, 2013AJ....146..161M, 2018MNRAS.481.3548S}
that have been generally superseded by \gaia\ for the brightest objects, $\gaiag \lessapprox 20$\,mag.
In the case of most late-L and T dwarfs, ground-based astrometry is still the predominant
source~\citep[e.g.\ ][]{2004AJ....127.2948V, 2012ApJS..201...19D, 2016ApJ...833...96L, 2018ApJS..234....1B}.
For dimmer objects, beyond mid-L dwarfs, parallaxes and proper motions are gathered by space-based infrared surveys
and are analysed in-depth by~\citet{2021ApJS..253....7K}.
Young moving groups are constrained using these complete kinematics.
See the \banyan\ series and references therein for detail on nearby young moving groups and
clusters~\citep[][to~\citealt{2018ApJ...862..138G}]{2014ApJ...783..121G} or similarly, the LACEwING
code~\citep{2017AJ....153...95R}, designed around young objects in the Solar neighbourhood.
Subdwarfs, meanwhile, are characterised by their statistically higher space
velocities indicative of the older
population~\citep[e.g.\ ][]{2005A&A...440.1061L, 2007ApJ...657..494B, 2017A&A...598A..92L, 2017MNRAS.464.3040Z}.

This is the fourth item in the \gaia\ UltraCool Dwarf Sample series~\citep[GUCDS,][]{GUCDS1, GUCDS2, GUCDS3}
and is an ongoing, international, multi-year programme aimed at characterising all of the UCDs visible to \gaia.
\gdrthree\ produced astrophysical parameters for ${\approx}470$ million sources~\citep{2023A&A...674A..26C},
including effective temperatures, \teff.
The ${\approx}$94\,000 \gdrthree\ \teff\ values relating to UCDs by~\citet{2023A&A...674A..26C}
were provided under the \teffespucd\ keyword.
The full sample of UCDs detected by \gaia\ with \gdrthree\ \teff\ values were documented
and analysed by~\citet{2023A&A...669A.139S}.
In our analysis, we will use the values from these \gdrthree\ derivative works to compare with the 
equivalent values directly measured by us.
There is significant overlap between the~\citet{2023A&A...669A.139S} sample and the GUCDS,
although the majority of UCD sources as seen by \gaia\ are as yet not characterised through spectroscopic follow-up.
A subset of this~\citet{2023A&A...669A.139S} sample has public \gaia\ RP spectra (see the 
\gaia\ \href{https://gea.esac.esa.int/archive/documentation/GDR3/Gaia_archive/chap_datamodel/sec_dm_spectroscopic_tables/ssec_dm_xp_summary.html}
{\texttt{xp\_summary}} table\footnote{\url{https://gea.esac.esa.int/archive/documentation/GDR3/Gaia_archive/chap_datamodel/sec_dm_spectroscopic_tables/ssec_dm_xp_summary.html}}), which covers
the \gaiarp\ passband~\citep[$\Delta \lambda \approx 6200\text{--}10420\,\angstrom$,][]{2021A&A...649A...3R}.
This subset from~\citet{2023A&A...669A.139S} was further analysed for spectroscopic outliers
by~\citet{2024MNRAS.527.1521C}.
The internally calibrated \gaia\ RP spectra and processing were discussed thoroughly
by~\citet{2021A&A...652A..86C},~\citet{2023A&A...674A...2D} and~\citet{2023A&A...674A...3M}.

The aim of this work is to complement the literature population with measurements
and inferences from low- and mid-resolution optical spectroscopy.
In Section~$\S$\ref{sec:observations} we explain the target selection ($\S$\ref{subsec:targetselection}) and
observation strategy ($\S$\ref{subsec:observations}).
Different reduction techniques with a test case are discussed in Section~$\S$\ref{sec:reduction}.
Section~$\S$\ref{sec:gtcanalysis} explains our techniques for determining
spectral types ($\S$\ref{subsec:spectypinganalysis}),
astrophysical parameters ($\S$\ref{subsec:astrophysical_parametersanalysis}),
and kinematics ($\S$\ref{subsec:kinematicsanalysis}) including membership in moving groups ($\S$\ref{subsec:movinggroupsanalysis}).
Section~$\S$\ref{sec:gtcresults} follows a discussion of our results for
spectral types ($\S$\ref{subsec:spectypingresults}),
kinematics ($\S$\ref{subsec:kinematicsresults})
and astrophysical parameters ($\S$\ref{subsec:astrophysical_parametersresults}).
We also discuss individual objects ($\S$\ref{subsubsec:individuals}) before summarising
the overall conclusions in Section~$\S$\ref{sec:gtcdiscussion}.

\section{Data collection} \label{sec:observations}
We obtained optical spectroscopy of 53 unique UCDs using the
OSIRIS~\citep[Optical System for Imaging and low-intermediate Resolution
Integrated Spectroscopy --][]{osiris} instrument on the 10.4\,m Gran Telescopio Canarias (GTC)
at El Roque de los Muchachos in the island of La Palma, Spain,
under proposal IDs GTC54-15A and GTC8-15ITP (PIs Caballero and Marocco, respectively)\@.
The objects were observed in semesters 2015A, 2015B and 2016A\@.

The observed data from the GTC were complemented with \gdrthree.
\gaia\ also carries a radial velocity spectrometer, although this was unsuitable for our purposes as all of our
targets were fainter than the \gaia\ selection limit~\citep[$\gaiag < 14$\,mag,]{2023A&A...674A...5K}.

\onecolumn
\begin{longtable}[c]{ll cc c cc l}
\caption{\label{table:target_list}
The 53 targets observed at the GTC with OSIRIS and presented in this work.
}\\
\hline
    Object     & \gdrthree\ & $\alpha$ & $\delta$ & $\varpi$ & $G$   & $J$   & Grism/VPHG \\
    short name & source ID  & [hms]     & [dms]    & [mas]     & [mag] & [mag] &  \\
\hline
\endfirsthead
\hline
    Object     & \gdrthree\ & $\alpha$ & $\delta$ & $\varpi$ & $G$   & $J$   & Grism/VPHG \\
    short name & source ID  & [hms]     & [dms]    & [mas]     & [mag] & [mag] &  \\
\hline
\endhead
\hline
\endfoot
\hline
\caption*{\hypertarget{posrefs}{References -- Positions all at 2016.5 except at the indicated epochs}: 1.\ ~\citet{2007MNRAS.379.1599L} -- 2008, 2.\ ~\citet{2006AJ....131.1163S} -- 1998--2000, 3.\ ~\citet{2016arXiv161205560C} -- 2012--2013, 4.\ ~\citet{2020AJ....159..257B} -- 2014--2018, 5.\ ~\citet{2016AJ....152...24W} -- 2007--2013.}
\endlastfoot
    J0028$-$1927 & 2363496283669200768 & 0 28 55.6 & -19 27 16 & 25.742 & 18.97 & 14.19 & R2500I\\
    J0235$-$0849 & 5176990610359832576 & 2 35 47.5 & -8 49 20 & 21.742 & 20.35 & 15.57 & R2500I\\
    J0428$-$2253 & 4898159654173165824 & 4 28 51.1 & -22 53 20 & 39.398 & 18.72 & 13.51 & R2500I\\
    J0453$-$1751 & 2979566285233332608 & 4 53 26.5 & -17 51 55 & 33.064 & 20.14 & 15.14 & R2500I\\
    J0502$+$1442 & 3392546632197477248 & 5 02 13.5 & +14 42 36 & 21.746 & 18.90 & 14.27 & R2500I\\
    J0605$-$2342 & 2913249451860183168 & 6 05 01.9 & -23 42 25 & 30.185 & 19.31 & 14.51 & R2500I\\
    J0741$+$2316 & 867083081644418688 & 7 41 04.4 & +23 16 38 & 13.019 & 20.83 & 16.16 & R2500I\\
    J0752$+$4136 & 920980385721808128 & 7 52 59.4 & +41 36 47 & 11.734 & 17.71 & 14.00 & R2500I\\
    J0809$+$2315 & \ldots & 8 09 10.7$\hyperlink{posrefs}{^1}$ & +23 15 16$\hyperlink{posrefs}{^1}$ & \ldots & \ldots & 16.72 & R2500I\\
    J0823$+$0240 & 3090298891542276352 & 8 23 03.1 & +2 40 43 & \ldots & 21.18 & 16.06 & R2500I\\
    J0823$+$6125 & 1089980859123284864 & 8 23 07.3 & +61 25 17 & 39.467 & 19.66 & 14.82 & R2500I\\
    J0847$-$1532 & 5733429157137237760 & 8 47 28.9 & -15 32 41 & 57.511 & 18.38 & 13.51 & R300R\\
    J0918$+$2134 & \ldots & 9 18 38.2$\hyperlink{posrefs}{^2}$ & +21 34 06$\hyperlink{posrefs}{^2}$ & \ldots & \ldots & 15.66 & R2500I\\
    J0935$-$2934 & 5632725432610141568 & 9 35 28.0 & -29 34 58 & 29.969 & 19.00 & 14.04 & R2500I\\
    J0938$+$0443 & 3851468354540078208 & 9 38 58.9 & +4 43 43 & 15.448 & 19.89 & 15.24 & R2500I\\
    J0940$+$2946 & 696581955256736896 & 9 40 47.7 & +29 46 52 & 17.961 & 20.30 & 15.29 & R2500I\\
    J0953$-$1014 & 3769934860057100672 & 9 53 21.2 & -10 14 22 & 28.022 & 18.44 & 13.47 & R2500I\\
    J1004$+$5022 & 824017070904063488 & 10 04 20.4 & +50 22 56 & 46.195 & 20.13 & 14.83 & R300R \& R2500I\\
    J1004$-$1318 & 3765325471089276288 & 10 04 40.2 & -13 18 22 & 40.438 & 19.84 & 14.68 & R2500I\\
    J1047$-$1815 & 3555963059703156224 & 10 47 30.7 & -18 15 57 & 35.589 & 19.01 & 14.20 & R300R \& R2500I\\
    J1058$-$1548 & 3562717226488303360 & 10 58 47.5 & -15 48 17 & 55.098 & 19.24 & 14.16 & R300R \& R2500I\\
    J1109$-$1606 & 3559504797109475328 & 11 09 26.9 & -16 06 56 & 24.161 & 19.65 & 14.97 & R2500I\\
    J1127$+$4705 & 785733068161334656 & 11 27 06.5 & +47 05 48 & 23.758 & 19.94 & 15.20 & R2500I\\
    J1213$-$0432 & 3597096309389074816 & 12 13 02.9 & -4 32 44 & 59.095 & 19.86 & 14.68 & R2500I\\
    J1216$+$4927 & 1547294197819487744 & 12 16 45.5 & +49 27 45 & \ldots & 20.92 & 15.59 & R2500I\\
    J1221$+$0257 & 3701479918946381184 & 12 21 27.6 & +2 57 19 & 53.812 & 17.86 & 13.17 & R2500I\\
    J1222$+$1407 & \ldots & 12 22 59.3$\hyperlink{posrefs}{^3}$ & +14 07 50$\hyperlink{posrefs}{^3}$ & \ldots & \ldots & \ldots & R300R\\
    J1232$-$0951 & 3579412039247581824 & 12 32 18.1 & -9 51 52 & 34.5$\hyperlink{posrefs}{^4}$ & 18.74 & 13.73 & R2500I\\
    J1246$+$4027 & 1521895105554830720 & 12 46 47.0 & +40 27 13 & 44.738 & 20.28 & 15.09 & R300R \& R2500I\\
    J1331$+$3407 & 1470080890679613696 & 13 31 32.6 & +34 07 55 & 34.791 & 19.01 & 14.33 & R300R \& R2500I\\
    J1333$-$0215 & 3637567472687103616 & 13 33 45.1 & -2 16 02 & 26.599 & 20.10 & 15.38 & R2500I\\
    J1346$+$0842 & 3725064104059179904 & 13 46 07.2 & +8 42 33 & 23.339 & 20.47 & 15.74 & R2500I\\
    J1412$+$1633 & 1233008320961367296 & 14 12 24.5 & +16 33 10 & 31.278 & 18.67 & 13.89 & R300R \& R2500I\\
    J1421$+$1827 & 1239625559894563968 & 14 21 30.6 & +18 27 38 & 52.862 & 17.84 & 13.23 & R2500I\\
    J1439$+$0039 & \ldots & 14 39 15.1$\hyperlink{posrefs}{^1}$ & +0 39 42$\hyperlink{posrefs}{^1}$ & \ldots & \ldots & 18.00 & R300R\\
    J1441$-$0945 & 6326753222355787648 & 14 41 36.9 & -9 46 00 & 32.505 & 19.22 & 14.02 & R300R \& R2500I\\
    J1527$+$0553 & \ldots & 15 27 22.5$\hyperlink{posrefs}{^1}$ & +5 53 16$\hyperlink{posrefs}{^1}$ & \ldots & \ldots & 17.63 & R300R\\
    J1532$+$2611 & 1222514886931289088 & 15 32 23.3 & +26 11 19 & \ldots & 21.08 & 16.12 & R2500I\\
    J1539$-$0520 & 4400638923299410048 & 15 39 42.6 & -5 20 41 & 59.266 & 18.98 & 13.92 & R2500I\\
    J1548$-$1636 & 6260966349293260928 & 15 48 58.1 & -16 36 04 & 37.535 & 18.54 & 13.89 & R2500I\\
    J1617$+$7733B & 1704566318127301120 & 16 17 06.5 & +77 34 03 & 13.705 & 16.55 & 13.10 & R300R \& R2500I\\
    J1618$-$1321 & 4329787042547326592 & 16 18 44.9 & -13 21 31 & 21.86$\hyperlink{posrefs}{^5}$ & 19.34 & 14.25 & R2500I\\
    J1623$+$1530 & 4464934407627884800 & 16 23 21.8 & +15 30 39 & 10.301 & 20.59 & 15.94 & R2500I\\
    J1623$+$2908 & \ldots & 16 23 07.4$\hyperlink{posrefs}{^2}$ & +29 08 28$\hyperlink{posrefs}{^2}$ & \ldots & \ldots & 16.08 & R2500I\\
    J1705$-$0516 & 4364462551205872000 & 17 05 48.5 & -5 16 48 & 53.122 & 18.19 & 13.31 & R300R\\
    J1707$-$0138 & 4367890618008483968 & 17 07 25.3 & -1 38 10 & 25.976 & 19.25 & 14.29 & R300R \& R2500I\\
    J1717$+$6526 & 1633752714121739264 & 17 17 14.5 & +65 26 20 & 45.743 & 20.26 & 14.95 & R300R \& R2500I\\
    J1724$+$2336 & 4569300467950928768 & 17 24 37.4 & +23 36 50 & 14.625 & 20.19 & 15.68 & R300R\\
    J1733$-$1654 & 4124397553254685440 & 17 33 42.4 & -16 54 51 & 54.935 & 18.50 & 13.53 & R300R\\
    J1745$-$1640 & 4123874907297370240 & 17 45 34.8 & -16 40 56 & 50.918 & 18.44 & 13.65 & R2500I\\
    J1750$-$0016 & 4371611781971072768 & 17 50 24.4 & -0 16 12 & 108.581 & 18.29 & 13.29 & R2500I\\
    J2155$+$2345 & 1795137592033253888 & 21 55 58.6 & +23 45 30 & \ldots & 20.93 & 15.99 & R2500I\\
    J2339$+$3507 & 2873220249284763392 & 23 39 25.5 & +35 07 16 & 36.230 & 20.46 & 15.36 & R2500I\\
\end{longtable}
\twocolumn

We acquired 63 spectra in which we observed 53 unique objects, shown in Table~\ref{table:target_list}.
These 63 observations are shown in Table~\ref{table:obslog}, including the 
air mass and humidity of the observation.
Of the 63 spectra, 46 were observed with the R2500I volume phased holographic grating
(hereafter VPHG), whilst 17 were observed with the R300R grism.
Ten of the 53 objects were observed with both dispersive elements.

Twenty of the 53 objects already had full 6D positional and kinematic information in the literature.
Fifty-one had proper motions, 43 had parallaxes, and two had only $\alpha\ \text{and}\ \delta$.
All values along with their provenance are given in Table~\ref{table:target_list}.
In the next sub-sections we discuss the target list selection and observations.

\subsection{Target selection} \label{subsec:targetselection}

Our targets were drawn from a combination of two samples:
benchmark systems~\citep[system with a star and a UCD,][]{2006MNRAS.368.1281P}
and known L dwarfs with poor or no available spectroscopy.
The targets were selected by~\citet{2017MNRAS.470.4885M} and~\citet{GUCDS3},
and here we briefly summarise their selection criteria.
Both samples were chosen with the aim of gathering low- and mid-resolution spectra,
mostly to achieve radial velocities and to confirm their status as L dwarfs.
Benchmark system selection used the procedure of~\citet[][their section~4]{2017MNRAS.470.4885M}.
To summarise, primary systems consisting of possibly metal-rich or metal-poor stars were selected
with metallicity cuts of [Fe/H] $< -0.3$ and [Fe/H] $> 0.2$\,dex from a number of
catalogues~\citep[][their table~2]{2017MNRAS.470.4885M}.
If more than one value of [Fe/H] was available, the one with the smallest uncertainty was used; 
\citet{2017MNRAS.470.4885M} did not investigate if there were any systematic offsets between different catalogues,
as this was beyond the scope of that work.
The companions to these systems were filtered by a series of colour, absolute magnitude and photometric quality
cuts from 2MASS, SDSS~\citep[the Sloan Digital Sky Survey,][]{2000AJ....120.1579Y} and 
ULAS~\citep[United Kingdom Infrared Telescope Deep Sky Survey, Large Area Survey,][]{2007MNRAS.379.1599L}
photometry in equation~(\ref{eq:colourselect}).
These colour cuts in equation~(\ref{eq:colourselect}) are taken directly from~\citet{2017MNRAS.470.4885M}
as that work created part of the target list used in this work.
Magnitudes from 2MASS were converted into UKIRT/WFCAM magnitudes via the equations of~\citet{2004PASP..116....9S}.

\begin{align}
    & \label{eq:colourselect} Y - J > 0.85; \\
    & \nonumber J - H > 0.50; \\
    & \nonumber z - J > 2.1; \\
    & \nonumber \sigma_J < 0.1; \\
    & \nonumber [2.5 \times (z - J) + 4] < M_J < [5 \times (z - J) + 1]; \\
    & \nonumber M_J > 11.5; \\
    & \nonumber 1.6 < i - z < 6.0; \\
    & \nonumber 11.5 < M_z < [3.5714 \times (i - z) + 9.286]; \\
    & \nonumber M_z \geq 15; \\
    & \nonumber M_z \geq [3.5714 \times (i - z) + 6.5]; \\
    & \nonumber i - z \leq 2.1.
\end{align}

These companions were determined as being candidate benchmark systems with a maximum matching radius of 3\,arcmin,
i.e.\ the maximum separation to the primary object.
The remaining targets, known L dwarfs, were already spectroscopically confirmed bright L dwarfs that were predicted
to be visible to the astrometry and photometry in (at the time, upcoming) \gaia\ data releases.
These known L dwarfs should be single systems.
They would, however, not be bright enough for the \gaia\ radial velocity spectrometer~\citep{2023A&A...674A...5K},
and thus were chosen to determine radial velocities for, as a complement to the 30\,pc volume-limited sample.
This list was complemented with additional targets too dim for \gaia\ photometry and astrometry,
which were detected in UKIDSS, and by a few well-known L dwarfs, such as G~196--3B,
which could serve as template standards.

\subsubsection{Cross-matching} \label{subsubsec:crossmatch}
All observed targets (Table~\ref{table:target_list}) were cross-matched with \gaia, 2MASS, and AllWISE\@.
These surveys were chosen because they are all-sky and we were aiming for completeness in this process.
The targets were also cross-matched with Pan-STARRS (50/53 successful matches), for the additional optical components for those sources within
the Pan-STARRS footprint.
This sample of 53 objects was then also cross-matched against the astrophysical
parameter and \texttt{xp\_summary} tables from \gdrthree\footnote{
These tables are logically distinct from the main \gaia\ table in terms of schema and completeness.
}.
Thirty-eight of these objects had a \teffespucd\ value, and 28 had a public RP spectrum.
Internally calibrated \gaia\ RP spectra were then extracted from the \gaia\ archive with
a linearly dispersed grid from 6000\,$\angstrom$ to 10500\,$\angstrom$ using the
\texttt{gaiaxpy.convert}~\citep{daniela_ruz_mieres_2022_6674521}
and \texttt{gaiaxpy-batch}~\citep{Cooper_gaiaxpy-batch_2022} codes.
We also searched for common proper motion systems within Simbad~\citep{2000A&AS..143....9W} with
the selection criteria given in the GUCDS, specifically equation~(1) of~\citet{GUCDS3}:

\begin{align}
    & \label{eq:cpmselection} \rho < 100 \varpi; \\
    & \nonumber \Delta \varpi < \max[3 \sigma_{\varpi}, 1]; \\
    & \nonumber \Delta \mu < 0.1 \mu; \\
    & \nonumber \Delta \theta < 15\,\deg.
\end{align}

In equation~(\ref{eq:cpmselection}), $\rho$ is the separation in arcseconds, $\theta$ is the proper motion position
angle in degrees, whilst $\varpi$ (milli-arcseconds) and $\mu$ (milli-arcseconds per year)
are our target list's \gdrthree\ parallax and proper motion, respectively.
Like with the photometric selection, equation~(\ref{eq:colourselect}), the common proper motion selection was
taken directly from~\citet{GUCDS3}.
This is because the target list in this work is drawn from the same wider target list used in the GUCDS\@.
In effect, this selection is creating a widest possible physical separation 
of 100\,000\,AU~\citep[see the discussion on binding energies by][]{2009A&A...507..251C}.

\subsection{Observations} \label{subsec:observations}

The OSIRIS instrument used a $2 \times 1$
mosaic of 2048\,$\times$\,4096 pixel (photosensitive area) red-optimised CCDs (Marconi MAT-44-82 type)
with a $7.8 \times 7.8$\,arcmin$^2$ unvignetted field of view.
We used the standard operational mode of $2 \times 2$ binning, which has a
physical pixel size of 0.254\,arcsec\,pixel$^{-1}$.
For our purposes, we used the 7.4\,arcmin long slit with a width of 1.2\,arcsec.
We had variable seeing between 0.6 and 2.5\,arcsec,
with the vast majority having seeing ${<}1.2\text{--}1.5$.
The undersampling of the Full Width at Half-Maximum (FWHM) when the seeing is significantly less
than the slit width would cause uncertainty in the wavelength calibration.
In the worst cases, this can approach the resolution element.
This was then included in the systematic uncertainty estimate on the radial velocities.
We used the R300R and R2500I grisms and purely read off CCD 2
due to the instrument calibration module having a strong gradient from CCD 1 to 2 in the flat fields.
The R300R grism has a wavelength range of ${\approx} 4800 \text{--} 10\,000\,\angstrom$ with a dispersion of
${\approx} 7.74$\,\angstrom\,pix$^{-1}$ for a resolution of ${\approx} 350$ whilst
the R2500I VPHG has a wavelength range of ${\approx} 7330 \text{--} 10\,000\,\angstrom$ with a dispersion of
${\approx} 1.36$\,\angstrom\,pix$^{-1}$ for a resolution of ${\approx} 2500$, as per the
\href{http://www.gtc.iac.es/instruments/osiris/osiris.php#Longslit_Spectroscopy}{online documentation}\footnote{\url{http://www.gtc.iac.es/instruments/osiris/osiris.php##Longslit_Spectroscopy}}.
Both dispersive elements experience an increase in fringing at wavelengths ${\gtrsim}9200\,\angstrom$ to ${\geq}5$\,per cent.
The R300R grism however, had second order light from
$4800$ to $4900\,\angstrom$ contaminating the $9600$ to $9800\,\angstrom$ region.
This is because standards, but not UCDs, have flux in the blue regime, hence affecting the flux calibration
in the red regime.
As a result, the R300R spectra were conservatively truncated to $9000\,\angstrom$.
Our standards were a selection of white dwarfs plus two well-studied bright main sequence dwarf stars,
all with literature flux calibrated spectra and spectral types:
Ross~640~\citep[DZA6,][]{1974ApJS...27...21O, 2020MNRAS.499.1890M};
Hilt~600~\citep[B1,][]{1992PASP..104..533H, 1994PASP..106..566H};
GD~153~\citep[DA1,][]{1995AJ....110.1316B, 2014PASP..126..711B};
G191-B2B~\citep[DA1,][]{1990AJ.....99.1621O, 1995AJ....110.1316B, 2014PASP..126..711B};
GD~248~\citep[DC5,][]{2011ApJ...730..128T, 2020MNRAS.499.1890M},
GD~140~\citep[DA2,][]{2011ApJ...730..128T, 2020MNRAS.499.1890M}
and G~158-100~\citep[dG-K,][]{1990AJ.....99.1621O}.
We took a series of short exposures for the brightest objects to avoid saturation and non-linearity.
The majority of observations had a bright moon whilst the sky condition varied from photometric 
to clear with humidity typically ${\lesssim} 50$\,per cent.
All calibration frames were taken at the start and end of each night,
the arc lamps being used to solve the wavelength solution were:
Hg-Ar, Ne and Xe.
The full observing log is given in Table~\ref{table:obslog}.

\section{Data reduction} \label{sec:reduction}
We aimed to determine spectral types, spectral indices and radial velocities from directly measuring the GTC spectra.
Furthermore, we inferred astrophysical parameters (effective temperature, \teff\,[K];
surface gravity, \logg\,[dex];
and metallicity, [Fe/H]\,[dex]) from comparisons with atmospheric models.

Our adopted \href{https://github.com/pypeit/PypeIt}{\texttt{PypeIt}}\footnote{\url{https://github.com/pypeit/PypeIt}}~\citep{pypeit:zenodo,
    pypeit:joss_pub} reduction procedure applied to every object was as follows:
master calibration files were created by median stacking the relevant flat, bias and arc frames.
Basic image processing was performed including bias subtraction, flat fielding,
spatial flexure correction and cosmic ray masking via the L.A.\ Cosmic Rejection algorithm~\citep{2001PASP..113.1420V}.
We then manually identified the arc lines using the median stacked master arc.
These arc lines were used to manually create a wavelength solution through \texttt{pypeit\_identify} with
typical RMS values of ${\approx}0.0804\,\angstrom$ for the R2500I VPHG and
${\approx}0.1394\,\angstrom$ for the R300R grism.
The R2500I wavelength calibration solution was a 6\thdate order polynomial,
whilst the R300R solution was only 3\rd.
The information inside the object headers (observation date,
object sky position, longitude and latitude of the observatory)
was used to heliocentric correct the wavelength solution.
The \texttt{PypeIt} wavelength solution was defined in vacuum.

The standard frames were median stacked before the global sky was subtracted and corrected for spectral flexure
(to account for fringing).
Both the stacked standard and object were then extracted using both boxcar (5\,pixel)
and optimal~\citep{1986PASP...98..609H} extraction methods, with the latter being the presented spectra.

We then fitted a function to account for the sensitivity, CCD quantum efficiency and zeropoint.
The telluric regions listed by~\citet{2007AandA...473..245R} and~\citet{2015AandA...576A..77S} were masked out.
We divided each standard by its corresponding flux calibrated spectrum from the literature, as listed above.
This sensitivity function was then applied to the reduced standard and object to flux calibrate the extracted spectra.
If an observation had more than one science frame, those were co-added after wavelength and flux calibration.

The standards observed under the R2500I VPHG were used to create a telluric model from a high resolution atmospheric
grid derived at Las Campanas.
This grid was interpolated through to find the best match across airmass and precipitable water vapour.
The telluric model was applied back to the flux calibrated standard and object.
This telluric corrected standard was visually checked to confirm that the telluric model
was behaving appropriately.
The configuration files used in our reduction procedure are given in Appendix~\ref{sec:pypeit}.

It is important to mention here that we made a comparison between this \texttt{PypeIt} reduction and that
of a customised reduction (both the full basic image and spectral reductions) using standard \texttt{IRAF} tasks.
This was done with the aim of validating the quality of the \texttt{PypeIt} data against that
from a well proven reference source.
In Appendices~\ref{subsec:comparisonroutines} and~\ref{subsubsec:meth_val} we describe this procedure in detail
for one suitably chosen test object from our selection sample,
and which is common to both independent reductions: J1745$-$1640.

\begin{figure*}
    \centering
    \includegraphics[width=\linewidth]{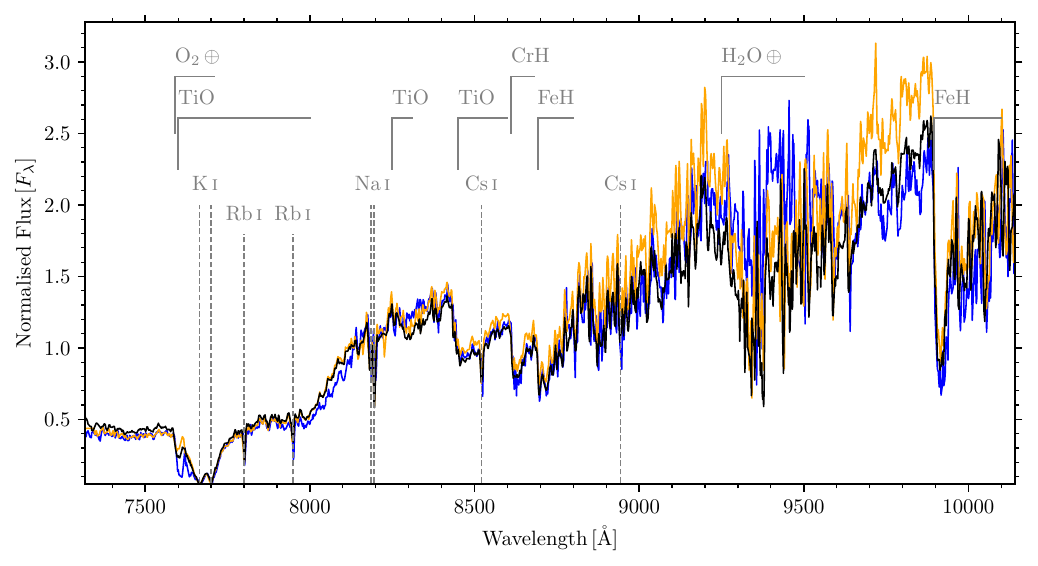}
    \caption{
        R2500I spectra for J1745$-$1640, normalised at $8100\text{--}8200\,\angstrom$,
        comparing two independent reduction procedures:
        \texttt{PypeIt} in black and \texttt{IRAF} in orange.
        In blue, the heliocentric corrected MagE spectra~\citep{2015ApJS..220...18B}
        for the same object is shown (which is not telluric corrected).
        The Earth symbol indicates the telluric bands present in the spectra.
    }
    \label{fig:j1745-1640-r2500i}
\end{figure*}

A comparison between the \texttt{PypeIt} reduction, and that which used standard \texttt{IRAF} routines,
is shown in the normalised spectra of J1745$-$1640 in Figure~\ref{fig:j1745-1640-r2500i}.
We show good agreement in the flux profile up to ${\sim}8900\,\angstrom$.
The \texttt{IRAF} reduced spectra is brighter in the broad H$_2$O region,
due to the differing telluric correction methods.
The MagE spectrum was not telluric corrected whilst the \texttt{IRAF} spectrum was telluric corrected using
a blackbody, instead of Ross~640 (the corresponding white dwarf standard).
This difference does not affect the model fitting of the spectra, as this is done in localised, small, chunks.
All spectra then agree at wavelengths ${\gtrapprox}9800\,\angstrom$.

\section{Analysis} \label{sec:gtcanalysis}
Here, we discuss the analysis of the reduced spectra, in order to produce spectral types, astrophysical parameters
and kinematics.
We discuss our measurements of astrophysical parameters first because the cross-correlation technique used to measure RV
requires the use of a best-fitting model derived template, obtained from the best fit of astrophysical parameters.
The code used for both estimating astrophysical parameters and calculating RV is
\texttt{rvfitter}~\citep{Cooper_rvfitter_2022}.
This program was developed to effectively recreate in \texttt{python} older codes (e.g.\ \texttt{IRAF.Fxcorr},
\texttt{IRAF.Splot}, \texttt{IDL.gaussfit}) designed for allowing a user to manually cross-correlate spectra
and fit line centres with different profiles.
All wavelengths discussed in this Section are in standard air, hence we converted our \texttt{PypeIt}
spectra from vacuum to air.
This was performed via the \texttt{specutils} package, using the corrections by~\citet{1953JOSA...43..339E}.

\subsection{Spectral typing} \label{subsec:spectypinganalysis}
We spectral typed both the R300R and R2500I spectra using the \texttt{classifyTemplate}
method of the \texttt{kastredux}~\citep{kastredux} package.
This compared each spectrum against SDSS standards~\citep{2007AJ....133..531B, 2010AJ....139.1808S,
    2017ApJS..230...16K},
from M0--T0, and selected the spectral type with the minimum difference in scaled
fluxes ($\Delta F$: equations~(\ref{eq:kasttype} -~\ref{eq:kastsf})) with equally weighted ($W$) points.

\begin{equation}
    \label{eq:kasttype} \Delta F = \sum \frac{W (F_{\mathrm{object}} - K F_{\mathrm{standard}})^2}
    {\sigma_{\mathrm{object}}^2}
\end{equation}

\begin{equation}
    \label{eq:kastsf} \left.K = \sum \frac{W F_{\mathrm{object}} F_{\mathrm{standard}}}{\sigma_{\mathrm{object}}^2} \middle/ \right.
    \sum \frac{W F_{\mathrm{standard}} F_{\mathrm{standard}}}{\sigma_{\mathrm{object}}^2}
\end{equation}

The spectra had all been smoothed in wavelength with a Gaussian $5\sigma$ kernel,
and we only compared the regions from $8000$ to $8500\,\angstrom$ for R2500I
and $7000$ to $8000\,\angstrom$ for R300R\@.
This was decided through experimentation, which deliberately excluded regions with telluric features,
as those features can cause poorer solutions.
Each object was also visually checked against known standards~\citep{1999ApJ...519..802K},
the spectral sub-types by which we refer to as `by eye'.
Any spectra with indicators of youth are given optical gravity classes as defined by~\citet{2009AJ....137.3345C},
from $\beta, \gamma, \delta$ in order of decreasing surface gravity.
The \texttt{kastredux} spectral types were our adopted spectral types.

\subsubsection{GTC spectral sequence} \label{subsubsec:spectralsequence}
\begin{figure*}
    \centering
    \includegraphics[width=\linewidth]{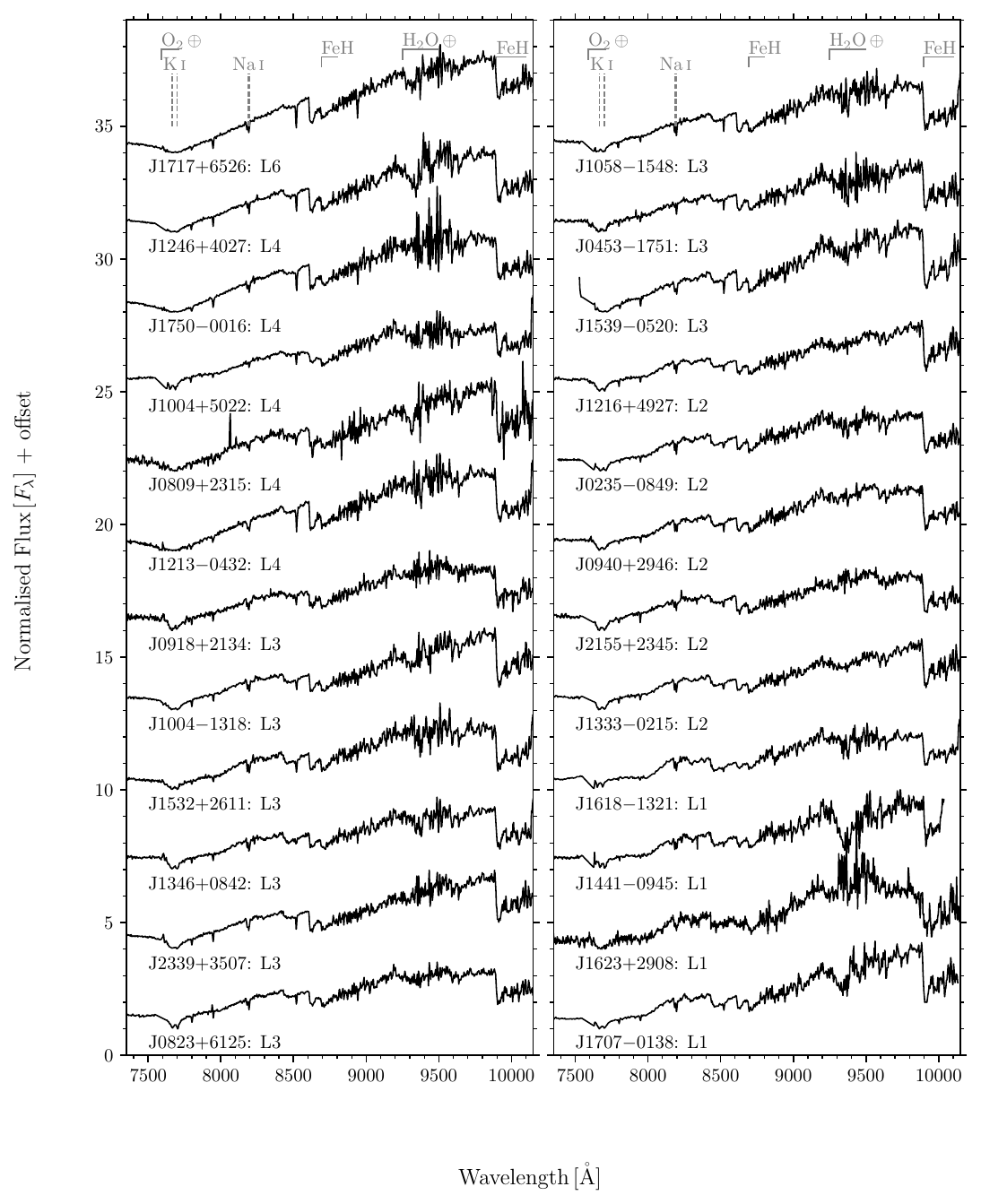}
    \caption{
        The first 24 of the R2500I VPHG spectra with a linear offset applied, sorted by spectral sub-type.
        We show the short names and the spectral sub-types from this work,
        attached to each spectrum.
        At the top of the figure are grey lines denoting a selection of spectral
        features typical to L dwarfs, plus the two main telluric bands.
    }
    \label{fig:r2500i-full-seq1}
\end{figure*}

\begin{figure*}
    \centering
    \includegraphics[width=\linewidth]{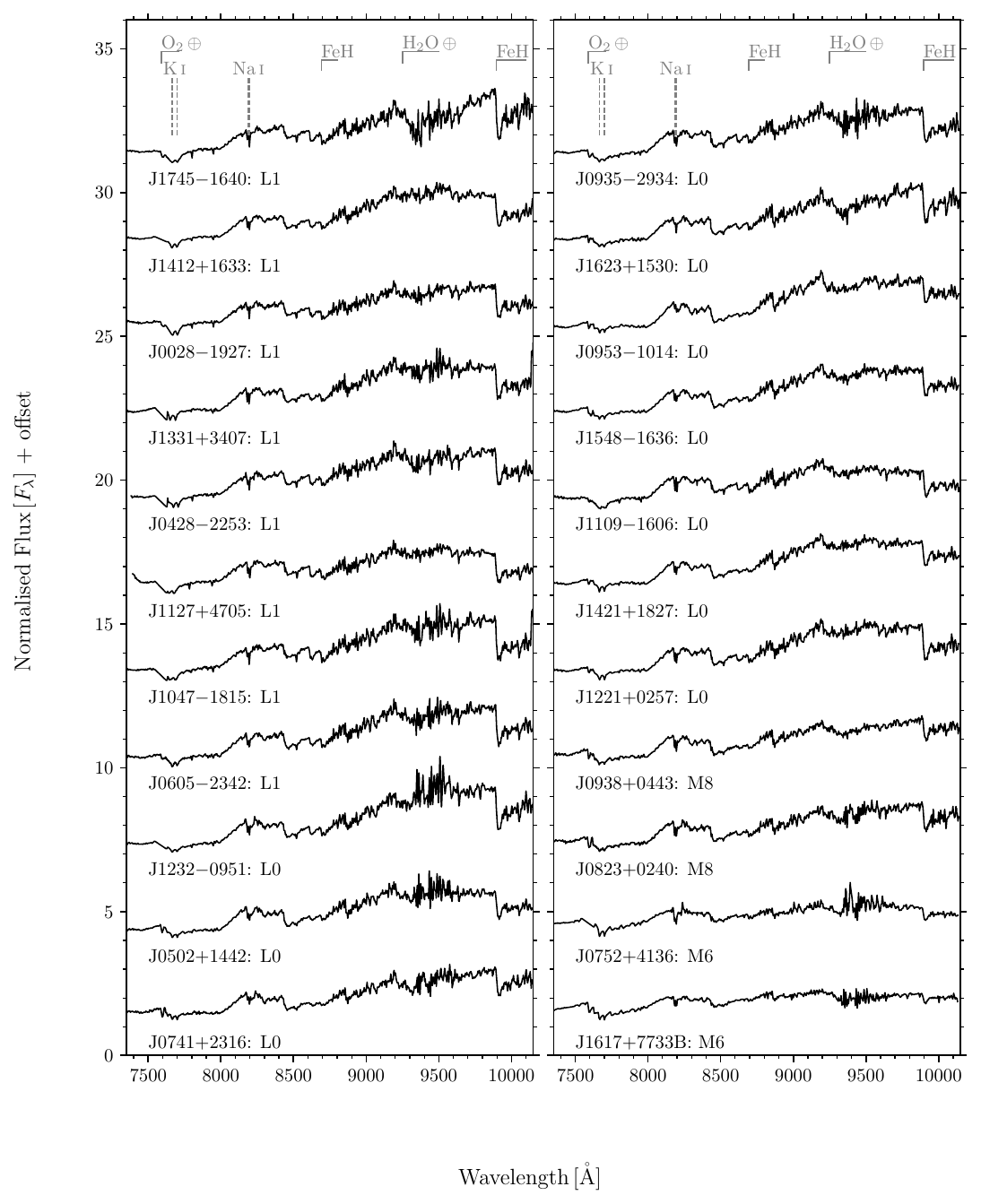}
    \caption{
        Same as Figure~\ref{fig:r2500i-full-seq1} but for the second half of the R2500I VPHG sample.
    }
    \label{fig:r2500i-full-seq2}
\end{figure*}

\begin{figure*}
    \centering
    \includegraphics[width=\linewidth]{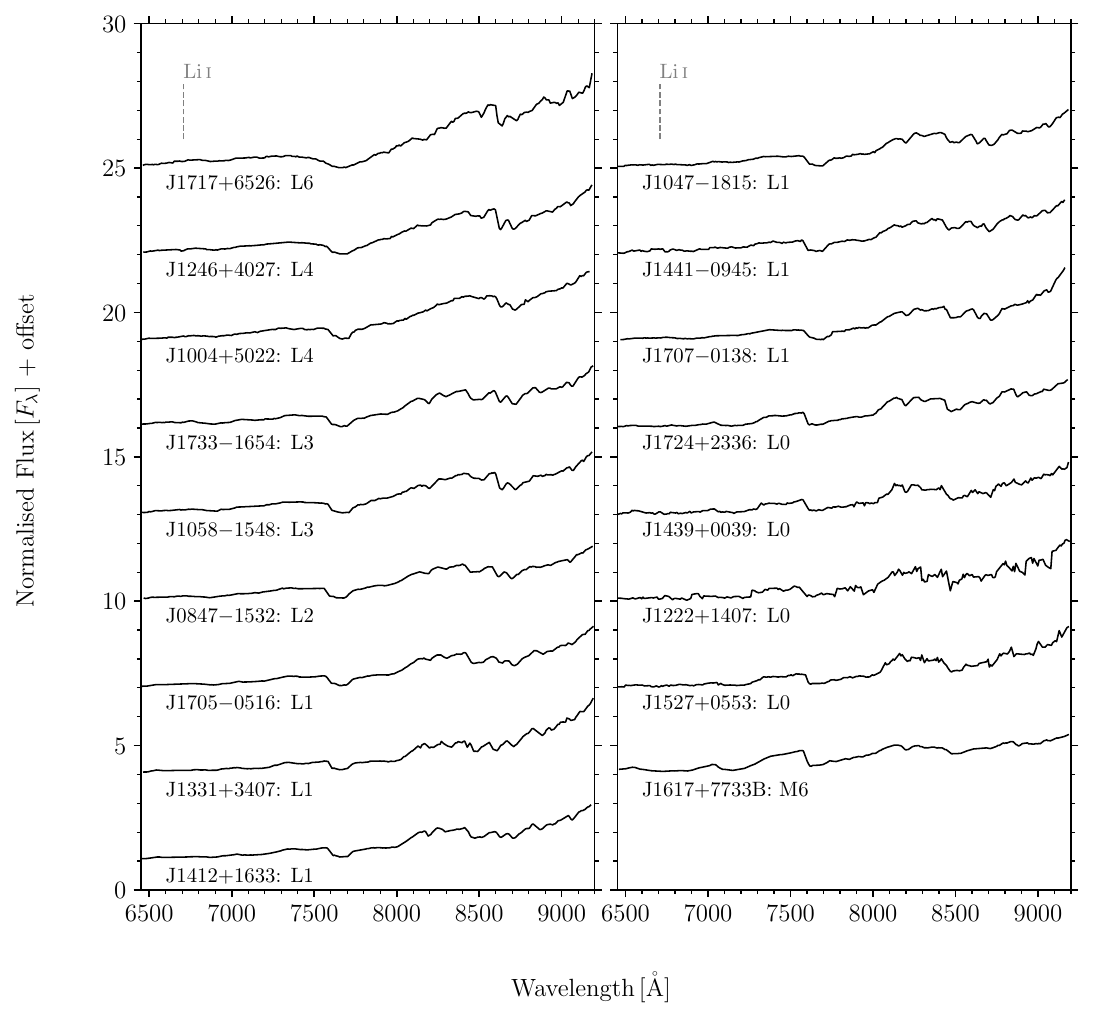}
    \caption{
        Same as Figure~\ref{fig:r2500i-full-seq1} but for the R300R grism spectra.
        Instead of the spectral features visible in Figures~\ref{fig:r2500i-full-seq1}
        and~\ref{fig:r2500i-full-seq2}, we only show where any lithium detection would be.
    }
    \label{fig:r300r-full-seq1}
\end{figure*}

The 46 spectra from the R2500I VPHG, ordered by our adopted spectral type,
are shown in Figures~\ref{fig:r2500i-full-seq1} and~\ref{fig:r2500i-full-seq2}.
All spectra are heliocentric corrected, such that the relative motion of the Earth has been removed.
Each spectrum shown had an outlier masking routine applied such that points
within a rolling ${\approx}15\,\angstrom$ (ten data points) chunk are removed
if they had a difference greater than the standard deviation from the median.
Additionally, some objects had problematic O$_2$ A-band tellurics.
In those cases, we interpolated over the region $7540 \text{--} 7630\,\angstrom$ from the maximum
of the first ${\approx}7.5\,\angstrom$ to minimum of the last ${\approx}7.5\,\angstrom$.
Where appropriate, spectra were co-added.
All spectra appear noisy in the primary H$_2$O band of ${\approx} 9200 \text{--} 9600\,\angstrom$.
The 17 heliocentric corrected, reduced spectra from the R300R grism
are shown in Figure~\ref{fig:r300r-full-seq1}.
The R300R spectra were trimmed from $6500 < \lambda < 9000\,\angstrom$ due to (a) the lack of
signal in the blue regime and (b) to constrain to purely the first order light.
Unlike the R2500I spectra, the R300R spectra were not telluric corrected.

\subsection{Fundamental astrophysical parameters} \label{subsec:astrophysical_parametersanalysis}
We used the \texttt{rvfitter.crosscorrelate} code on our R300R and R2500I spectra with
BT-Settl CIFIST model grids from $1200 \leq \teff \leq 4000$\,K and $4.5 \leq \logg \leq 5.5$\,dex~\citep{2011ASPC..448...91A}.
Lower surface gravity grids were available but not routinely used as the focus was on RV measurement with
an a priori expectation of field surface gravity, ${\approx}5$\,dex.
These models assume a solar metallicity with no variation and are linearly dispersed in steps
of $100$\,K and $0.5$\,dex.
This code allowed us to visually select the best fitting model from the array of model grids and
for each spectral line from Table~\ref{table:spectrallines}.

\begin{table}
    \centering
    \begin{tabular}{lc}
        \hline
        \noalign{\smallskip}
        Line & $\lambda$\,[$\angstrom$] \\ 
        \hline
        K\,\textsc{i}-a & 7664.8991 \\
        K\,\textsc{i}-b & 7698.9646 \\
        Rb\,\textsc{i}-a & 7800.268 \\
        Rb\,\textsc{i}-b & 7947.603 \\
        Na\,\textsc{i}-a & 8183.2556 \\
        Na\,\textsc{i}-b & 8194.824 \\
        Cs\,\textsc{i}-a & 8521.13165 \\
        Cs\,\textsc{i}-b & 8943.47424 \\
        \hline
    \end{tabular}
    \caption{
        The list of atomic alkali metal lines used when estimating astrophysical parameters
        and calculating radial velocities.
        Wavelengths are as measured by~\citet{nistdb} and are defined in standard air.
    }
    \label{table:spectrallines}
\end{table}

We used these chosen lines rather than correlating against the entire model because the models
do not exactly match the flux profile of ground based spectra.
It was also known that the BT-Settl grids were generated using a different line list to our selected alkali
lines, taken from the NIST database~\citep{nistdb}.
For efficiency purposes, each model when being loaded into the code, was interpolated onto the wavelength
array of the object being compared against.
The models could optionally be Gaussian smoothed, which was helpful for fitting any `messy' regions of models
(e.g.\ telluric bands in models with $\teff\ \gtrapprox 2000$\,K).
We normalised the model and data by their respective medians in a given variably sized
`chunk' around each spectral line.
We noted that around certain lines, particular models appeared almost identical to each other,
e.g.\ around 7000--8000$\,\angstrom$,
the $1900$ and $2000$\,K models are not visually distinct.
This means there is a higher uncertainty for effective temperatures within the 1900--2000\,K region.
Not every spectral line was used for each object as some have poorly resolved features or low signal-to-noise.
Our selected \teff\ was the mean \teff\ from each line measurement, as was \logg.
To determine the error on each \teff\ and \logg\ final value, we chose to use the standard deviation
from each independent line fit divided by square root of the number of lines used.
This error was added in quadrature with half of the separation between each grid, i.e.\ $50$\,K for \teff\ and
$0.25$\,dex for \logg.

Additionally, we created an `expected' effective temperature, \teffexpect, using the Filippazzo, sixth order 
field \teff\ relation~\citep{2015ApJ...810..158F} and our adopted spectral types.
The errors on \teffexpect\ correspond with the mean difference in \teff\ across $\pm 0.5$ spectral sub-types
(our spectral sub-type uncertainty), plus the quoted relation RMS of 113\,K\@.

\subsection{Calculating the radial velocities} \label{subsec:kinematicsanalysis}

Only our R2500I spectra were used to determine RVs as the features in R300R spectra
are mostly blended/unresolved.
We used two methods by which to measure an adopted RV: line centre fitting and cross correlation.
We note that our seeing (Table~\ref{table:obslog}, corrected for airmass) was almost always smaller than the slit width,
which affects the RV offset as the slit is not fully illuminated.
The full width at half-maximum was typically 3--4\,pixels, corresponding to ${\approx}0.75\text{--}1$\,arcseconds.
Most observations were seeing-limited, whilst a few, taken in poorer conditions, were slit-limited.
The following methods were performed only on heliocentric corrected spectra, hence any quoted RV values are
heliocentric corrected.

\subsubsection{Line centre fitting} \label{subsubsec:linecentrefitting}
\begin{figure*}
    \centering
    \includegraphics[width=\linewidth]{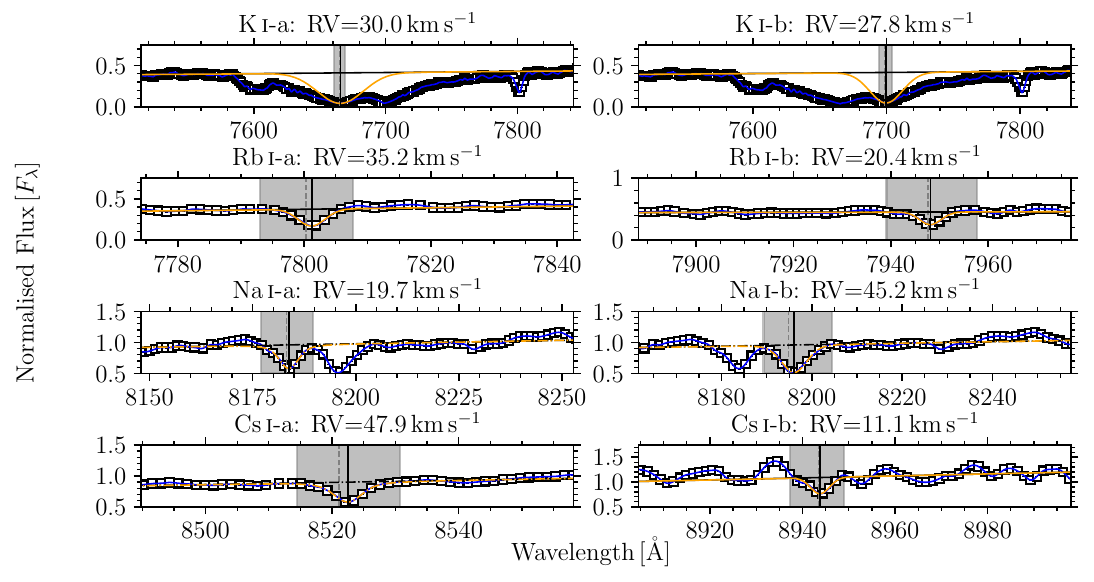}
    \caption{
        J1745$-$1640 RV calculation via different line profiles
        (orange: solid -- Gaussian; dash-dot -- Voigt) against the data
        (black squares) and fifth order spline fit (blue) in the regime around the eight listed line centres.
        The flux uncertainty is smaller than the height of each square.
        The shift from the laboratory line position (vertical dashed grey line) is shown as the
        vertical solid black line.
        The horizonal black line (solid or dash-dot, depending on the fitted line profile as above) is the continuum,
        as is subtracted from the data.
        A grey band is given, corresponding to the region of data the line profiles are fitted to.
        The shown region is between the inner edges of the continuum regions.
    }
    \label{fig:j1745-1640-rv}
\end{figure*}

Using the same atomic absorption lines listed in Table~\ref{table:spectrallines},
we applied the \texttt{rvfitter.linecentering} code to interactively fit Gaussian, Lorentzian and Voigt
profiles with the minimum possible width.
This minimum possible width is equal to the number of free parameters plus one
(although this does not guarantee a successful fit).
We used these different profiles to obtain the best fit for a particular line given its underlying absorption
characteristics and the available signal-to-noise of the spectral region.
The fitting technique used was
\href{https://docs.scipy.org/doc/scipy/reference/generated/scipy.optimize.leastsq.html#scipy.optimize.leastsq}
{least-mean-square}\footnote{\url{https://docs.scipy.org/doc/scipy/reference/generated/scipy.optimize.leastsq.html##scipy.optimize.leastsq}} minimisation.
For each spectral line, we subtracted a linear continuum from the data.
The continuum corresponds to the medians of selected regions to the blueward and redward sides of the spectral line.
Each continuum region is chosen to follow the shape of the spectra with a minimum width of ${\approx}50\,\angstrom$
within 100--200\,$\angstrom$ of the spectral line.
Also shown during the fitting routine is a fifth order spline, as a visual aid;
the minima of the spline does not necessarily correspond to the line position.
A example of this routine is given for J1745$-$1640 in Figure~\ref{fig:j1745-1640-rv}.
The fits were only accepted if they appeared to accurately represent the spectral lines profile upon visual inspection.
In general, the most consistently reliable lines were the rubidium lines, sodium doublet and first caesium line.
The potassium doublet often was affected by rotational broadening whilst the second caesium line
was often affected by neighbouring features.
The uncertainty for each line, was the value in the diagonal of the covariance matrix corresponding
to centroid position from the least-squares fit, plus the wavelength calibration RMS for that object,
Doppler shifted into RV space.

After measuring every line, we then calculated the overall weighted mean ($\mu_{\text{LC}}$) and
weighted standard deviation ($\sigma_{\text{LC}}$),
the weights were the inverse of the uncertainties of each line used, squared.
The uncertainty from the vacuum to air conversion was negligible (${\ll}0.1$\,\kms) compared to the
fitting uncertainties calculated from the eight (or less, if rejected) aforementioned lines.
The final line centre RV standard error was the weighted standard deviation divided
by the square root of the number of lines fit.

\subsubsection{Cross-correlation} \label{subsubsec:xcorr}
\begin{figure*}
    \centering
    \includegraphics[width=\linewidth]{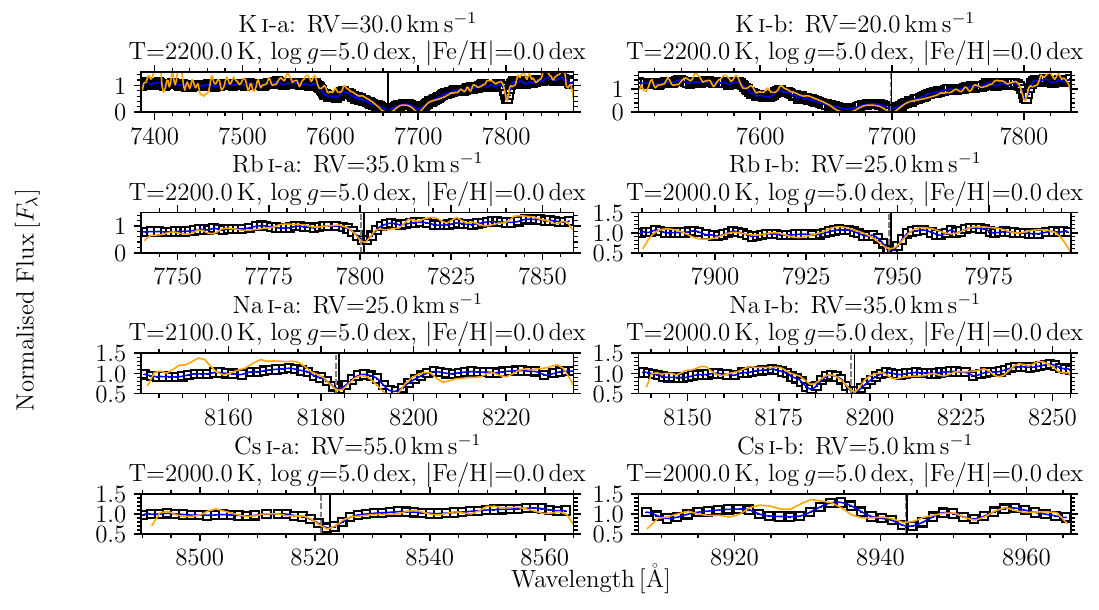}
    \caption{
        J1745$-$1640 RV calculation via the manually shifted BT-Settl model (orange)
        against the data (black squares) and fifth order spline fit (blue).
        The flux uncertainty is smaller than the height of each square.
        The laboratory line position (vertical dashed grey line) has been manually shifted by the RV given
        on the sub-plot title (vertical solid black line).
        Effective temperature, gravity and metallicity are also indicated on
        each features title.
    }
    \label{fig:xcorr_fit}
\end{figure*}

In addition to estimating the astrophysical parameters with \texttt{rvfitter.crosscorrelate} in
Section~$\S$\ref{subsec:astrophysical_parametersanalysis}, we also used the same package to measure RV by
manually shifting the best fitting BT-Settl model as a template.
No smoothing was applied to the model template to match the spectral resolution of the object spectrum.
This was because smoothing could confuse where the centroid of a line was, when looking by-eye.
Likewise, there was no continuum subtraction applied to the object spectrum.
The RV shift was in steps of $5,\ 10,\ 100$\,\kms, which in turn defined the RV uncertainty on each line
($2.5,\ 5,\ 50$\,\kms, i.e.\ the margin of error).
These RV errors are added to the wavelength calibration RMS for the given object (Doppler shifted into an RV error).
Not all atomic lines were always used, only in the cases where the model appeared to closely match
the apparent line profile.
The typical technique was to select a broad region ($\Delta \lambda = 100\text{--}200\,\angstrom$)
around each spectral line, find the best fitting template in terms of \teff\ and \logg,
then narrow that region ($\Delta \lambda \approx 50\,\angstrom$) to then find an RV\@.
This was a predominantly by-eye technique, although root-mean-square deviation divided by interquartile range
(RMSDIQR) values were computed as a numerical guide when comparing models.
We also show a fifth order spline, as with the line centering method, as a visual aid.
This initial broad region is shown for J1745$-$1640 in Figure~\ref{fig:xcorr_fit}.

As in Section~$\S$\ref{subsubsec:linecentrefitting}, the overall cross-correlated weighted mean RV value ($\mu_{\text{XC}}$)
and weighted standard deviation ($\sigma_{\text{XC}}$) was calculated using all of the manually selected lines.
We used the same method to estimate the uncertainty in final cross-correlation derived RVs as for
the line centre results, by finding the standard error of the mean.

\subsubsection{Adopted RV} \label{subsubsec:final_rv}
\begin{figure}
    \centering
    \includegraphics[width=\linewidth]{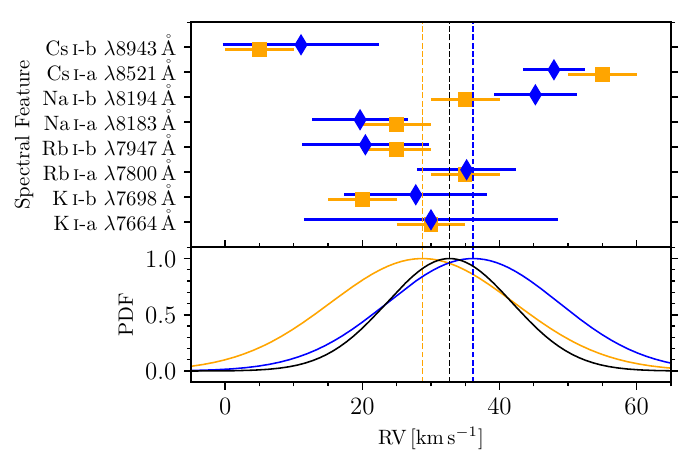}
    \caption{
        J1745$-$1640 RV values for each given line.
        In the top panel, orange squares are cross-correlated RVs, blue diamonds are line centre RVs;
        each spectral feature has been indicated on the $y$ axis.
        In the bottom panel, the orange curve is the cross-correlated PDF;
        the blue curve is the line centre PDF;
        and the black curve is the adopted PDF.
        The dotted vertical lines are the mean RV values as associated with each PDF.
    }
    \label{fig:adoptedrv}
\end{figure}

We created an adopted RV by constructing a weighted mean, using the deviation in each method as the weighting.
The different RV values for each line, method and the corresponding probability distribution functions (PDFs) are
shown in Figure~\ref{fig:adoptedrv}, for J1745$-$1640.
We also note that our final adopted RV for J1745$-$1640 obtained from combining the results of the two
measurement techniques ($32.7\pm6.5$\,\kms) is in agreement with the values obtained from both the customised IRAF reduced
data and the value reported by~\citet{2015ApJS..220...18B}, within their respective uncertainties.
See Appendix~\ref{subsubsec:meth_val} for a full description.

The adopted RV was the mean ($\mu_{\text{RV}}$) whilst the standard error ($\delta_{\text{RV}}$)
was equal to the standard deviation ($\sigma_{\text{RV}}$) divided by $\sqrt{2}$.
The mean and standard deviation was calculated through the inverse variance
weighting equations~(\ref{eq:muposterior} and~\ref{eq:sigposterior}).
Typically, we found that the cross-correlation technique was more precise (being more controlled by-eye)
and robust.
The line centre fitting was often more accurate, however, and performed best on the higher quality spectra.

\begin{equation}
    \label{eq:muposterior} \mu_{\text{RV}} = \frac{\mu_{\text{LC}} \sigma_{\text{XC}}^2 +
    \mu_{\text{XC}} \sigma_{\text{LC}}^2}{\sigma_{\text{LC}}^2 + \sigma_{\text{XC}}^2}
\end{equation}
\begin{equation}
    \label{eq:sigposterior} \sigma_{\text{RV}} = \sqrt{\frac{\sigma_{\text{LC}}^2
    \sigma_{\text{XC}}^2}{\sigma_{\text{LC}}^2 + \sigma_{\text{XC}}^2}}
\end{equation}

\subsection{Kinematics} \label{subsec:movinggroupsanalysis}
Galactic UVW velocities were calculated using our adopted RVs plus \gaia\ astrometric measurements,
using the equations from \texttt{\hyperlink{https://github.com/segasai/astrolibpy}{astrolibpy}}.
We corrected for the Local Standard of Rest (LSR) using the values from~\citet{2011MNRAS.412.1237C}
where U, V, W = ($-8.50, +13.38, +6.49$)\,\kms.
These equations follow the work by~\citet{1987AJ.....93..864J}, except that U is orientated towards the
Galactic anti-centre.
We also used \banyan~\citep{2015ApJS..219...33G, 2018ApJ...856...23G}, which provided moving group
classification with associated probability.
When using \banyan, we checked the resultant probabilities both with and without RV\@.
This was because RV has by far the lowest precision, thus could reduce a likely membership candidate into
a field object in error.
Our final values are the ones which include RV\@.
Notably, when using velocities in the Galactic reference frame, one can select a Galactic component with
\vtotal\ (where $\vtotal$ is the total space velocity,
$\vtotal = \sqrt {U^2 + V^2 + W^2}$).
We followed the work by~\citet{2010A&A...511L..10N} and define thick disc and halo objects as having
$\vtotal > 70\,\kms$ and $\vtotal > 180\,\kms$ respectively.
This definition, especially for separating the thin and thick disc, is indicative of metallicity;
see the Besan\c{c}on Galaxy models~\citep{2014A&A...564A.102C, 2021A&A...654A..13L}.

\section{Results} \label{sec:gtcresults}
In this Section, we present the spectral types, radial velocities and astrophysical parameters.
In Table~\ref{table:photometry}, we provide photometry from the \gaia, 2MASS and ALLWISE catalogues.
We discuss individually interesting objects and objects where
our measured results differ significantly from published values.

\subsection{Spectral types} \label{subsec:spectypingresults}
In Table~\ref{table:our_spts} we list published spectral types based on optical spectra,
near-infrared spectra and the `by eye' and \texttt{kastredux}
methods discussed in Section~$\S$\ref{subsec:spectypinganalysis}.
This work has produced the first spectral type estimates for six of the 53 objects.

\setlength{\tabcolsep}{3pt}
\begin{table*}
\centering
\caption{
Our spectral types compared with the literature optical and near-infrared types for each object.
}
\label{table:our_spts}
\begin{tabular}{l llll | l llll}
    \hline
    Object     & Lit Opt  & Lit NIR   &  By eye   & \texttt{kastredux} & 
    Object     & Lit Opt  & Lit NIR   &  By eye   & \texttt{kastredux} \\
    short name & sp. type & sp. type  &  sp. type &     sp. type        &
    short name & sp. type & sp. type  &  sp. type &     sp. type        \\
    \hline
J0028$-$1927 & L0:$\hyperlink{litspts}{^{1}}$ & L0.5$\hyperlink{litspts}{^{2}}$ & L0.5 & L1& J0235$-$0849 & L2$\hyperlink{litspts}{^{3}}$ & L2:$\hyperlink{litspts}{^{2}}$ & L2 & L2\\
    J0428$-$2253 & L0.5$\hyperlink{litspts}{^{4}}$ & L0$\hyperlink{litspts}{^{2}}$ & L0.5 & L1& J0453$-$1751 & L3:$\hyperlink{litspts}{^{5}}$ & L3$\hyperlink{litspts}{^{2}}$ & L3$\beta$ & L3\\
    J0502$+$1442 & L0$\hyperlink{litspts}{^{6}}$ & M9$\hyperlink{litspts}{^{2}}$ & M9$\beta$ & L0& J0605$-$2342 & L0:$\hyperlink{litspts}{^{7}}$ & L1:$\hyperlink{litspts}{^{2}}$ & L0.5 & L1\\
    J0741$+$2316 & L1$\hyperlink{litspts}{^{8}}$ & \ldots & L0 & L0& J0752$+$4136 & M7$\hyperlink{litspts}{^{9}}$ & \ldots & M6 & M6\\
    J0809$+$2315 & \ldots & \ldots & L4: & L4& J0823$+$0240 & \ldots & \ldots & M9 & M8\\
    J0823$+$6125 & L2:$\hyperlink{litspts}{^{1}}$ & L2.5$\hyperlink{litspts}{^{2}}$ & L3 & L3& J0847$-$1532 & L2$\hyperlink{litspts}{^{5}}$ & \ldots & L2 & L2\\
    J0918$+$2134 & L2.5$\hyperlink{litspts}{^{10}}$ & L2.5$\hyperlink{litspts}{^{2}}$ & L3 & L3& J0935$-$2934 & L0$\hyperlink{litspts}{^{1}}$ & L0.5$\hyperlink{litspts}{^{2}}$ & L0 & L0\\
    J0938$+$0443 & L0$\hyperlink{litspts}{^{6}}$ & M8$\hyperlink{litspts}{^{2}}$ & M9 & M8& J0940$+$2946 & L1$\hyperlink{litspts}{^{6}}$ & L0.5$\hyperlink{litspts}{^{2}}$ & \ldots & L2\\
    J0953$-$1014 & L0$\hyperlink{litspts}{^{7}}$ & M9.5$\hyperlink{litspts}{^{2}}$ & M9.5$\beta$ & L0& J1004$+$5022 & L3Vl-G$\hyperlink{litspts}{^{11}}$ & L3Int-G$\hyperlink{litspts}{^{12}}$ & L3$\beta$ & L4\\
    J1004$-$1318 & L0$\hyperlink{litspts}{^{13}}$ & L1:$\hyperlink{litspts}{^{14}}$ & L3.5$\beta$ & L3& J1047$-$1815 & L2.5$\hyperlink{litspts}{^{15}}$ & L0.5$\hyperlink{litspts}{^{2}}$ & L1 & L1\\
    J1058$-$1548 & L3$\hyperlink{litspts}{^{10}}$ & L3$\hyperlink{litspts}{^{16}}$ & L3$\beta$ & L3& J1109$-$1606 & L0$\hyperlink{litspts}{^{6}}$ & \ldots & L1 & L0\\
    J1127$+$4705 & L1$\hyperlink{litspts}{^{6}}$ & \ldots & L1 & L1& J1213$-$0432 & L5$\hyperlink{litspts}{^{5}}$ & L4$\hyperlink{litspts}{^{2}}$ & L5$\beta$ & L4\\
    J1216$+$4927 & L1$\hyperlink{litspts}{^{6}}$ & \ldots & L2: & L2& J1221$+$0257 & L0.5$\hyperlink{litspts}{^{17}}$ & M9p$\hyperlink{litspts}{^{18}}$ & M9.5 & L0\\
    J1222$+$1407 & M9$\hyperlink{litspts}{^{8}}$ & \ldots & M9:: & L0& J1232$-$0951 & L0$\hyperlink{litspts}{^{1}}$ & M9.5$\hyperlink{litspts}{^{2}}$ & M9.5 & L0\\
    J1246$+$4027 & L4$\hyperlink{litspts}{^{19}}$ & L4$\hyperlink{litspts}{^{2}}$ & L4 w/ Li & L4& J1331$+$3407 & L0$\hyperlink{litspts}{^{1}}$ & L1p(red)$\hyperlink{litspts}{^{20}}$ & L0 & L1\\
    J1333$-$0215 & L3$\hyperlink{litspts}{^{6}}$ & L2$\hyperlink{litspts}{^{2}}$ & \ldots & L2& J1346$+$0842 & L2$\hyperlink{litspts}{^{6}}$ & \ldots & L2.5 & L3\\
    J1412$+$1633 & L0.5$\hyperlink{litspts}{^{19}}$ & L0$\hyperlink{litspts}{^{2}}$ & L0 & L1& J1421$+$1827 & L0$\hyperlink{litspts}{^{1}}$ & M9$\hyperlink{litspts}{^{2}}$ & M9.5 & L0\\
    J1439$+$0039 & \ldots & \ldots & \ldots & L0& J1441$-$0945 & L0.5$\hyperlink{litspts}{^{11}}$ & L0.5$\hyperlink{litspts}{^{2}}$ & L0.5 & L1\\
    J1527$+$0553 & \ldots & \ldots & \ldots & L0& J1532$+$2611 & L1$\hyperlink{litspts}{^{6}}$ & \ldots & \ldots & L3\\
    J1539$-$0520 & L4:$\hyperlink{litspts}{^{11}}$ & L2$\hyperlink{litspts}{^{21}}$ & L4.5 & L3& J1548$-$1636 & \ldots & L2:$\hyperlink{litspts}{^{22}}$ & M9.5 & L0\\
    J1617$+$7733B & \ldots & \ldots & \ldots & M6& J1618$-$1321 & L0:$\hyperlink{litspts}{^{11}}$ & M9.5$\hyperlink{litspts}{^{2}}$ & L0 & L1\\
    J1623$+$1530 & L0$\hyperlink{litspts}{^{6}}$ & \ldots & M9 & L0& J1623$+$2908 & L1$\hyperlink{litspts}{^{6}}$ & \ldots & L1:: & L1\\
    J1705$-$0516 & L0.5$\hyperlink{litspts}{^{1}}$ & L1$\hyperlink{litspts}{^{12}}$ & L1 & L1& J1707$-$0138 & L0.5$\hyperlink{litspts}{^{13}}$ & L2$\hyperlink{litspts}{^{23}}$ & L1 & L1\\
    J1717$+$6526 & L4$\hyperlink{litspts}{^{3}}$ & L6$\hyperlink{litspts}{^{2}}$ & L6 & L5& J1724$+$2336 & \ldots & \ldots & \ldots & L0\\
    J1733$-$1654 & L0.5:$\hyperlink{litspts}{^{24}}$ & L1$\hyperlink{litspts}{^{2}}$ & L2 & L3& J1745$-$1640 & L1.5:$\hyperlink{litspts}{^{24}}$ & L1.5$\hyperlink{litspts}{^{2}}$ & L1 & L1\\
    J1750$-$0016 & \ldots & L5.5$\hyperlink{litspts}{^{22}}$ & L5.5 & L4& J2155$+$2345 & \ldots & L2$\hyperlink{litspts}{^{20}}$ & L3 & L2\\
    J2339$+$3507 & L3.5$\hyperlink{litspts}{^{1}}$ & \ldots & L3.5 & L3&  &  &  &  &  \\ 

    \hline
\end{tabular}
\newline
\hypertarget{litspts}{Literature Spectral Types}: 1.\ ~\citet{2008AJ....136.1290R}, 2.\ ~\citet{2014ApJ...794..143B}, 3.\ ~\citet{2002AJ....123.3409H}, 4.\ ~\citet{2003A&A...403..929K}, 5.\ ~\citet{2003AJ....126.2421C}, 6.\ ~\citet{2010AJ....139.1808S}, 7.\ ~\citet{2007AJ....133..439C}, 8.\ ~\citet{2017MNRAS.470.4885M}, 9.\ ~\citet{2011AJ....141...97W}, 10.\ ~\citet{1999ApJ...519..802K}, 11.\ ~\citet{2008ApJ...689.1295K}, 12.\ ~\citet{2013ApJ...772...79A}, 13.\ ~\citet{2010A&A...517A..53M}, 14.\ ~\citet{2013AJ....146..161M}, 15.\ ~\citet{1999AJ....118.2466M}, 16.\ ~\citet{2004AJ....127.3553K}, 17.\ ~\citet{2014AJ....147...34S}, 18.\ ~\citet{2015ApJS..219...33G}, 19.\ ~\citet{2000AJ....120..447K}, 20.\ ~\citet{2010ApJS..190..100K}, 21.\ ~\citet{2004A&A...416L..17K}, 22.\ ~\citet{2007A&A...466.1059K}, 23.\ ~\citet{2011ApJ...735...14P}, 24.\ ~\citet{2008MNRAS.383..831P}.

The `:' after a spectral type indicates uncertainty of $\pm 1$ whilst `p' indicates peculiarity. 
The surface gravity flag $\beta$ is given when appropriate,
 and is discussed in Section~$\S$\ref{subsubsec:individuals}.
The adopted spectral type is the \texttt{kastredux} method,
 only overwritten where there are gravity flags in the `by eye' method.
In addition, J1246$+$4027 has been typed as having a potential Li\,\textsc{i} detection ($\lambda6708\,\angstrom$),
 which can only be seen in the R300R spectra.

\end{table*}

\begin{figure}
    \centering
    \includegraphics[width=\linewidth]{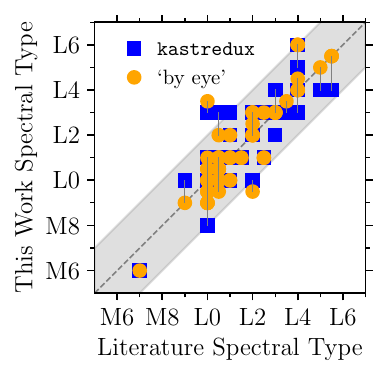}
    \caption{
        Comparison between this works spectral types and the literature spectral types.
        Blue squares are spectral types from our adopted, \texttt{kastredux} method whilst orange circles
        are from the manual `by eye' method.
        Grey lines connect these two methods and we show a one-to-one dashed grey line with associated $\pm2$
        spectral sub-types confidence bands.
    }
    \label{fig:spectypes}
\end{figure}

The 47 objects with known spectral types have a standard deviation of 0.5 sub-types between
the published values and the `by eye'/\texttt{kastredux} results, which we adopt as the error on the new
spectral sub-types.
When the literature values for a given object differ we adopted the optical spectral type.
Our spectral types across the two methods are displayed against the adopted literature spectral types
in Figure~\ref{fig:spectypes}.

All objects except J1004$-$1318 have sub-type differences between the spectral type derived in this work
and the adopted literature spectral type of less than two sub-types.
J1004$-$1318, has an optical (Opt) spectral sub-type of L0~\citep{2010A&A...517A..53M}
whilst~\citet{2013AJ....146..161M} found a sub-type of L1 using near-infrared (NIR) spectra;
we find a sub-type of L3.
However, a more recent study,~\citet{2016ApJ...830..144R}, found a sub-type of L4 (NIR),
which is more consistent with our result.
The fit statistic from \texttt{kastredux} is about twice larger for L1 than for L3.
In Figure~\ref{fig:r2500i-full-seq1}, J1004$-$1318 does not seem dissimilar to the neighbouring objects,
whereas the L0/L1 spectra appear different (e.g.\ weaker alkali lines).
The different spectral typing of J1004$-$1318 may be due to lower signal-to-noise (S/N) ratios of some observations.
For example,~\citet{2010A&A...517A..53M} exposed for 2400\,s at the 2.56\,m Nordic Optical Telescope,
while we exposed for 1500\,s, and with moderately good seeing and low airmass,
with a telescope with an aperture over 16 times larger.

\subsection{Radial velocity analysis} \label{subsec:kinematicsresults}

\begin{figure*}
    \centering
    \includegraphics[width=\linewidth]{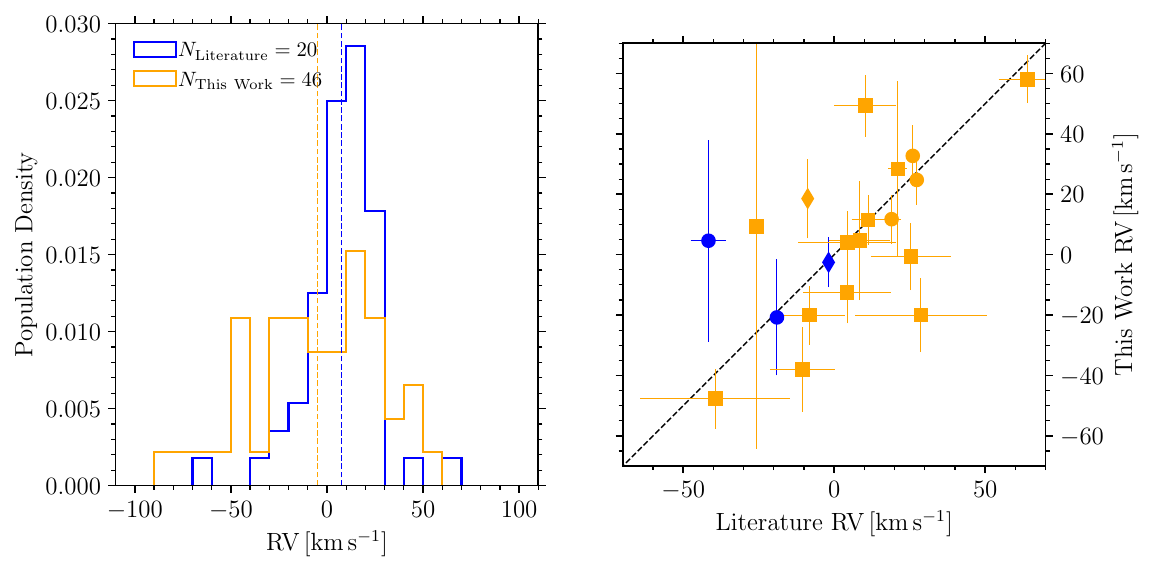}
    \caption{
        [Left Panel]: Histograms of the RVs calculated in this work (orange)
        and from the literature (blue) to show the relevant population densities.
        The dashed vertical lines indicate the means of the associated distributions.
        [Right Panel]: The RV values from the literature on the $x$ axis with our adopted RV values, on the $y$.
        We show a one-to-one relation, over which our 20 comparison RVs are plotted.
        Squares are from low-resolution literature measurements, whereas circles and diamonds
        are mid- and high-resolution literature measurements respectively.
        Orange points are like-for-like comparisons and blue points are for the three benchmark systems,
        i.e., comparisons between our measured secondary RV against the literature RV of the primary.
    }
    \label{fig:rvcomparison-full}
\end{figure*}

We have derived RVs for 46 of the observed 53 objects, the seven objects that we did not measure RVs
were only observed with the R300R grism.
For 20 of the 53 objects, there are published RVs and for 17 of these we have measured RVs.
The objects J1004$+$5022, J1441$-$0945 and J1617$+$7733B are candidate members of benchmark systems
(Section~$\S$\ref{subsec:targetselection}), and we adopt the RVs of their primary stars
as a comparison with our measured values for the secondary, for a total of 20 comparison RVs.
In Figure~\ref{fig:rvcomparison-full}, we plot histograms of the 20 published and
the 46 measured values.
We also show the difference between the published and measured values of the 20 overlapping objects.
If there is more than one literature value, we take the weighted mean RV and standard error on the mean,
to compare against the adopted RV from this work.
We show literature measurements with respect to their resolutions and define these as:
low, $R<2\,500$; mid, $2\,500\leq R \leq 25\,000$; high, $R>25\,000$.
The error used to define $\sigma$ are the quadrature summed errors from the literature and our adopted RV\@.

Our 46 RVs in the heliocentric reference frame are presented in Table~\ref{table:radial_velocities}.
This reference frame has been experimented with, in that the heliocentric/barycentric correction
via \texttt{pypeit} has been compared with a manual barycentric correction
using \texttt{barycorrpy}~\citep{2018RNAAS...2....4K}.
Resultant RV differences from the manual barycentric correction to the pipeline barycentric
correction differ by ${\approx}0.1$\,\kms.
The difference between heliocentric and barycentric correction is $0.5$\,\kms\ in the case of J1745$-$1640.

\begin{table*}
\centering
\caption{
    RVs measured in this work and compared to the literature.
}
\label{table:radial_velocities}
\begin{tabular}{l c ccc}
    \hline
    Object     & Literature RV       & Line Centre RV       & Cross Correlation RV       & Adopted RV \\
    short name & [kms$^{-1}$]        & [kms$^{-1}$]         &      [kms$^{-1}$]          &    [kms$^{-1}$]  \\
    \hline
    J0028$-$1927 & \ldots & $18.6\pm7.7^{\mathrm{11111110}}$ & $26.3\pm13.2^{\mathrm{11111111}}$ & $20.4\pm14.9$\\
    J0235$-$0849 & $15.3\pm11.2\hyperlink{litrvs}{^{1}}$, $22.8\pm6.1\hyperlink{litrvs}{^{2}}$ & $30.9\pm27.8^{\mathrm{00110111}}$ & $26.7\pm21.0^{\mathrm{00111111}}$ & $28.4\pm29.1$\\
    J0428$-$2253 & \ldots & $69.3\pm36.3^{\mathrm{11111110}}$ & $16.0\pm6.1^{\mathrm{00011111}}$ & $17.0\pm12.3$\\
    J0453$-$1751 & \ldots & $15.1\pm2.3^{\mathrm{00010110}}$ & $13.3\pm5.9^{\mathrm{00111111}}$ & $15.0\pm8.3$\\
    J0502$+$1442 & \ldots & $41.3\pm5.8^{\mathrm{11101110}}$ & $41.2\pm5.4^{\mathrm{11111111}}$ & $41.3\pm10.7$\\
    J0605$-$2342 & \ldots & $23.4\pm6.3^{\mathrm{00011110}}$ & $25.0\pm8.8^{\mathrm{11111111}}$ & $23.7\pm11.1$\\
    J0741$+$2316 & \ldots & $31.1\pm0.2^{\mathrm{00011000}}$ & $38.0\pm3.3^{\mathrm{11001110}}$ & $31.1\pm7.8$\\
    J0752$+$4136 & $8.5\pm10.1\hyperlink{litrvs}{^{1}}$ & $-3.0\pm19.9^{\mathrm{01001100}}$ & $14.2\pm15.7^{\mathrm{11111100}}$ & $4.7\pm19.7$\\
    J0809$+$2315 & \ldots & $-48.2\pm3.2^{\mathrm{00011110}}$ & $-38.0\pm9.5^{\mathrm{00011111}}$ & $-47.4\pm8.9$\\
    J0823$+$0240 & \ldots & $-26.6\pm3.8^{\mathrm{00001110}}$ & $-4.3\pm4.5^{\mathrm{11111110}}$ & $-21.4\pm8.8$\\
    J0823$+$6125 & \ldots & $-22.6\pm11.3^{\mathrm{00011110}}$ & $-12.9\pm12.7^{\mathrm{10111111}}$ & $-19.6\pm15.4$\\
    J0918$+$2134 & \ldots & $-92.9\pm7.3^{\mathrm{00001110}}$ & $-80.0\pm7.5^{\mathrm{00111110}}$ & $-88.2\pm10.6$\\
    J0935$-$2934 & \ldots & $-3.7\pm13.2^{\mathrm{10001110}}$ & $-22.5\pm6.1^{\mathrm{11111111}}$ & $-16.9\pm12.8$\\
    J0938$+$0443 & $25.4\pm13.3\hyperlink{litrvs}{^{1}}$ & $2.7\pm7.2^{\mathrm{00011110}}$ & $-5.7\pm6.7^{\mathrm{11111101}}$ & $-0.7\pm11.1$\\
    J0940$+$2946 & $27.3\pm11.8\hyperlink{litrvs}{^{1}}$, $4.1\pm7.1\hyperlink{litrvs}{^{2}}$ & $51.6\pm5.1^{\mathrm{00011110}}$ & $35.0\pm9.2^{\mathrm{11111111}}$ & $49.4\pm10.3$\\
    J0953$-$1014 & \ldots & $63.2\pm7.9^{\mathrm{01000011}}$ & $10.0\pm7.5^{\mathrm{11111111}}$ & $47.6\pm11.3$\\
    J1004$+$5022 & $-1.7\pm1.5^{\dagger}\hyperlink{litrvs}{^{3}}$, $-1.6\pm0.9^{\dagger}\hyperlink{litrvs}{^{4}}$, $-0.7\pm1.2^{\dagger}\hyperlink{litrvs}{^{5}}$, $-2.8\pm0.9^{\dagger}\hyperlink{litrvs}{^{6}}$ & $1.1\pm11.4^{\mathrm{00000100}}$ & $-3.0\pm1.8^{\mathrm{00011111}}$ & $-2.6\pm8.2$\\
    J1004$-$1318 & \ldots & $22.4\pm5.9^{\mathrm{00111110}}$ & $13.3\pm8.4^{\mathrm{00111111}}$ & $19.7\pm11.1$\\
    J1047$-$1815 & \ldots & $-17.2\pm4.6^{\mathrm{00001111}}$ & $-18.0\pm6.6^{\mathrm{00011111}}$ & $-17.4\pm9.6$\\
    J1058$-$1548 & \ldots & $-0.5\pm9.9^{\mathrm{00011111}}$ & $-1.0\pm5.7^{\mathrm{00011111}}$ & $-0.9\pm11.1$\\
    J1109$-$1606 & $48.7\pm16.1\hyperlink{litrvs}{^{1}}$, $69.9\pm10.0\hyperlink{litrvs}{^{2}}$ & $58.5\pm1.2^{\mathrm{00000011}}$ & $48.7\pm2.8^{\mathrm{11111111}}$ & $58.1\pm7.9$\\
    J1127$+$4705 & $-23.7\pm11.1\hyperlink{litrvs}{^{1}}$, $-26.4\pm6.5\hyperlink{litrvs}{^{2}}$ & $8.9\pm62.3^{\mathrm{00011110}}$ & $10.0\pm69.3^{\mathrm{00011110}}$ & $9.4\pm73.7$\\
    J1213$-$0432 & \ldots & $-20.6\pm17.0^{\mathrm{00011110}}$ & $-40.0\pm24.7^{\mathrm{00111111}}$ & $-25.3\pm22.4$\\
    J1216$+$4927 & $4.3\pm16.2\hyperlink{litrvs}{^{1}}$ & $2.2\pm4.1^{\mathrm{11111111}}$ & $8.8\pm6.9^{\mathrm{11111111}}$ & $3.9\pm10.6$\\
    J1221$+$0257 & $2.0\pm10.1\hyperlink{litrvs}{^{1}}$, $-8.0\pm3.0\hyperlink{litrvs}{^{7}}$, $-12.6\pm4.1\hyperlink{litrvs}{^{2}}$, $-8.8\pm0.1\hyperlink{litrvs}{^{8}}$ & $17.5\pm7.0^{\mathrm{11111111}}$ & $20.0\pm8.1^{\mathrm{11111111}}$ & $18.6\pm13.2$\\
    J1232$-$0951 & \ldots & $1.8\pm8.1^{\mathrm{11111111}}$ & $-8.6\pm7.3^{\mathrm{11111110}}$ & $-4.2\pm13.1$\\
    J1246$+$4027 & \ldots & $-46.7\pm12.5^{\mathrm{00111111}}$ & $-46.7\pm15.0^{\mathrm{00111111}}$ & $-46.7\pm18.3$\\
    J1331$+$3407 & $4.1\pm10.2\hyperlink{litrvs}{^{1}}$, $15.4\pm7.8\hyperlink{litrvs}{^{2}}$ & $-5.6\pm24.0^{\mathrm{00000100}}$ & $12.0\pm1.8^{\mathrm{00011111}}$ & $11.5\pm8.3$\\
    J1333$-$0215 & $28.7\pm21.8\hyperlink{litrvs}{^{1}}$ & $-29.2\pm7.2^{\mathrm{00111111}}$ & $-7.5\pm7.2^{\mathrm{11111111}}$ & $-20.0\pm12.2$\\
    J1346$+$0842 & $-67.9\pm12.2\hyperlink{litrvs}{^{1}}$, $-17.7\pm10.6\hyperlink{litrvs}{^{2}}$ & $-50.7\pm4.0^{\mathrm{00111111}}$ & $-35.6\pm7.0^{\mathrm{11111111}}$ & $-47.7\pm10.0$\\
    J1412$+$1633 & \ldots & $-63.4\pm15.9^{\mathrm{11111100}}$ & $-81.4\pm20.5^{\mathrm{11111111}}$ & $-70.8\pm25.8$\\
    J1421$+$1827 & \ldots & $-12.6\pm9.6^{\mathrm{11111110}}$ & $-10.0\pm9.1^{\mathrm{11011110}}$ & $-11.2\pm14.2$\\
    J1441$-$0945 & $-41.6\pm5.9^{\dagger}\hyperlink{litrvs}{^{4}}$ & $-1.3\pm53.8^{\mathrm{00001001}}$ & $8.0\pm25.8^{\mathrm{00111011}}$ & $4.6\pm33.4$\\
    J1532$+$2611 & $-38.8\pm36.6\hyperlink{litrvs}{^{1}}$, $9.2\pm12.4\hyperlink{litrvs}{^{2}}$ & $-15.6\pm9.4^{\mathrm{00011111}}$ & $-11.7\pm4.4^{\mathrm{00111111}}$ & $-12.5\pm10.3$\\
    J1539$-$0520 & $27.3\pm0.2\hyperlink{litrvs}{^{9}}$, $27.0\pm4.0\hyperlink{litrvs}{^{7}}$ & $36.7\pm7.4^{\mathrm{00011011}}$ & $24.0\pm1.7^{\mathrm{00011111}}$ & $24.8\pm8.2$\\
    J1548$-$1636 & \ldots & $11.8\pm6.3^{\mathrm{11111111}}$ & $21.3\pm7.4^{\mathrm{11111111}}$ & $15.8\pm12.4$\\
    J1617$+$7733B & $-19.0\pm0.8^{\dagger}\hyperlink{litrvs}{^{4}}$ & $-31.6\pm24.5^{\mathrm{00011111}}$ & $-18.0\pm12.5^{\mathrm{10011110}}$ & $-20.8\pm19.2$\\
    J1618$-$1321 & \ldots & $-39.5\pm9.8^{\mathrm{00011111}}$ & $-75.0\pm48.0^{\mathrm{00101101}}$ & $-41.2\pm17.0$\\
    J1623$+$1530 & $-17.8\pm11.5\hyperlink{litrvs}{^{1}}$, $5.4\pm17.2\hyperlink{litrvs}{^{2}}$ & $-50.2\pm10.4^{\mathrm{00111111}}$ & $-28.7\pm7.8^{\mathrm{11111111}}$ & $-38.0\pm14.1$\\
    J1623$+$2908 & $-8.1\pm11.5\hyperlink{litrvs}{^{2}}$ & $-18.8\pm5.3^{\mathrm{00000111}}$ & $-26.0\pm9.2^{\mathrm{00011111}}$ & $-20.0\pm9.8$\\    J1707$-$0138 & \ldots & $25.2\pm7.3^{\mathrm{11111111}}$ & $18.3\pm9.5^{\mathrm{00111111}}$ & $22.2\pm13.5$\\    J1717$+$6526 & \ldots & $-62.6\pm3.3^{\mathrm{00111111}}$ & $-76.7\pm6.1^{\mathrm{00111111}}$ & $-64.4\pm8.7$\\    J1745$-$1640 & $26.0\pm2.0\hyperlink{litrvs}{^{7}}$ & $36.2\pm4.4^{\mathrm{11111111}}$ & $28.8\pm4.7^{\mathrm{11111111}}$ & $32.7\pm10.1$\\    J1750$-$0016 & $19.0\pm3.0\hyperlink{litrvs}{^{7}}$ & $1.5\pm2.9^{\mathrm{00110110}}$ & $16.0\pm1.7^{\mathrm{00110111}}$ & $11.7\pm8.1$\\    J2155$+$2345 & \ldots & $-47.6\pm12.6^{\mathrm{00111111}}$ & $-46.7\pm11.0^{\mathrm{00111111}}$ & $-47.1\pm16.3$\\    J2339$+$3507 & \ldots & $-60.0\pm10.9^{\mathrm{00011110}}$ & $-47.1\pm10.4^{\mathrm{01111111}}$ & $-55.1\pm14.4$\\
    \hline
\end{tabular}
\newline
\hypertarget{litrvs}{Literature Radial Velocities}: 1.\ ~\citet{2019AJ....157..231K}, 2.\ ~\citet{2010AJ....139.1808S}, 3.\ ~\citet{2016MNRAS.455.3345B}, 4.\ ~\citet{2018A&A...616A...1G}, 5.\ ~\citet{2012ApJ...758...56S}, 6.\ ~\citet{2012AJ....144..109S}, 7.\ ~\citet{2015ApJS..220...18B}, 8.\ ~\citet{2021ApJS..257...45H}, 9.\ ~\citet{2010ApJ...723..684B}.\\
Indices: 1 if line from Table~\ref{table:spectrallines} used, 0 otherwise.\\

Quoted RVs are already heliocentric corrected.
A `$\dagger$' symbol next to an RV means the RV is that of the primary star in the common proper motion system a
given object is part of.    

\end{table*}

The median difference between our adopted RVs and the literature RVs was 7.8\,\kms.
This 7.8\,\kms\ was then added in quadrature to our adopted RV error.
We used this value to account for systematic uncertainties such as night-to-night instrumental drift and any
FWHM undersampling.
A S/N ratio of 20--30 also correlates with an RV uncertainty of ${\approx}8$\,\kms, which was the typical S/N ratio seen
around our alkali lines.
Some lines, such as the potassium doublet, had lower S/N ratios and lower local resolutions due to a combination
of wider features and lower flux values.
All objects except J0940$+$2946 and J1221$+$0257
have an adopted and literature RV difference less than twice the sum of the respective errors in quadrature.
J0940$+$2946 was $2.69 \sigma$ from the weighted mean literature value.
Of the two literature values constructing this weighted mean, our value is ${<}2 \sigma$ from the value
from~\citet{2019AJ....157..231K}, which is notably larger than the value from~\citet{2010AJ....139.1808S}.
J1221$+$0257 was $2.08 \sigma$ from the weighted mean literature value.
Our RV value was closest to the value from~\citet{2019AJ....157..231K}, with less agreement shown with
the value from~\citet{2010AJ....139.1808S}, which itself was most similar to the values
from~\citet{2015ApJS..220...18B} and~\citet{2021ApJS..257...45H}.
We note for both of these objects that the RV values from~\citet{2010AJ....139.1808S}
utilised considerably lower resolution spectra, hence a worse agreement being shown.
Any objects in Table~\ref{table:radial_velocities} which have known
primary stars with literature RVs are discussed below:

\begin{description}
    \item \textbf{J1004$+$5022}:
        G 196--3B is the binary companion to G 196--3A~\citep{2008ApJ...689.1295K}\@.
        G 196--33A has a mean RV of $-1.6 \pm 0.4$\,\kms~\citep{2012ApJ...758...56S, 2012AJ....144..109S,
        2016MNRAS.455.3345B, 2018A&A...616A...1G}.
        This mean RV of the primary is $0.1 \sigma$ away from the RV of the secondary companion from this work.
    \item \textbf{J1441$-$0945}:
        DENISJ144137.2$-$094558 is the binary companion to G 124--62~\citep{2003AJ....126.1526B, 2005A&A...440..967S}.
        G 124--62 has an RV of $-41.65 \pm 5.91$\,\kms~\citep{2018A&A...616A...1G}, which is within $1.4 \sigma$
        of the companion (which had large uncertainties).
    \item \textbf{J1617$+$7733B}:
        TYC4571-1414-1B is the binary companion of TYC4571-1414-1A\@~\citep{2015A&A...583A..85A}.
        TYC4571-1414-1A has an RV of $-19 \pm 0.8$\,\kms~\citep{2018A&A...616A...1G}, this RV is $0.1 \sigma$
        from the companion RV\@.
\end{description}

\subsubsection{Moving groups} \label{subsubsec:movinggroupsresults}

Our results for UVW Galactic kinematic components are presented in Table~\ref{table:uvw-kinematics}
with each object's moving group classification and associated probability from \banyan.
When accounting for RV in \banyan, the resultant probability was often lower than the calculation
without RV\@.
This was due to the Bayesian probabilities being designed for a higher 
recovery rate (moving from 82\,per cent to 90\,per cent) when accounting for the RV~\citep[see the
\banyan\ \href{http://www.exoplanetes.umontreal.ca/banyan/banyansigma.php}{cautionary note}\footnote{\url{http://www.exoplanetes.umontreal.ca/banyan/banyansigma.php}},][]{2018ApJ...856...23G}.
In addition, the RV uncertainties from this work are much higher than
proper motion or parallax uncertainties from \gaia.

\begin{table*}
\centering
\caption{
    The UVW velocities and \banyan\ classification (with associated probability) from this work.
}
\label{table:uvw-kinematics}
\begin{tabular}{l cc cc c ll lc}
    \hline
    Object     & $V_{\mathrm{tan}}$ & $V_r$        & U              & V             & W            &
     \vtotal            &Galaxy    & \banyan        & Probability \\
    short name &      [kms$^{-1}$]  & [kms$^{-1}$]   & [kms$^{-1}$]  & [kms$^{-1}$] &
     [kms$^{-1}$]        &component  & component                  & classification  &    [per cent]    \\
    \hline
    J0028$-$1927 & 20.5 & 20.4 & 10.7 & 10.1 & -14.8 & 20.9 & Thin & Field & 100.0\\
    J0235$-$0849 & 6.0 & 28.4 & 3.0 & 18.0 & -19.8 & 26.9 & Thin & Field & 100.0\\
    J0428$-$2253 & 23.3 & 17.0 & 19.5 & 7.0 & 9.7 & 22.9 & Thin & Field & 100.0\\
    J0453$-$1751 & 7.0 & 15.0 & 0.6 & 0.2 & 2.2 & 2.3 & Thin & $\beta$ Pictoris & 98.9\\
    J0502$+$1442 & 17.1 & 41.3 & 32.1 & -5.1 & 4.7 & 32.8 & Thin & Hyades & 99.1\\
    J0605$-$2342 & 19.4 & 23.7 & 18.3 & 9.9 & -7.8 & 22.2 & Thin & Field & 100.0\\
    J0741$+$2316 & $10.1\hyperlink{kinerefs}{^{1}}$ & 31.1 & 22.9 & 4.3 & 8.1 & 24.7 & Thin & Field & 99.9\\
    J0752$+$4136 & 10.3 & 4.7 & -0.7 & 19.8 & 1.5 & 19.9 & Thin & Field & 100.0\\
    J0823$+$6125 & 61.6 & -19.6 & 17.3 & -9.8 & -48.1 & 52.1 & Thin & Field & 100.0\\
    J0847$-$1532 & 19.8 & -1.0$\hyperlink{kinerefs}{^{2}}$ & -26.2 & 4.5 & 5.6 & 27.2 & Thin & Field & 100.0\\
    J0935$-$2934 & 11.1 & -16.9 & -3.6 & 32.9 & 8.8 & 34.3 & Thin & Field & 100.0\\
    J0938$+$0443 & 13.1 & -0.7 & -13.4 & 3.2 & -0.3 & 13.8 & Thin & Field & 100.0\\
    J0940$+$2946 & 38.2 & 49.4 & 46.3 & -14.0 & 18.7 & 51.8 & Thin & Field & 100.0\\
    J0953$-$1014 & 18.0 & 47.6 & 11.1 & -32.3 & 17.3 & 38.3 & Thin & Field & 98.4\\
    J1004$+$5022 & 25.3 & -2.6 & 0.7 & -9.6 & 0.7 & 9.6 & Thin & Field & 99.8\\
    J1004$-$1318 & 27.1 & 19.7 & -4.9 & -17.9 & -5.0 & 19.2 & Thin & Field & 66.5\\
    J1047$-$1815 & 49.0 & -17.4 & 34.9 & 17.6 & -21.8 & 44.8 & Thin & Field & 100.0\\
    J1058$-$1548 & 22.4 & -0.9 & 11.9 & 8.9 & -1.7 & 14.9 & Thin & Argus & 93.1\\
    J1109$-$1606 & 105.0 & 58.1 & 43.3 & -90.0 & -25.7 & 103.1 & Thick & Field & 100.0\\
    J1127$+$4705 & 13.2 & 9.4 & 2.8 & 4.2 & 13.6 & 14.5 & Thin & Field & 100.0\\
    J1213$-$0432 & 29.6 & -25.3 & 19.5 & 10.8 & -20.5 & 30.2 & Thin & Carina Near & 72.0$^{\dagger}$\\
    J1221$+$0257 & 13.3 & 18.6 & -1.4 & -3.4 & 20.3 & 20.6 & Thin & Field & 100.0\\
    J1232$-$0951 & $30.2\hyperlink{kinerefs}{^{3}}$ & -4.2 & 6.6 & -8.3 & -8.9 & 13.8 & Thin & Field & 99.8\\
    J1246$+$4027 & 17.6 & -46.7 & -32.3 & 3.0 & -36.1 & 48.5 & Thin & Field & 100.0\\
    J1331$+$3407 & 55.8 & 11.5 & 14.6 & -33.9 & 28.4 & 46.6 & Thin & Field & 100.0\\
    J1333$-$0215 & 52.5 & -20.0 & 27.3 & -25.2 & -13.2 & 39.4 & Thin & Field & 100.0\\
    J1346$+$0842 & 52.6 & -47.7 & 35.2 & -25.8 & -33.5 & 55.0 & Thin & Field & 100.0\\
    J1412$+$1633 & 17.9 & -70.8 & 4.5 & -2.0 & -63.7 & 63.9 & Thin & Field & 100.0\\
    J1421$+$1827 & 69.2 & -11.2 & 32.2 & -42.8 & 16.7 & 56.1 & Thin & Field & 100.0\\
    J1441$-$0945 & 30.7 & 4.6 & 6.5 & -9.9 & 20.6 & 23.8 & Thin & Field & 63.1\\
    J1539$-$0520 & 47.9 & 24.8 & -46.2 & 51.0 & -2.1 & 68.8 & Thin & Field & 100.0\\
    J1548$-$1636 & 30.3 & 15.8 & -14.9 & -17.0 & 20.8 & 30.7 & Thin & Field & 100.0\\
    J1617$+$7733B & 18.7 & -20.8 & 1.5 & -11.4 & -1.8 & 11.7 & Thin & Field & 96.0\\
    J1618$-$1321 & $29.3\hyperlink{kinerefs}{^{4}}$ & -41.2 & 31.5 & -15.5 & -4.6 & 35.4 & Thin & Field & 100.0\\
    J1623$+$1530 & 14.7 & -38.0 & 14.3 & -14.9 & -11.8 & 23.8 & Thin & Field & 100.0\\
    J1705$-$0516 & 14.9 & 12.2$\hyperlink{kinerefs}{^{5}}$ & -25.4 & 14.2 & -2.8 & 29.2 & Thin & Field & 100.0\\
    J1707$-$0138 & 5.8 & 22.2 & -30.5 & 19.1 & 9.7 & 37.3 & Thin & Field & 100.0\\
    J1717$+$6526 & 20.1 & -64.4 & -27.4 & -30.5 & -41.2 & 58.1 & Thin & Field & 100.0\\
    J1733$-$1654 & 7.0 & 17.0$\hyperlink{kinerefs}{^{2}}$ & -26.1 & 16.5 & 2.1 & 31.0 & Thin & Field & 100.0\\
    J1745$-$1640 & 13.7 & 32.7 & -42.5 & 16.7 & -3.2 & 45.7 & Thin & Field & 100.0\\
    J1750$-$0016 & 19.4 & 11.7 & -15.0 & 15.9 & 28.1 & 35.6 & Thin & Field & 100.0\\
    J2339$+$3507 & 23.4 & -55.1 & -4.1 & -44.6 & 20.6 & 49.3 & Thin & Field & 100.0\\
    \hline
\end{tabular}
\newline
\hypertarget{kinerefs}{Literature astrometry used to generate UVWs}: 1.\ ~\citet{2014MNRAS.437.3603S}, 2.\ ~\citet{2015ApJS..220...18B}, 3.\ ~\citet{2020AJ....159..257B}, 4.\ ~\citet{2016AJ....152...24W}, 5.\ ~\citet{2010ApJ...723..684B}.\\

U is in the direction of the Galactic anti-centre. 
Derived using this work's adopted radial velocity in combination with
\gdrthree\ kinematics unless otherwise indicated.
We also show the predicted Galaxy component, taken from the UVW velocities
and \vtotal\ cuts in~\citet{2010A&A...511L..10N}.\\
$\dagger$: J1213$-$0432 had an additional probability (26\,per cent) of being a member of Argus, for a total non-field probability of 98\,per cent.

\end{table*}

We find four objects are members of the following young moving groups and clusters:
Argus~\citep[30--50\,Myr,][]{2000MNRAS.317..289M};
$\beta$~Pictoris~\citep{2001ApJ...562L..87Z}, 
20--26\,Myr~\citep[][and references therein]{2014MNRAS.445.2169M, 2023ApJ...946....6C};
Carina-Near~\citep[${\sim}200$\,Myr,][]{2006ApJ...649L.115Z};
and the Hyades cluster~\citep[600--800\,Myr,][]{1998A&A...331...81P, 2018ApJ...856...40M, 2018A&A...615L..12L}.
These objects (J1058$-$1548, J0453$-$1751, J1213$-$0432, and J0502$+$1442 -- respectively)
are discussed below in Section~$\S$\ref{subsubsec:individuals}.

\subsubsection{Galactic components} \label{subsubsec:galcomp}
\begin{figure}
    \centering
    \includegraphics[width=\linewidth]{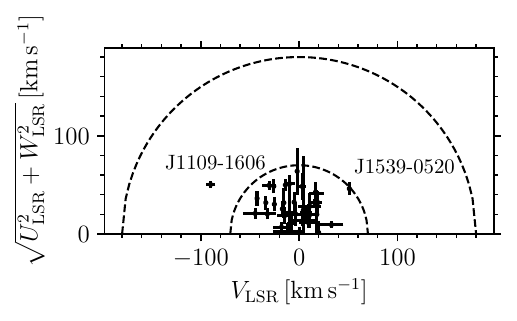}
    \caption{
        Toomre diagram, as done by~\citet{2005AandA...433..185B}, using \gdrthree\ astrometry
        in combination with our calculated RVs.
        V is on the $x$ axis, against the velocity dispersion ($\sqrt{\mathrm{U}^2 + \mathrm{W}^2}$) on the $y$ axis.
        Black circles are UVW velocities calculated with the RVs from this work, with associated error-bars given.
        We show the respective thick disc and halo selection lines at
        $\vtotal > 70$\,\kms\ and $\vtotal > 180$\,\kms\ respectively.
    }
    \label{fig:toomre}
\end{figure}

Thin disc objects were differentiated from thick disc and halo objects
using the LSR corrected UVW Galactic velocities;
the thick disc and halo objects were those with $\vtotal > 70$\,\kms\ and
$\vtotal > 180$\,\kms\ respectively~\citep{2010A&A...511L..10N}.
$\vtotal$ is the total space velocity.
We calculated upper and lower bounds for UVW Galactic velocities using the propagated parallax,
proper motion, and RV errors;
these UVW velocities with associated uncertainties are shown in Figure~\ref{fig:toomre}.
The objects J1109$-$1606 ($\vtotal = 103\pm5$\,\kms) and
J1539$-$0520 ($\vtotal = 69\pm4$\,\kms) are found using the above criteria to be most likely thick disc objects,
and are highlighted in Figure~\ref{fig:toomre}.
J1539$-$0520, is a borderline thick disc object, within $1 \sigma$ of the thick disc cut-off.
Considering that a nearby object is most likely within the thin disc~\citep{2009A&A...501..941H},
J1539$-$0520 is a reasonable thick disc candidate, hence the inclusion here.
It was also assigned a 64.6\,per cent probability of being in the thick disc by~\citet{2024MNRAS.527.1521C},
although it did not pass the conservative subdwarf candidate selection criteria in that work.
Without metallicity information, an object being in the thick disc is not a direct inference on age.
These objects are worth visiting with higher resolution spectroscopy to gain metallicity information,
to confirm any potential subdwarf candidacy.
This future work would also involved gathering NIR spectra, as in work 
by~\citet[and references therein]{2018csss.confE..44Z, 2018MNRAS.480.5447Z}.

\subsection{Astrophysical parameters} \label{subsec:astrophysical_parametersresults}

\begin{figure}
    \centering
    \includegraphics[width=\linewidth]{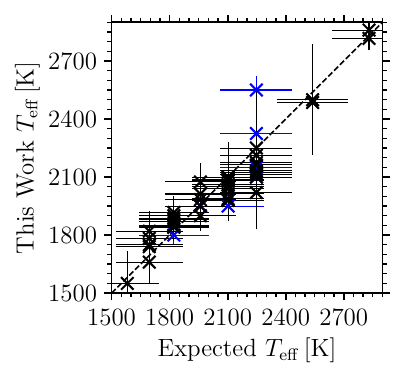}
    \caption{
        The expected \teffexpect~\citep[calculated via spectral
        type through a Filippazzo relation,][]{2015ApJ...810..158F} on the $x$ axis and the best-fitting BT-Settl
        model mean \teff\ on the $y$ axis.
        Blue crosses are for objects with a fit to the R300R spectra whilst black crosses are objects with a fit
        to the R2500I spectra.
    }
    \label{fig:teffcomparison}
\end{figure}

We present the \teff\ and \logg\ values from the model fitting
(Section~$\S$\ref{subsec:astrophysical_parametersanalysis}) in Table~\ref{table:astrophysical_parameters}
along with \teffexpect, assuming our adopted spectral type and equation~(4)
by~\citet{2015ApJ...810..158F} and \teffespucd\ values from \gdrthree.
In Figure~\ref{fig:teffcomparison}, we plot the difference between our value and the expected value.
In the cases of objects with both R2500I and R300R spectra available, we default to the higher resolution result.

\begin{table*}
\centering
\caption{
    Effective temperatures and surface gravities from this work.
}
\label{table:astrophysical_parameters}
\begin{tabular}{l ccc c|l ccc c}
    \hline
    \noalign{\smallskip}
    Object     & \teffexpect & \teff & \gaia\ \teff    & \logg    &
    Object     & \teffexpect & \teff & \gaia\ \teff    & \logg    \\
    short name &        [K]         &  [K]   &      [K]         &  [dex]   &
    short name &        [K]         &  [K]   &      [K]         &  [dex]   \\
    \hline
    J0028$-$1927 & 2102$\pm185$ & 1988$\pm116$ & 2115$\pm112$ & 4.8$\pm0.4$&     J0235$-$0849 & 1959$\pm183$ & 1983$\pm62$ & 2035$\pm190$ & 5.0$\pm0.3$\\
        J0428$-$2253 & 2102$\pm185$ & 1980$\pm64$ & 2322$\pm71$ & 5.0$\pm0.3$&     J0453$-$1751 & 1822$\pm179$ & 1850$\pm70$ & 1921$\pm176$ & 5.0$\pm0.3$\\
        J0502$+$1442 & 2249$\pm186$ & 2212$\pm126$ & 2285$\pm80$ & 4.9$\pm0.3$&     J0605$-$2342 & 2102$\pm185$ & 2088$\pm136$ & 2121$\pm82$ & 4.8$\pm0.4$\\
        J0741$+$2316 & 2249$\pm186$ & 2020$\pm190$ & \ldots & 5.0$\pm0.3$&     J0752$+$4136 & 2831$\pm189$ & 2817$\pm62$ & \ldots & 4.9$\pm0.4$\\
        J0809$+$2315 & 1695$\pm173$ & 1820$\pm64$ & \ldots & 5.5$\pm0.3$&     J0823$+$0240 & 2539$\pm184$ & 2500$\pm287$ & \ldots & 5.1$\pm0.4$\\
        J0823$+$6125 & 1822$\pm179$ & 1843$\pm70$ & 1951$\pm93$ & 4.7$\pm0.4$&     J0847$-$1532 & 1959$\pm183$ & 1950$\pm70$ & 2040$\pm50$ & 5.0$\pm0.3$\\
        J0918$+$2134 & 1822$\pm179$ & 1880$\pm110$ & \ldots & 5.2$\pm0.4$&     J0935$-$2934 & 2249$\pm186$ & 2162$\pm121$ & 2316$\pm39$ & 5.0$\pm0.3$\\
        J0938$+$0443 & 2539$\pm184$ & 2486$\pm228$ & 2364$\pm88$ & 5.1$\pm0.4$&     J0940$+$2946 & 1959$\pm183$ & 1950$\pm70$ & 2144$\pm164$ & 4.6$\pm0.4$\\
        J0953$-$1014 & 2249$\pm186$ & 2100$\pm150$ & 2181$\pm70$ & 4.6$\pm0.4$&     J1004$+$5022 & 1695$\pm173$ & 1740$\pm70$ & 1899$\pm100$ & 4.5$\pm0.3$\\
        J1004$-$1318 & 1822$\pm179$ & 1850$\pm70$ & 1886$\pm197$ & 5.0$\pm0.3$&     J1047$-$1815 & 2102$\pm185$ & 1980$\pm64$ & 2103$\pm81$ & 5.0$\pm0.3$\\
        J1058$-$1548 & 1822$\pm179$ & 1900$\pm102$ & 1834$\pm109$ & 5.0$\pm0.3$&     J1109$-$1606 & 2249$\pm186$ & 2175$\pm82$ & 2104$\pm112$ & 5.0$\pm0.3$\\
        J1127$+$4705 & 2102$\pm185$ & 2060$\pm94$ & 2136$\pm120$ & 4.9$\pm0.4$&     J1213$-$0432 & 1695$\pm173$ & 1783$\pm143$ & 1580$\pm152$ & 5.0$\pm0.3$\\
        J1216$+$4927 & 1959$\pm183$ & 2012$\pm59$ & \ldots & 4.8$\pm0.4$&     J1221$+$0257 & 2249$\pm186$ & 2250$\pm295$ & 2210$\pm41$ & 5.0$\pm0.3$\\
        J1222$+$1407 & 2249$\pm186$ & 2150$\pm70$ & \ldots & 5.0$\pm0.3$&     J1232$-$0951 & 2249$\pm186$ & 2114$\pm144$ & \ldots & 5.0$\pm0.3$\\
        J1246$+$4027 & 1695$\pm173$ & 1750$\pm91$ & 1780$\pm162$ & 4.6$\pm0.4$&     J1331$+$3407 & 2102$\pm185$ & 2040$\pm70$ & 2170$\pm71$ & 4.9$\pm0.4$\\
        J1333$-$0215 & 1959$\pm183$ & 2075$\pm96$ & 2104$\pm76$ & 4.8$\pm0.4$&     J1346$+$0842 & 1822$\pm179$ & 1888$\pm78$ & 1889$\pm349$ & 4.8$\pm0.4$\\
        J1412$+$1633 & 2102$\pm185$ & 2014$\pm97$ & 2104$\pm55$ & 4.6$\pm0.4$&     J1421$+$1827 & 2249$\pm186$ & 2133$\pm157$ & 2233$\pm69$ & 4.9$\pm0.4$\\
        J1439$+$0039 & 2249$\pm186$ & 2325$\pm139$ & \ldots & 5.0$\pm0.3$&     J1441$-$0945 & 2102$\pm185$ & 2060$\pm94$ & 2240$\pm60$ & 4.9$\pm0.4$\\
        J1527$+$0553 & 2249$\pm186$ & 2100$\pm50$ & \ldots & 5.0$\pm0.3$&     J1532$+$2611 & 1822$\pm179$ & 1917$\pm84$ & \ldots & 4.8$\pm0.4$\\
        J1539$-$0520 & 1822$\pm179$ & 1840$\pm70$ & 1804$\pm109$ & 5.4$\pm0.4$&     J1548$-$1636 & 2249$\pm186$ & 2125$\pm147$ & 2272$\pm82$ & 4.9$\pm0.3$\\
        J1617$+$7733B & 2831$\pm189$ & 2860$\pm94$ & \ldots & 4.9$\pm0.4$&     J1618$-$1321 & 2102$\pm185$ & 2050$\pm100$ & \ldots & 5.0$\pm0.3$\\
        J1623$+$1530 & 2249$\pm186$ & 2112$\pm105$ & 2339$\pm147$ & 4.8$\pm0.4$&     J1623$+$2908 & 2102$\pm185$ & 2080$\pm90$ & \ldots & 5.2$\pm0.4$\\
        J1705$-$0516 & 2102$\pm185$ & 1950$\pm70$ & 2065$\pm35$ & 5.0$\pm0.3$&     J1707$-$0138 & 2102$\pm185$ & 2100$\pm180$ & 2019$\pm78$ & 5.0$\pm0.3$\\
        J1717$+$6526 & 1581$\pm166$ & 1550$\pm168$ & 1589$\pm63$ & 4.7$\pm0.4$&     J1724$+$2336 & 2249$\pm186$ & 2550$\pm70$ & 2320$\pm88$ & 5.0$\pm0.3$\\
        J1733$-$1654 & 1822$\pm179$ & 1800$\pm50$ & 2055$\pm63$ & 4.8$\pm0.4$&     J1745$-$1640 & 2102$\pm185$ & 2088$\pm105$ & 2008$\pm49$ & 5.0$\pm0.3$\\
        J1750$-$0016 & 1695$\pm173$ & 1660$\pm113$ & 1542$\pm71$ & 5.1$\pm0.4$&     J2155$+$2345 & 1959$\pm183$ & 1900$\pm76$ & \ldots & 5.0$\pm0.3$\\
        J2339$+$3507 & 1822$\pm179$ & 1871$\pm86$ & 1855$\pm138$ & 5.0$\pm0.3$&  &  &  &  \\ 

    \hline
\end{tabular}
\newline
These \teff\ values are generated using fits to preferentially R2500I spectra if available, else R300R.
Model fits assume solar metallicities.
\teffexpect\ represents the expected effective temperature, based on an object's spectral type.
\gaia\ \teff\ are the \teffespucd\ effective temperatures from \gdrthree.
\end{table*}

Although the best-fitting surface gravity values can be indicative of youth,
they are quite degenerate and without corresponding metallicity values,
therefore they are not relied upon in our discussion below.
The best fitting spectral sub-types and BT-Settl models are shown in a spectral sequence for R2500I VPH spectra
in Figures~\ref{fig:r2500i-full-seq1-comp} and~\ref{fig:r2500i-full-seq2-comp}.

\begin{figure*}
    \centering
    \includegraphics[width=\linewidth]{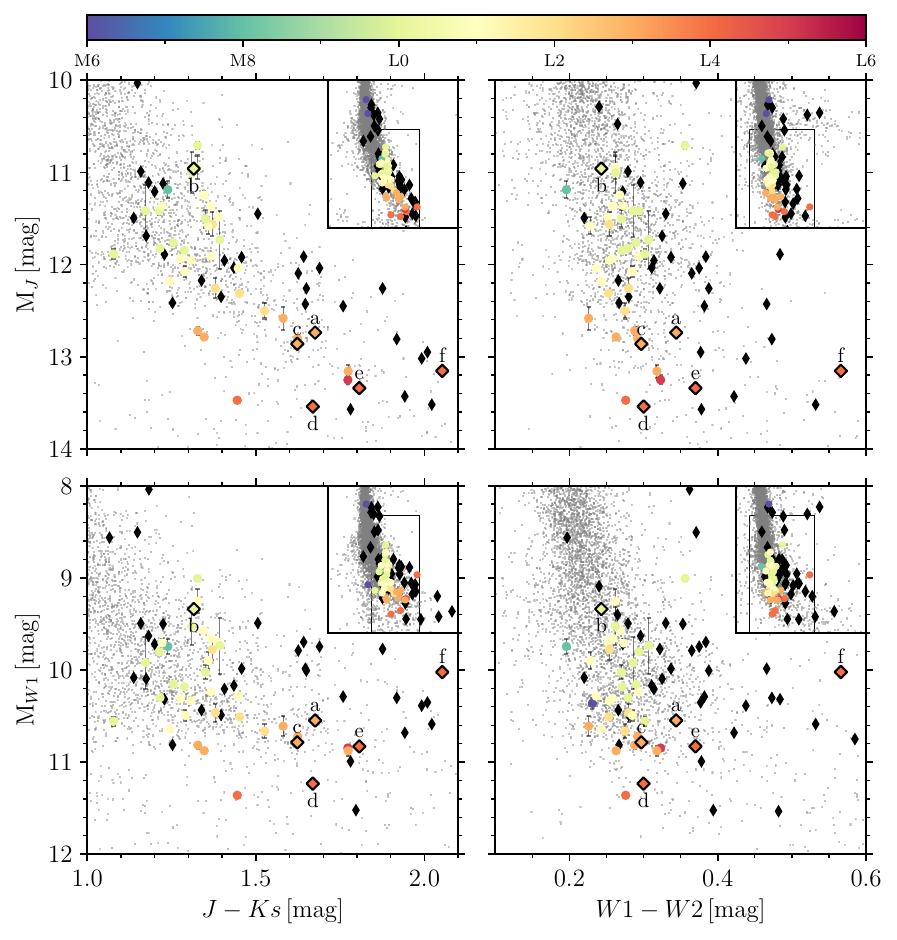}
    \caption{
        CAMDs of 2MASS and AllWISE photometry, focused on the majority of this sample
        (an inset of the full sequence is shown in the upper right).
        The 2MASS\,$J - K_s$ colour is on the $x$ axis for the first column,
        with the AllWISE\,$W1 - W2$ colour on the $x$ axis on the second column.
        Absolute 2MASS\,$J$ magnitude is on the $y$ axis for the first row whilst AllWISE\,$M_{W1}$
        is the $y$ axis of the second row.
        Underlying the plots as grey points is the full UCD sequence from the GUCDS\@.
        Known young objects from~\citet{2016ApJS..225...10F} and~\citet{2016ApJ...833...96L}
        are displayed as black diamonds.
        Each object is coloured by our adopted spectral type, with absolute magnitude error shown.
        Coloured diamonds are the young candidates discussed in Section~$\S$\ref{subsubsec:individuals}.
        Key: a--J0453$-$1751, b--J0502$+$1442, c--J1058$-$1548, d--J1213$-$0432, e--J1246$+$4027,
        f--J1004$+$5022.
    }
    \label{fig:empirical-sequence}
\end{figure*}

Figure~\ref{fig:empirical-sequence} shows a set of colour-absolute magnitude diagrams (CAMD),
2MASS\,$J - K_s$, 2MASS\,$M_{J}$, AllWISE\,$W1 - W2$ and AllWISE\,$M_{W1}$.
Parallaxes from \gaia\ were used to generate the absolute magnitudes.
Highlighted here are the objects with spectral features that are indicative of youth.
These are compared to known young UCDs from~\citet{2016ApJS..225...10F}
and~\citet[`VL-G' or `Young']{2016ApJ...833...96L},
as well as the full sample from the GUCDS\@.
These young objects tend to be over-bright, although the effect varies across filters and is
further complicated by intrinsic scatter plus variability.

\subsubsection{Individual objects} \label{subsubsec:individuals}
We further discuss here objects we have indicated as being non-typical, with interesting features or results.
We check for any age classifications, based on the moving group membership from \banyan\ and location on the CAMD
in Figure~\ref{fig:empirical-sequence}.
There are additional objects which exist in the same colour space as our highlighted objects in 
Figure~\ref{fig:empirical-sequence} which are not discussed below.
This is because there can be large implicit colour scatter due to unresolved binarity, metallicity and dust.
Hence, only objects which are interesting either spectrally or kinematically are discussed.
The following four objects were found to be members of the moving groups listed above,
in Section~$\S$\ref{subsubsec:movinggroupsresults}.

\begin{description}

    \item \textbf{J0453$-$1751}:
        This L3 object, 2MASS J04532647$-$1751543, is a probable member of $\beta$~Pictoris with a 99\,per cent confidence,
        this is an increase on the 55\,per cent categorisation by~\citet[using \gdrtwo\ data]{2020AJ....159..166U}.
        \citet{2015ApJ...798...73G} by comparison found this object as a member (96\,per cent) of the similarly aged
        Columba association~\citep[20--40\,Myr,][]{2008hsf2.book..757T}.
        We have used \gdrthree\ kinematics, which are consistent with the values from \gdrtwo\ but with reduced uncertainties.
        In \gdrtwo~\citep{2018A&A...616A...1G}, this was 
        $\varpi=33.2\pm0.6$\,mas, $\mu_{\alpha}\cos{\delta}=+44.6\pm0.7$\,\masyr and $\mu_{\delta}=-20.8\pm0.8$\,\masyr.
        In \gdrthree~\citep{2023A&A...674A...1G}, 
        $\varpi=33.1\pm0.5$\,mas, $\mu_{\alpha}\cos{\delta}=+44.4\pm0.4$\,\masyr and $\mu_{\delta}=-20.6\pm0.4$\,\masyr.
        The work by~\citet{2020AJ....159..257B} is in broad agreement, with larger uncertainties,
        $\pi=37.4\pm5.7$\,mas, $\mu_{\alpha}\cos{\delta}=+34.7\pm4.9$\,\masyr and $\mu_{\delta}=-24.0\pm3.9$\,\masyr.
        The change of from \gdrtwo\ to \gdrthree\ in isolation did not alter the confidence (99.2\,per cent),
        whereas the inclusion of our adopted RV value dropped this to 98.9\,per cent.
        Our adopted RV was $15.0\pm8.3$\,\kms, which is within $1 \sigma$ of the `optimal' RV from \banyan,
        $21.5\pm1.5$\,\kms.
        From Figure~\ref{fig:empirical-sequence}, we see J0453$-$1751 (a) is photometrically similar to known young objects.
        Its \teff\ of $1850\pm70$\,K is in good agreement with \teffexpect\ and \teffespucd, although is cooler than
        the $2100$\,K from~\citet{2015ApJS..219...33G}.
        We can conclude that this object is an L3 within $\beta$~Pictoris.
        
    \item \textbf{J0502$+$1442}:
        2MASS J05021345$+$1442367, an L0, we find as a member of the Hyades cluster with a 99\,per cent probability.
        This improves the membership confidence by~\citet[][75\,per cent]{2018ApJ...862..138G} and
        concurs with the classifications
        by~\citet[100\,per cent confidence,][using the Melotte~25 name]{2018A&A...616A..10G, 2020A&A...640A...1C}.
        Works by~\citet{2020MNRAS.498.1920O} and~\citet{2021MNRAS.503.3279S} also placed this object
        in Melotte~25 with 96\,per cent and 99\,per cent confidences, respectively.
        It also agrees with the classification by~\citet{2019A&A...623A..35L}, which had a `\texttt{c} parameter' 
        of 5.88, well within their Hyades membership limit, \texttt{c}\,$<25.9$.
        Figure~\ref{fig:empirical-sequence}, places J0502$+$1442 (b) also as photometrically similar to known young objects,
        being somewhat over-bright, although there is considerable overlap with standard M-L sequence.
        With a \teff\ of $2212\pm126$\,K, J0502$+$1442 is an L0 object in the Hyades cluster.
        
    \item \textbf{J1058$-$1548}:
        Another L3 object, SIPS J1058$-$1548, is classified with 93\,per cent confidence as a member of Argus.
        \citet{2015ApJ...798...73G} had the same classification with a much lower probability (35\,per cent).
        \gdrtwo\ astrometry in isolation gave a confidence of 96.3\,per cent,
        whilst \gdrthree\ reduced this to 94.8\,per cent,
        the inclusion of our adopted RV value further dropped this to 93.1\,per cent.
        Our adopted RV was $-0.9\pm11.1$\,\kms, which is within $1 \sigma$ of the `optimal' RV from \banyan,
        $8.5\pm1.4$\,\kms.
        Specifically, in \gdrtwo, this was
        $\varpi=54.6\pm0.5$\,mas, $\mu_{\alpha}\cos{\delta}=-258.1\pm0.8$\,\masyr and $\mu_{\delta}=+31.1\pm0.7$\,\masyr.
        In \gdrthree,
        $\varpi=55.1\pm0.3$\,mas, $\mu_{\alpha}\cos{\delta}=-258.6\pm0.3$\,\masyr and $\mu_{\delta}=+30.8\pm0.3$\,\masyr.
        These values are in broad agreement with non-\textit{Gaia} works, where $\pi$ ranges from 49.2\,mas--66.5\,mas,
        $\mu_{\alpha}\cos{\delta}$ from $-60$\,\masyr ($\pm160$\,\masyr) to $-276$\,\masyr and $\mu_{\delta}$ from
        $+14$\,\masyr to $+210$\,\masyr ($\pm150$\,\masyr);
        c.f.~\citet{2002AJ....124.1170D, 2007ApJ...667..520C, 2007A&A...468..163D, 2007AJ....133.2258S,
                    2009AJ....137....1F, 2012ApJ...752...56F, 2016AJ....152...24W, 2017AJ....154..147D,
                    2018MNRAS.481.3548S}.
        J1058$-$1548 has a \teff\ ${=}1900\pm102$\,K~\citep[in exact agreement with][]{2015ApJS..219...33G},
        but is not as convincingly over-bright as neighbouring known young objects,
        see (c) in Figure~\ref{fig:empirical-sequence}.
        \citet{2023ApJ...959...63S} conclude that for J1058$-$1548, ``it is probable that the YMG assignment [Argus] is incorrect",
        because their spectrum well matched L-dwarf FLD-G standards, although the \logg\ value of $4.27$\,dex was an outlier
        and more typical of a VL-G object (their figure~(21)).
        The \logg\ value in this work was $5.0\pm0.3$\,dex, although this less robust than that from~\citet{2023ApJ...959...63S},
        who also had a much lower \teff ${=}1570$\,K, which itself is more akin to a cooler object, ${\approx}$L5.
        We would argue that this a probable L3 member of Argus but more high resolution spectra and modelling is required to ascertain
        youth.
        
    \item \textbf{J1213$-$0432}:
        2MASS J12130336$-$0432437 (L4) we classify as a member of Carina-Near or Argus (98\,per cent),
        which is an update on the 75\,per cent classification of being in Carina-Near
        by~\citet{2018ApJ...862..138G}.
        Just using \gdrtwo\ astrometry gave a confidence of 68.5\,per cent
        (with a 30.6\,per cent likelihood of being in Argus),
        whilst \gdrthree\ increased this to 74.3\,per cent (24.7\,per cent for Argus),
        the inclusion of our adopted RV value (with large uncertainty) updated this to 72.0\,per cent,
        with a 26.0\,per cent likelihood of being in Argus.
        Our adopted RV was $-25.3\pm22.4$\,\kms, which is within $1.5 \sigma$ of the `optimal' RV from \banyan,
        $2.4\pm0.8$\,\kms.
        In \gdrtwo, it was
        $\varpi=59.5\pm1.0$\,mas, $\mu_{\alpha}\cos{\delta}=-368.1\pm2.2$\,\masyr and $\mu_{\delta}=-34.6\pm1.4$\,\masyr.
        In \gdrthree,
        $\varpi=59.1\pm0.6$\,mas, $\mu_{\alpha}\cos{\delta}=-367.9\pm0.7$\,\masyr and $\mu_{\delta}=-34.0\pm0.5$\,\masyr.
        The work by~\citet{2020AJ....159..257B} is also in good agreement,
        $\pi=56.3\pm3.8$\,mas, $\mu_{\alpha}\cos{\delta}=-380.9\pm2.7$\,\masyr and $\mu_{\delta}=-33.4\pm2.4$\,\masyr.
        Figure~\ref{fig:empirical-sequence} (d) shows this object as being under-bright compared with known young objects,
        with a \teff\ of $1783\pm143$\,K\@.
        Being the age of Carina-Near could explain this relative under-brightness,
        as it should be tending towards field-like behaviour.
        This object can be classified then as an L4 member of Carina-Near.
        
\end{description}

There are two further field objects that we have highlighted as interesting due to their spectral features:

\begin{description}

    \item \textbf{J1246$+$4027}:
        The L4 dwarf, 2MASSW J1246467$+$402715, observed at the two resolutions,
        is of interest due to the potential Li\,\textsc{i} detection at ${\approx}6708\,\angstrom$.
        As this feature is only in the wavelength regime of the R300R spectra, this is not definitive enough
        a detection to confirm 
        lithium~\citep[see discussion by][using the equation from~\citet{1988IAUS..132..345C}]{2018ApJ...856...40M}.
        Higher resolution (R ${\gtrapprox}2000$) spectra would be required for confirmation~\citep{2014MNRAS.439.3890G}.
        Assuming a true detection, employing the lithium test~\citep{1992ApJ...389L..83R} alongside our fitted effective
        temperature of $\teff = 1750\pm91$\,K would identify this object as being substellar.
        This \teff\ is in good agreement with the expected temperature of $\teffexpect = 1717\pm116$\,K
        and the \gdrthree\ \teff\ of $1780\pm162$\,K\@.
        This substellar argument is in line with discussion
        by~\citet{1998ASPC..134..394B},~\citet{1999AJ....118.1005M} and~\citet{1999ApJ...519..802K},
        because our \teff\ is in the range $2670 > \teff > 1400$\,K\@.
        Figure~\ref{fig:empirical-sequence} suggests J1246$+$4027 (e) neighbours some known young objects.
        The best fitting model had a surface gravity of $\logg = 4.6\pm0.3$\,dex,
        although we have no complementary metallicity information.
        \banyan\ finds no correlation with any known young moving groups.
        J1246$+$4027 could be classed as an L4$\beta$ field object.

    \item \textbf{J1004$+$5022}:
        G 196--3B is known to be a low gravity brown dwarf~\citep{1998Sci...282.1309R, 2008ApJ...689.1295K, 2013ApJ...772...79A},
        to which we concur, with a spectral sub-type of L3$\beta$\@.
        Our \logg\ value is $4.5\pm0.2$\,dex ($\teff = 1740\pm113$\,K), as would be expected from the
        already known young nature.
        This object sits extremely red and over-bright
        in Figure~\ref{fig:empirical-sequence} (f), even more extremely than most known young objects.
        It is a companion to the well known G 196--3A M3 star, to which we compared our kinematics in
        Section~$\S$\ref{subsec:kinematicsresults}, finding a $0.1\sigma$ difference.
        There is much deeper discussion on this benchmark system by~\citet{2010ApJ...715.1408Z},
        which measures an angular separation of $\rho = 15.99\pm0.06\,\arcsec$.
        Combined with a \gdrthree\ parallax of $\varpi = 46.1952\pm0.5452$\,mas
        ~\citep[in agreement with the $49.0\pm2.3$\,mas and $41.0\pm4.1$\,mas measurements by][respectively]{2016ApJ...833...96L, 2014A&A...568A...6Z},
        this implies a projected separation of $s = 739\pm1$\,AU\@.
        This is slightly more than the projected physical separation range calculated by~\citet{2010ApJ...715.1408Z},
        285--640\,AU\@.
        We found a probability of the secondary being a field object of 99.9\,per cent, which is an increase
        on the 32\,per cent probability of being a member of AB~Doradus by~\citet{2014ApJ...783..121G}.
        \citet{2016ApJ...833...96L} kinematically confirmed that G 196--3B is a young field object.
        This is also in agreement with the 50\,per cent classification of the primary being a member of AB~Doradus
        by~\citet{2012AJ....143...80S}, which was later downgraded to 0\,per cent by~\citet{2016MNRAS.455.3345B};
        however, the primary was also classified as being a member of the controvertible Castor
        moving group~\citep{1998A&A...339..831B} with 75\,per cent confidence~\citep{2014A&A...567A..52K}.
        The Castor moving group was not included in \banyan, hence not being included in our analysis.
        We classify this object as an L3$\beta$ object.

\end{description}

\section{Summary and Conclusions} \label{sec:gtcdiscussion}
We have presented the low and mid resolution optical GTC/OSIRIS spectra of 53 objects observed
between 2015 and 2016\@.
Our data reduction was non-standard, using a pipeline package, \texttt{PypeIt};
this reduction was validated with an independent \texttt{IRAF} spectral extraction and
calibration for one of the objects.
We used \texttt{kastredux} to create 53 automated spectral types, six of which are for objects not yet
spectrally typed, alongside the established technique of comparing against spectral standard template spectra.
We found that our chosen spectral reduction package, \texttt{PypeIt}, introduced some non-optimal
artefacts during reduction.
One example is a spike appearing near the O$_2$ A band from the telluric correction procedure,
which required interpolating over for visualisation purposes (it does not affect wavelength solutions).

In addition to using new data reduction software, we also used novel analysis software, \texttt{rvfitter},
that we developed to perform manual line centering and cross-correlation (against BT-Settl CIFIST models).
The \texttt{rvfitter} code also used an uncertainty-weighted mean to create an adopted RV\@.
This produced 46 RVs, 29 of which are new,
which we have validated against standard \texttt{IRAF} and \texttt{IDL} software techniques.
There were 17 RVs which were compared against literature values, showing good agreement with a median
difference of 7.8\,\kms, adopted as our systematic uncertainty.
Our median RV uncertainty was 11.2\,\kms, indicating that further high-resolution spectroscopy would be necessary
to validate our RV values and conclusions.
The cross-correlation also produced mean \teff\ and \logg\ values for all 53 objects.

In this work, we performed further analysis on our spectral types, RVs and \teff\ values
by making comparisons to the literature where appropriate and ensuring all results were
within two spectral sub-types, $\Delta \text{RV} < 2\sigma$ and
$\Delta \teff < 2\sigma$ (against \teffexpect\ and \gdrthree\ \teffespucd).
We then discussed any measurements which did not conform with these standards,
such as J0940$+$2946, which had a $\Delta \text{RV} = 2.69\sigma$.
There were four objects that we classified through \banyan\ as being a member of a young moving group:
SIPS J1058-1548 (J1058$-$1548), 2MASS J04532647-1751543 (J0453$-$1751), 2MASS J12130336-0432437 (J1213$-$0432),
and 2MASS J05021345+1442367 (J0502$+$1442).
There were two objects we placed as members of the thick disc:
SIPS J1109-1606 (J1109$-$1606) and 2MASS J15394189-0520428 (J1539$-$0520).

Finally, by relating to gravity sensitive alkali lines and the aforementioned young moving group members,
we discuss the interesting young candidates J1246$+$4027 and J1004$+$5022.
2MASSW J1246467$+$402715 (J1246$+$4027) has a potential lithium indication and is otherwise an L4$\beta$ field object.
G 196--3B (J1004$+$5022) is confirmed as a young object, as was known from its primary companion.

In conclusion, this work was part of the GUCDS series of papers.
A search of the GUCDS yields 145 known L dwarfs with measured RVs, excluding those from the SDSS\@.
The 29 new L dwarf RVs presented in this work are therefore an ${\approx}20$\,per cent
increase to the number of 6-D complete L dwarfs.
A number of interesting objects were identified or confirmed, either into young moving groups or
young field objects.
We used novel open-source techniques at all stages of our procedure, which we make available to the astronomical community.
These techniques have been compared with established and accepted techniques in order to
generate a baseline of trust.
The observation campaign to complete the 30\,pc sample is ongoing, with predominantly NIR spectrographs.
This campaign will continue to produce work discussing, expanding and exploring this 30\,pc sample.

\subsection*{Data availability}
The data underlying this article will be available in \href{https://vizier.cds.unistra.fr/viz-bin/VizieR}{CDS VizieR}\footnote{\url{https://vizier.cds.unistra.fr/viz-bin/VizieR}},
the \href{https://gucds.inaf.it}{GUCDS Data Browser}\footnote{\url{https://gucds.inaf.it}}, and the \href{https://simple-bd-archive.org/}{SIMPLE Database}\footnote{\url{https://simple-bd-archive.org/}}.
The code used to generate the reduced spectra and analysis is available either through open-source 
repositories~\citep[see][and the acknowledgements]{Cooper_rvfitter_2022} or upon any reasonable request.

\section*{Acknowledgements}
We would like to thank the anonymous referees for their very useful and much appreciated feedback,
which has much improved this manuscript.
Based on observations made with the Gran Telescopio Canarias (GTC),
installed in the Spanish Observatorio del Roque de los Muchachos of the Instituto de Astrof\'isica de Canarias,
on the island of La Palma.
This work is based on data obtained with the instrument OSIRIS, built by a Consortium led by the
Instituto de Astrof\'isica de Canarias in collaboration with the Instituto de Astronom\'ia
of the Universidad Aut\'onoma de M\'exico.
OSIRIS was funded by GRANTECAN and the National Plan of Astronomy and Astrophysics of the Spanish Government.

This research has made use of the SIMBAD database, operated at CDS, Strasbourg, France.
This work presents results from the European Space Agency (ESA) space mission Gaia.
Gaia's data are processed by the Gaia Data Processing and Analysis Consortium (DPAC).
Funding for the DPAC is provided by national institutions, in particular the institutions
participating in the Gaia Multi Lateral Agreement (MLA).
This publication makes use of data products from the Two Micron All Sky Survey,
which is a joint project of the University of Massachusetts and the
Infrared Processing and Analysis Center/California Institute of Technology,
funded by the National Aeronautics and Space Administration and the National Science Foundation.
This publication makes use of data products from the Wide-field Infrared Survey Explorer,
which is a joint project of the University of California, Los Angeles,
and the Jet Propulsion Laboratory/California Institute of Technology, and NEOWISE,
which is a project of the Jet Propulsion Laboratory/California Institute of Technology.
WISE and NEOWISE are funded by the National Aeronautics and Space Administration.

IRAF is distributed by the National Optical Astronomy Observatory,
which is operated by the Association of Universities for Research in Astronomy (AURA)
under a cooperative agreement with the National Science Foundation.
We have made use of the on-line resources available from the \texttt{IDL} Astronomy Library
hosted by the NASA Goddard Space Flight Center, in particular the \texttt{DeFringeFlat.pro} routine.
We acknowledge the relevant open source packages used in our \texttt{python}~\citep{python} codes:
\texttt{astropy}~\citep{astropy:2013, astropy:2018},
\texttt{barycorrpy}~\citep{2018RNAAS...2....4K},
\texttt{kastredux}~\citep{kastredux},
\texttt{matplotlib}~\citep{Hunter:2007},
\texttt{numpy}~\citep{harris2020array},
\texttt{pandas}~\citep{mckinney-proc-scipy-2010, reback2020pandas},
\texttt{scipy}~\citep{2020SciPy-NMeth},
\texttt{specutils}~\citep{nicholas_earl_2021_5721652},
\texttt{splat}~\citep{2017ASInC..14....7B} and
\texttt{tqdm}~\citep{casper_da_costa_luis_2021_5517697}.
This research also made use of \texttt{PypeIt} (v.1.4.0),\footnote{\url{https://pypeit.readthedocs.io/en/latest/}}
a \texttt{Python} package for semi-automated reduction of astronomical
slit-based spectroscopy~\citep{pypeit:zenodo, pypeit:joss_pub}.

WJC is funded by a University of Hertfordshire studentship.
WJC, HRAJ, SF, BB and DJP recognise the computing infrastructure provided via STFC grant ST/R000905/1
at the University of Hertfordshire.
RLS has been supported by a STSM grant from COST Action
CA18104: MW-Gaia.
Funded in part by Chinese Academy of Sciences President’s International Fellowship Initiative,
Grant No. 2020VMA0033.
DM, JAC, MCGO and NL acknowledge financial support from the Spanish Agencia Estatal
de Investigación of the Ministerio de Ciencia e Innovación
(AEI/10.13039/501100011033) and the ERDF “A way of making
Europe” through projects PID2019-109522GB-C51, -53 and -54 and PID2022-137241NB-C41, -42 and -44.

\bibliographystyle{mnras}
\bibliography{references}

\appendix

\section{Additional Information} \label{sec:extratables}
\subsection{Supplementary Tables} \label{subsec:extratables}

\onecolumn
\begin{longtable}[c]{ll ll c ccc}
\caption{ \label{table:obslog}
Additional information for all observations carried out as part of the two programmes presented here.
Note, multiple objects were observed multiple times,
 with either the same grism or the other.
Seeing is given as a range corresponding to reverse wavelength, and is corrected for airmass.}\\
\hline
    Object    &  Object     & Resolution & Programme &    UT Date    & Airmass &
     Humidity  & Seeing                                 \\
    Full Name &  short name & Grism/ VPH &     ID    & yyyy-mm-dd &   (z)   &
        [per cent]   &   $\lambda_{\max}$ -- $\lambda_{\min}$     \\
\hline
\endfirsthead
\hline
    Object    &  Object     & Resolution & Programme &    UT Date    & Airmass &
     Humidity  & Seeing                                 \\
    Full Name &  short name & Grism/ VPH &     ID    & yyyy-mm-dd &   (z)   &
        [per cent]   &   $\lambda_{\max}$ -- $\lambda_{\min}$     \\
\hline
\endhead
\hline
\endfoot
\hline
\endlastfoot
    2MASS J00285545$-$1927165 & J0028$-$1927 & R2500I & GTC8-15ITP & 2015-08-30 & 1.54 & 27 & 0.90 -- 0.96\\
    2MASS J02354756$-$0849198 & J0235$-$0849 & R2500I & GTC8-15ITP & 2015-08-31 & 1.49 & 30 & 0.89 -- 0.94\\
    2MASS J04285096$-$2253227 & J0428$-$2253 & R2500I & GTC8-15ITP & 2015-08-31 & 1.82 & 32 & 1.12 -- 1.19\\
    2MASS J04532647$-$1751543 & J0453$-$1751 & R2500I & GTC8-15ITP & 2015-10-01 & 1.51 & 11 & 0.67 -- 0.71\\
    2MASS J05021345$+$1442367 & J0502$+$1442 & R2500I & GTC8-15ITP & 2015-09-29 & 1.04 & 13 & 0.71 -- 0.76\\
    2MASSI J0605019$-$234226 & J0605$-$2342 & R2500I & GTC8-15ITP & 2015-11-30 & 1.66 & 58 & 1.77 -- 1.88\\
    2MASS J07410440$+$2316377 & J0741$+$2316 & R2500I & GTC8-15ITP & 2015-12-31 & 1.05 & 6 & 0.90 -- 0.95\\
    SDSS  J075259.48$+$413646.8 & J0752$+$4136 & R2500I & GTC8-15ITP & 2015-11-28 & 1.04 & 17 & 0.98 -- 1.04\\
    ULAS J080910.65$+$231515.7 & J0809$+$2315 & R2500I & GTC8-15ITP & 2015-12-31 & 1.10 & 7 & 1.20 -- 1.27\\
    2MASS J08230316$+$0240426 & J0823$+$0240 & R2500I & GTC8-15ITP & 2015-12-31 & 1.12 & 7 & 0.84 -- 0.89\\
    2MASS J08230838$+$6125208 & J0823$+$6125 & R2500I & GTC8-15ITP & 2015-11-30 & 1.21 & 51 & 1.27 -- 1.35\\
    2MASS J08472872$-$1532372 & J0847$-$1532 & R300R & GTC54-15A & 2015-04-04 & 1.40 & 13 & 1.49 -- 1.73\\
    2MASSW J0918382$+$213406 & J0918$+$2134 & R2500I & GTC8-15ITP & 2015-11-30 & 1.03 & 52 & 0.98 -- 1.04\\
    2MASS J09352803$-$2934596 & J0935$-$2934 & R2500I & GTC8-15ITP & 2015-11-30 & 1.90 & 47 & 1.79 -- 1.90\\
    2MASS J09385888$+$0443438 & J0938$+$0443 & R2500I & GTC8-15ITP & 2015-12-31 & 1.18 & 7 & 0.67 -- 0.72\\
    2MASS J09404793$+$2946534 & J0940$+$2946 & R2500I & GTC8-15ITP & 2016-02-26 & 1.27 & 13 & 1.01 -- 1.07\\
    2MASSI J0953212$-$101420 & J0953$-$1014 & R2500I & GTC54-15A & 2015-03-31 & 1.37 & 16 & 1.26 -- 1.34\\
    G196$-$3B & J1004$+$5022 & R2500I & GTC54-15A & 2015-04-27 & 1.09 & 2 & 0.83 -- 0.88\\
    G196$-$3B & J1004$+$5022 & R300R & GTC54-15A & 2015-04-27 & 1.08 & 2 & 0.82 -- 0.95\\
    2MASS J10044030$-$1318186 & J1004$-$1318 & R2500I & GTC8-15ITP & 2015-12-31 & 1.36 & 8 & 1.26 -- 1.34\\
    DENIS J104731.1$-$181558 & J1047$-$1815 & R300R & GTC54-15A & 2015-04-27 & 1.50 & 2 & 1.33 -- 1.54\\
    DENIS J104731.1$-$181558 & J1047$-$1815 & R2500I & GTC54-15A & 2015-04-27 & 1.55 & 2 & 1.36 -- 1.44\\
    DENIS J1058.7$-$1548 & J1058$-$1548 & R300R & GTC54-15A & 2015-04-27 & 1.52 & 2 & 1.12 -- 1.29\\
    DENIS J1058.7$-$1548 & J1058$-$1548 & R2500I & GTC54-15A & 2015-04-27 & 1.61 & 1 & 1.16 -- 1.23\\
    2MASS J11092745$-$1606515 & J1109$-$1606 & R2500I & GTC8-15ITP & 2015-12-30 & 1.42 & 21 & 1.18 -- 1.26\\
    2MASS J11270661$+$4705481 & J1127$+$4705 & R2500I & GTC8-15ITP & 2015-12-30 & 1.05 & 21 & 0.63 -- 0.67\\
    2MASS J12130336$-$0432437 & J1213$-$0432 & R2500I & GTC54-15A & 2015-04-28 & 1.29 & 2 & 0.81 -- 0.86\\
    2MASS J12164560$+$4927452 & J1216$+$4927 & R2500I & GTC8-15ITP & 2015-12-31 & 1.07 & 8 & 0.73 -- 0.77\\
    2MASS J12212770$+$0257198 & J1221$+$0257 & R2500I & GTC54-15A & 2015-04-01 & 1.24 & 7 & 0.79 -- 0.84\\
    ULAS J122259.30$+$140750.1 & J1222$+$1407 & R300R & GTC8-15ITP & 2016-01-19 & 1.04 & 6 & 1.16 -- 1.34\\
    DENIS J123218.3$-$095149 & J1232$-$0951 & R2500I & GTC54-15A & 2015-05-31 & 1.32 & 23 & 2.06 -- 2.19\\
    2MASSW J1246467$+$402715 & J1246$+$4027 & R2500I & GTC54-15A & 2015-04-29 & 1.05 & 2 & 0.63 -- 0.67\\
    2MASSW J1246467$+$402715 & J1246$+$4027 & R300R & GTC54-15A & 2015-04-29 & 1.03 & 2 & 0.53 -- 0.61\\
    2MASS J13313310$+$3407583 & J1331$+$3407 & R2500I & GTC54-15A & 2015-04-28 & 1.03 & 2 & 0.80 -- 0.85\\
    2MASS J13313310$+$3407583 & J1331$+$3407 & R300R & GTC54-15A & 2015-04-28 & 1.01 & 2 & 0.79 -- 0.91\\
    2MASS J13334540$-$0215599 & J1333$-$0215 & R2500I & GTC8-15ITP & 2015-12-31 & 1.23 & 7 & 1.28 -- 1.36\\
    2MASS J13460746$+$0842346 & J1346$+$0842 & R2500I & GTC8-15ITP & 2016-01-06 & 1.09 & 6 & 1.01 -- 1.07\\
    2MASSW J1412244$+$163312 & J1412$+$1633 & R2500I & GTC8-15ITP & 2016-01-19 & 1.06 & 6 & 1.26 -- 1.34\\
    2MASSW J1412244$+$163312 & J1412$+$1633 & R2500I & GTC54-15A & 2015-04-29 & 1.04 & 2 & 0.62 -- 0.66\\
    2MASSW J1412244$+$163312 & J1412$+$1633 & R300R & GTC54-15A & 2015-04-29 & 1.03 & 2 & 0.62 -- 0.72\\
    2MASSW J1421314$+$182740 & J1421$+$1827 & R2500I & GTC54-15A & 2015-04-01 & 1.03 & 6 & 0.71 -- 0.75\\
    ULAS J143915.10$+$003941.7 & J1439$+$0039 & R300R & GTC8-15ITP & 2016-03-29 & 1.16 & 10 & 0.57 -- 0.66\\
    DENIS J144137.2$-$094558 & J1441$-$0945 & R300R & GTC54-15A & 2015-05-05 & 1.28 & 11 & 1.01 -- 1.17\\
    DENIS J144137.2$-$094558 & J1441$-$0945 & R2500I & GTC54-15A & 2015-05-05 & 1.28 & 11 & 1.01 -- 1.07\\
    ULAS J152722.48$+$055316.2 & J1527$+$0553 & R300R & GTC8-15ITP & 2016-03-29 & 1.15 & 11 & 0.76 -- 0.88\\
    2MASS J15322338$+$2611189 & J1532$+$2611 & R2500I & GTC8-15ITP & 2016-01-29 & 1.08 & 17 & 0.82 -- 0.87\\
    2MASS J15394189$-$0520428 & J1539$-$0520 & R2500I & GTC8-15ITP & 2016-02-27 & 1.46 & 44 & 1.42 -- 1.51\\
    2MASS J15485834$-$1636018 & J1548$-$1636 & R2500I & GTC54-15A & 2015-04-01 & 1.47 & 10 & 1.10 -- 1.17\\
    2MASS J16170673$+$7734028 & J1617$+$7733B & R2500I & GTC54-15A & 2015-05-29 & 1.53 & 47 & 2.25 -- 2.39\\
    2MASS J16170673$+$7734028 & J1617$+$7733B & R300R & GTC54-15A & 2015-05-28 & 1.58 & 36 & 2.29 -- 2.65\\
    DENIS J161845.0$-$132129 & J1618$-$1321 & R2500I & GTC54-15A & 2015-04-28 & 1.61 & 1 & 0.93 -- 0.98\\
    2MASS J16232185$+$1530393 & J1623$+$1530 & R2500I & GTC8-15ITP & 2015-09-02 & 1.29 & 38 & 1.02 -- 1.08\\
    2MASS J16230740$+$2908281 & J1623$+$2908 & R2500I & GTC8-15ITP & 2016-02-12 & 1.03 & 21 & 1.51 -- 1.60\\
    2MASS J17054834$-$0516462 & J1705$-$0516 & R300R & GTC54-15A & 2015-04-01 & 1.21 & 7 & 1.08 -- 1.24\\
    2MASS J17072529$-$0138093 & J1707$-$0138 & R300R & GTC54-15A & 2015-05-29 & 1.21 & 45 & 1.76 -- 2.03\\
    2MASS J17072529$-$0138093 & J1707$-$0138 & R2500I & GTC54-15A & 2015-05-29 & 1.18 & 36 & 1.73 -- 1.84\\
    2MASS J17171408$+$6526221 & J1717$+$6526 & R2500I & GTC8-15ITP & 2015-08-30 & 1.51 & 31 & 0.89 -- 0.95\\
    2MASS J17171408$+$6526221 & J1717$+$6526 & R300R & GTC54-15A & 2015-05-03 & 1.42 & 4 & 0.75 -- 0.87\\
    2MASS J17171408$+$6526221 & J1717$+$6526 & R2500I & GTC54-15A & 2015-06-01 & 1.27 & 39 & 2.01 -- 2.14\\
    Gaia DR2 4569300467950928768 & J1724$+$2336 & R300R & GTC8-15ITP & 2015-09-01 & 2.13 & 28 & 1.36 -- 1.58\\
    DENIS J173342.3$-$165449 & J1733$-$1654 & R300R & GTC54-15A & 2015-06-26 & 1.62 & 26 & 1.74 -- 2.01\\
    DENIS J174534.6$-$164053 & J1745$-$1640 & R2500I & GTC54-15A & 2015-08-04 & 1.88 & 17 & 1.27 -- 1.35\\
    2MASS J17502484$-$0016151 & J1750$-$0016 & R2500I & GTC54-15A & 2015-04-02 & 1.15 & 19 & 0.95 -- 1.01\\
    2MASS J21555848$+$2345307 & J2155$+$2345 & R2500I & GTC8-15ITP & 2015-08-30 & 1.04 & 28 & 0.62 -- 0.66\\
    2MASS J23392527$+$3507165 & J2339$+$3507 & R2500I & GTC8-15ITP & 2015-08-06 & 1.05 & 38 & 0.90 -- 0.95\\
\end{longtable}
\clearpage
\twocolumn

\setlength{\tabcolsep}{3pt}
\begin{table*}
\centering
\caption{
Cross-matched absolute photometry from \gaia, 2MASS \& WISE, using \gaia\ parallaxes.
}
\label{table:photometry}
\begin{tabular}{l cc ccc ccc}
    \hline
    Object     & $M_{G}$ & $M_{RP}$ & $M_{J}$ & 
    $M_{H}$ & $M_{K_s}$ &
    $M_{W1}$ & $M_{W2}$ & $M_{W3}$ \\
    short name & [mag] & [mag] & [mag] & [mag] & [mag] & [mag] & [mag] & [mag] \\
    \hline
        J0028$-$1927 & $16.03\pm0.02$ & $14.45\pm0.03$ & $11.24\pm0.04$ & $10.38\pm0.04$ & $9.90\pm0.04$ & $9.57\pm0.03$ & $9.31\pm0.03$ & $8.88\pm0.27$\\
    J0235$-$0849 & $17.04\pm0.09$ & $15.43\pm0.10$ & $12.26\pm0.11$ & $11.50\pm0.11$ & $10.88\pm0.11$ & $10.47\pm0.10$ & $10.19\pm0.10$ & \ldots\\
    J0428$-$2253 & $16.70\pm0.02$ & $14.79\pm0.02$ & $11.48\pm0.03$ & $10.65\pm0.03$ & $10.10\pm0.03$ & $9.70\pm0.03$ & $9.45\pm0.03$ & $8.93\pm0.13$\\
    J0453$-$1751 & $17.73\pm0.04$ & $16.15\pm0.04$ & $12.74\pm0.05$ & $11.66\pm0.05$ & $11.06\pm0.05$ & $10.55\pm0.04$ & $10.20\pm0.04$ & $9.62\pm0.21$\\
    J0502$+$1442 & $15.59\pm0.03$ & $14.01\pm0.04$ & $10.96\pm0.04$ & $10.08\pm0.04$ & $9.64\pm0.04$ & $9.34\pm0.05$ & $9.10\pm0.04$ & $7.91\pm0.22$\\
    J0605$-$2342 & $16.71\pm0.02$ & $15.16\pm0.02$ & $11.91\pm0.04$ & $11.13\pm0.04$ & $10.54\pm0.04$ & $10.24\pm0.03$ & $9.94\pm0.03$ & $10.05\pm0.49$\\
    J0741$+$2316 & $16.40\pm0.30$ & $14.87\pm0.32$ & $11.73\pm0.31$ & $10.75\pm0.31$ & $10.34\pm0.31$ & $9.74\pm0.30$ & $9.43\pm0.31$ & ${>}7.16$\\
    J0752$+$4136 & $13.06\pm0.03$ & $11.59\pm0.03$ & $9.35\pm0.04$ & $8.79\pm0.04$ & $8.44\pm0.04$ & $10.36\pm0.06$ & $10.13\pm0.09$ & ${>}7.64$\\
    J0823$+$6125 & $17.65\pm0.02$ & $16.09\pm0.03$ & $12.80\pm0.04$ & $11.80\pm0.04$ & $11.18\pm0.04$ & $10.72\pm0.03$ & $10.43\pm0.03$ & $10.39\pm0.39$\\
    J0847$-$1532 & $17.18\pm0.01$ & $15.60\pm0.01$ & $12.31\pm0.03$ & $11.43\pm0.03$ & $10.86\pm0.02$ & $10.51\pm0.03$ & $10.26\pm0.02$ & $9.76\pm0.11$\\
    J0935$-$2934 & $16.39\pm0.02$ & $14.57\pm0.03$ & $11.42\pm0.04$ & $10.70\pm0.04$ & $10.21\pm0.04$ & $9.80\pm0.03$ & $9.51\pm0.03$ & $9.08\pm0.21$\\
    J0938$+$0443 & $15.84\pm0.07$ & $14.31\pm0.08$ & $11.19\pm0.09$ & $10.44\pm0.09$ & $9.95\pm0.10$ & $9.75\pm0.08$ & $9.55\pm0.08$ & ${>}7.90$\\
    J0940$+$2946 & $16.57\pm0.11$ & $14.93\pm0.12$ & $11.56\pm0.13$ & $10.61\pm0.13$ & $10.19\pm0.12$ & $9.78\pm0.11$ & $9.52\pm0.12$ & ${>}8.78$\\
    J0953$-$1014 & $15.68\pm0.02$ & $14.03\pm0.02$ & $10.71\pm0.03$ & $9.88\pm0.03$ & $9.38\pm0.03$ & $9.01\pm0.03$ & $8.65\pm0.03$ & $7.97\pm0.12$\\
    J1004$+$5022 & $18.45\pm0.03$ & $16.86\pm0.03$ & $13.15\pm0.05$ & $11.97\pm0.05$ & $11.10\pm0.04$ & $10.02\pm0.03$ & $9.46\pm0.04$ & $8.60\pm0.07$\\
    J1004$-$1318 & $17.87\pm0.03$ & $16.29\pm0.04$ & $12.72\pm0.05$ & $11.92\pm0.05$ & $11.39\pm0.05$ & $10.82\pm0.04$ & $10.53\pm0.04$ & $10.21\pm0.43$\\
    J1047$-$1815 & $16.76\pm0.02$ & $15.24\pm0.02$ & $11.96\pm0.04$ & $11.18\pm0.04$ & $10.65\pm0.04$ & $10.34\pm0.03$ & $10.08\pm0.03$ & $10.05\pm0.47$\\
    J1058$-$1548 & $17.95\pm0.01$ & $16.39\pm0.02$ & $12.86\pm0.04$ & $11.93\pm0.03$ & $11.24\pm0.03$ & $10.79\pm0.03$ & $10.49\pm0.03$ & $10.40\pm0.27$\\
    J1109$-$1606 & $16.56\pm0.04$ & $15.01\pm0.05$ & $11.89\pm0.06$ & $11.26\pm0.06$ & $10.81\pm0.07$ & $10.56\pm0.05$ & $10.26\pm0.05$ & $9.57\pm0.48$\\
    J1127$+$4705 & $16.82\pm0.05$ & $15.23\pm0.05$ & $12.08\pm0.06$ & $11.38\pm0.06$ & $10.79\pm0.06$ & $10.49\pm0.05$ & $10.21\pm0.06$ & $9.57\pm0.50$\\
    J1213$-$0432 & $18.71\pm0.02$ & $17.15\pm0.03$ & $13.54\pm0.04$ & $12.51\pm0.03$ & $11.87\pm0.04$ & $11.23\pm0.03$ & $10.93\pm0.03$ & $9.91\pm0.22$\\
    J1221$+$0257 & $16.52\pm0.01$ & $14.93\pm0.01$ & $11.82\pm0.02$ & $11.06\pm0.03$ & $10.61\pm0.03$ & $10.30\pm0.02$ & $10.02\pm0.02$ & $9.48\pm0.15$\\
    J1232$-$0951 & $16.43\pm0.28$ & $14.59\pm0.28$ & $11.42\pm0.28$ & $10.76\pm0.28$ & $10.24\pm0.28$ & $9.92\pm0.28$ & $9.64\pm0.28$ & $9.34\pm0.39$\\
    J1246$+$4027 & $18.53\pm0.03$ & $16.95\pm0.04$ & $13.34\pm0.06$ & $12.20\pm0.05$ & $11.53\pm0.05$ & $10.83\pm0.04$ & $10.46\pm0.04$ & $10.21\pm0.25$\\
    J1331$+$3407 & $16.72\pm0.02$ & $15.14\pm0.02$ & $12.04\pm0.03$ & $11.11\pm0.04$ & $10.59\pm0.03$ & $10.28\pm0.03$ & $10.05\pm0.03$ & $9.46\pm0.23$\\
    J1333$-$0215 & $17.23\pm0.07$ & $15.60\pm0.08$ & $12.50\pm0.08$ & $11.49\pm0.08$ & $10.98\pm0.09$ & $10.66\pm0.08$ & $10.39\pm0.08$ & $9.69\pm0.45$\\
    J1346$+$0842 & $17.32\pm0.10$ & $15.78\pm0.11$ & $12.58\pm0.13$ & $11.63\pm0.13$ & $11.00\pm0.12$ & $10.61\pm0.11$ & $10.38\pm0.11$ & ${>}9.08$\\
    J1412$+$1633 & $16.15\pm0.02$ & $14.55\pm0.02$ & $11.36\pm0.03$ & $10.63\pm0.04$ & $10.00\pm0.03$ & $9.66\pm0.03$ & $9.40\pm0.03$ & $8.89\pm0.13$\\
    J1421$+$1827 & $16.45\pm0.01$ & $14.86\pm0.01$ & $11.85\pm0.02$ & $11.04\pm0.02$ & $10.56\pm0.02$ & $10.18\pm0.02$ & $9.91\pm0.02$ & $8.84\pm0.05$\\
    J1441$-$0945 & $16.78\pm0.09$ & $14.86\pm0.09$ & $11.58\pm0.09$ & $10.75\pm0.09$ & $10.22\pm0.09$ & $9.90\pm0.09$ & $9.67\pm0.09$ & $9.67\pm0.35$\\
    J1539$-$0520 & $17.85\pm0.01$ & $16.26\pm0.02$ & $12.79\pm0.03$ & $11.92\pm0.03$ & $11.44\pm0.03$ & $10.88\pm0.03$ & $10.61\pm0.03$ & $10.54\pm0.28$\\
    J1548$-$1636 & $16.41\pm0.02$ & $14.87\pm0.02$ & $11.76\pm0.03$ & $10.98\pm0.03$ & $10.51\pm0.03$ & $10.16\pm0.03$ & $9.87\pm0.03$ & $9.35\pm0.24$\\
    J1617$+$7733B & $12.23\pm0.01$ & $10.87\pm0.01$ & $8.79\pm0.02$ & $8.21\pm0.02$ & $7.91\pm0.02$ & $7.62\pm0.02$ & $7.37\pm0.02$ & $7.13\pm0.10$\\
    J1618$-$1321 & $16.04\pm0.13$ & $14.14\pm0.13$ & $10.95\pm0.13$ & $10.10\pm0.13$ & $9.62\pm0.13$ & $9.25\pm0.13$ & $8.99\pm0.13$ & $8.55\pm0.36$\\
    J1623$+$1530 & $15.65\pm0.20$ & $14.10\pm0.20$ & $11.00\pm0.22$ & $10.20\pm0.22$ & $9.69\pm0.22$ & $9.53\pm0.20$ & $9.26\pm0.21$ & $7.53\pm0.54$\\
    J1705$-$0516 & $16.81\pm0.01$ & $15.22\pm0.01$ & $11.94\pm0.03$ & $11.18\pm0.03$ & $10.66\pm0.02$ & $10.31\pm0.03$ & $10.05\pm0.03$ & $9.67\pm0.21$\\
    J1707$-$0138 & $16.33\pm0.03$ & $14.72\pm0.03$ & $11.36\pm0.04$ & $10.64\pm0.04$ & $10.14\pm0.05$ & $9.71\pm0.04$ & $9.43\pm0.04$ & $9.32\pm0.49$\\
    J1717$+$6526 & $18.56\pm0.03$ & $16.90\pm0.03$ & $13.25\pm0.05$ & $12.14\pm0.04$ & $11.48\pm0.04$ & $10.85\pm0.03$ & $10.52\pm0.03$ & $9.82\pm0.07$\\
    J1724$+$2336 & $16.02\pm0.07$ & $14.45\pm0.07$ & $11.50\pm0.09$ & $10.95\pm0.11$ & $10.15\pm0.11$ & $10.03\pm0.08$ & $9.76\pm0.08$ & ${>}7.98$\\
    J1733$-$1654 & $17.20\pm0.01$ & $15.46\pm0.01$ & $12.23\pm0.05$ & $11.50\pm0.06$ & $11.05\pm0.03$ & \ldots & \ldots & \ldots\\
    J1745$-$1640 & $16.98\pm0.01$ & $15.38\pm0.01$ & $12.18\pm0.03$ & $11.41\pm0.02$ & $10.94\pm0.02$ & $10.64\pm0.03$ & $10.40\pm0.03$ & $10.82\pm0.46$\\
    J1750$-$0016 & ${>}18.47$ & $16.86\pm0.01$ & $13.47\pm0.02$ & $12.59\pm0.02$ & $12.03\pm0.02$ & $11.36\pm0.02$ & $11.08\pm0.02$ & $10.47\pm0.07$\\
    J2339$+$3507 & $18.26\pm0.05$ & $16.74\pm0.06$ & $13.16\pm0.07$ & $12.15\pm0.07$ & $11.38\pm0.06$ & $10.88\pm0.05$ & $10.56\pm0.05$ & $10.32\pm0.53$\\
    \hline
\end{tabular}
\newline
\end{table*}

\subsection{Comparison with standard routines} \label{subsec:comparisonroutines}
In the reduction we use two procedures based on \texttt{IRAF} and \texttt{Python} packages
with a comparison target~\citep[J1745$-$1640, DENIS J174534.6$-$164053,][]{2008MNRAS.383..831P} as a sanity check.
A full image and spectral reduction was carried out using standard tasks within the \texttt{IRAF}
package on one of our target objects (J1745--1640) plus complimentary flux standard (Ross~640).
This was done to assess both the quality of the data and to ascertain the necessary required reduction steps
to maximise data quality.
The results from this bespoke reduction method served as a reliable reference by which to
measure the performance of a \texttt{python} pipeline (with support for the GTC/OSIRIS instrument recently added),
which was later applied to all objects within our sample.

\subsubsection{Bespoke \texttt{IRAF} Reduction} \label{subsubsec:irafreduction}
Our \texttt{IRAF} reduction was applied to the science and calibration frames of J1745$-$1640 (L1--1.5)
and Ross~640 (DZA6) as appropriate
using the following tasks, beginning with basic image reduction:

\begin{description}
    \item []{{\textbf{\texttt{CCDPROC}}}:} Pre-scan bias level and bias structure removal;
        flat-fielding;
        illumination correction;
        data section trimming.
    \item []{{\textbf{\texttt{RESPONSE}}}:} Spectroscopic flat-field lamp colour removal (normalisation).
    \item []{{\textbf{\texttt{Illumination}} and \textbf{\texttt{CCDPROC}}}:} Correction for spatial axis
        illumination gradients, made from the extensive sky lines of a well exposed object frame.
    \item []{{\textbf{\texttt{IDENTIFY}} , \textbf{\texttt{FITCOORDS}} and \textbf{\texttt{TRANSFORM}}}:} Correction for geometric image
        distortion (curvature) along the spatial axis sky background.
\end{description}

For the spectral reduction:

\begin{description}
    \item []{{\textbf{\texttt{APALL}}}:} Trace and extraction using both optimal and fixed-width
        aperture summing using image distortion corrected arc frames.
    \item []{{\textbf{\texttt{IDENTIFY}} and \textbf{\texttt{DISPCOR}}:}} Wavelength calibration to a linear wavelength
        dispersion using image distortion corrected arc frames.
    \item []{{\textbf{\texttt{STANDARD}}, \textbf{\texttt{SENSFUNC}} and \textbf{\texttt{CALIBRATE}}:}} Flux calibration from the flux
        standard Ross 640 taken on same night as the target.
\end{description}

In addition to the \texttt{IRAF} tasks mentioned above, two extra reduction
software tools were utilised during the reduction process:

\begin{description}
    \item []{{\textbf{\texttt{DeFringFlat}}}:} An \texttt{IDL} routine aquired from the
        NASA \texttt{IDL} Astronomy library~\citep{nasa_idl} was used to provide capabilities in de-fringing
        the flat field frames~\citep[\href{https://ui.adsabs.harvard.edu/
        abs/2006ApJ...649..553R/abstract}{\texttt{DeFringFlat.pro;}}][]{2006ApJ...649..553R}.
    \item []{{\textbf{\texttt{SKYCALC}} }:} \href{http://www.eso.org/observing/etc/bin/
        gen/form?INS.MODE=swspectr+INS.NAME=SKYCALC}{ESO Sky Model Calculator} provides additional telluric
        correction during flux calibration.
        A telluric sky model was queried using meteorological (e.g.\ moon phase, precipitable water vapour) and
        astrometric parameters (e.g.\ altitude, angular separation)
        appropriate for the object in question.
\end{description}

During the bias subtraction we discovered that the pre-scan region of the second CCD
containing the spectrum displayed a gradient across it in ADU\@.
A carefully chosen restricted section of the pre-scan region was used (${\sim}3$\,pixels wide),
which was found to be reliable for row-by-row bias level subtraction, before the 2D image bias structure was removed.

To correct for illumination gradients evident along the spatial axis of the 2D image
introduced by the slit illumination function, we utilised the extensive sky lines of the well
exposed object frames as a pseudo twilight sky flat (no sky flats were available).
The \texttt{IRAF} \texttt{Illumination} task provided this functionality for correction,
and we estimate that, after the correction was applied, the error introduced by the slit
illumination gradient was reduced to a maximum of ${\sim}1.5$\,per cent in the flat-field frames.

The latter, longer wavelength half of the flat-field frames showed evidence
of fringing between wavelengths of approximately $8500\,\angstrom$ to $10,000\,\angstrom$,
coincident with the area of the CCD containing the spectra of interest.
We used the \texttt{IDL} routine \texttt{DeFringFlat} as mentioned above to attempt
to remove as much of the fringing as possible and found the best fit using the Morlet `wavemother' model,
and near default parameters.
We estimate from measuring the cleaned flat-fielded image that the amplitude of the
fringing was reduced from an original level of approximately $7$\,per cent, to a maximum of about $1.7$\,per cent.

A combined arc frame was made from the three arcs available from the night
of observation to cover the entire wavelength region of the spectrum.
An initial wavelength solution was created and applied as part of the geometric
image distortion correction, which resulted in a wavelength solution with an RMS error of $0.016\,\angstrom$.
A second wavelength calibration was subsequently made after
additional reduction steps to ensure no systematic errors had been introduced,
resulting in a more reasonable final RMS to the fitted wavelength solution of $0.025\,\angstrom$.
The final wavelength corrected spectrum had a linear dispersion $1.396\,\angstrom$\,pixel$^{-1}$
over the entire extracted range of $7339\,\angstrom\,$--$10,155\,\angstrom$.

Two separate flux calibrations were then made:
one used a blackbody to represent the DZ white dwarf flux standard with an
effective temperature $8070$\,K~\citep{new_temp} and with an $I$-band magnitude of 13.66\,mag~\citep{i_mag};
the second used the low resolution calibrated flux
standard spectrum of Ross~640 contained in the \texttt{IRAF} database.
In both cases, the sensitivity functions were created by interpolating
over the affected telluric regions, and regions of intrinsic absorption features.
Both of these sensitivity functions provided flux calibrations with almost identical results.
A correction for atmospheric extinction and telluric features to the target was included during the flux calibration.
An initial extinction correction was made from using a file containing
tabulated extinction magnitudes as a function of wavelength applicable to the observatory site, that was provided on the
\href{https://www.ing.iac.es//Astronomy/observing/manuals/ps/tech_notes/tn031.pdf}{GTC instrument website}.
However, an improved extinction correction was obtained from the
much higher spectral resolution telluric sky model mentioned above (via the ESO Sky Model Calculator).
The improvement is particularly evident over the wavelength regions
containing the potassium K\,\textsc{i} $\lambda\lambda$7665,7699\,{\AA} doublet 
and the H$_2$O band at about $9500\,\angstrom$.

\subsection{Radial velocity method validation} \label{subsubsec:meth_val}
In keeping with our strategy outlined in Section~\ref{subsec:comparisonroutines}
we again invoked an independent check, this time to validate our methods by helping
to identify any problems with our RV measurements relating to the \texttt{PypeIt} reduced data set.
The techniques used to measure RVs via the
centres of atomic neutral alkali lines and through cross-correlation of spectra were employed
by~\citet{2015ApJS..220...18B}, and we adopt a similar twin measurement approach to derive our final RVs.
We achieved this through the use of both \texttt{IRAF} and custom
prepared routines within \texttt{IDL} to measure the RV via
the Fourier cross-correlation and the line centre fitting methods.
This analysis was conducted on the bespoke \texttt{IRAF} reduced data of our test object J1745$-$1640\@.
We then used our validated RVs to classify any objects into young moving groups and stellar associations.

\subsubsection{Line centres}\label{subsubsec:meth_val_line}

Two interactive methods were employed here: the first using routines in
\texttt{IDL} to measure the 1D centroids of fitted Gaussian profiles to the atomic lines of J1745$-$1640,
while the second used the \texttt{IRAF} task \texttt{Splot}
to again measure the same lines but via fitting Voigt profiles.

In the first case, sub-sections of the spectrum surrounding the line features
to be measured were extracted and interpolated onto a ten times finer wavelength grid,
to facilitate the manual fitting of Gaussian profiles with a different number of terms via the \texttt{Gaussfit.pro} routine.
Best fitting model profiles to spectral features were initially determined by eye,
and determined by how closely the profile matched the feature with more emphasis being
given around the line centre region.
The reported RMS error and FWHM of fitted profiles were also
taken into account for when the different Gaussian profiles produced similar results,
such that the number of terms which fitted with the least error and narrowest FWHM were chosen.
The measured wavelength shifts from laboratory rest-frame
line centres~\citep[in standard air:][]{nistdb} were then converted to Doppler RVs.

Secondly, and by using \texttt{Splot}, Voigt profiles were fitted
to the same line features of appropriately pseudo-continuum subtracted sub-sections of the
spectrum, and Doppler RVs were then found in the same manner as previously from the reported line centres.
We obtained results for all eight line features from both measurement sets.
However, it was apparent that four of the measurements gave the least error and
particularly consistent results between both sets, these being Rb\,\textsc{i}-a,
    Rb\,\textsc{i}-b, Na\,\textsc{i}-a, Cs\,\textsc{i}-a with mean values for RV
found from these four selected for each measurement set.
The RV derived from the Gaussian fitted profiles (\texttt{IDL}) was found to be 35.1\,\kms,
and via Voigt profiles (\texttt{Splot}) 29.0\,\kms\ (all test results are Heliocentric:
barycentric correction calculated using \texttt{baryvel.pro}).
Typically, we found that Gaussian profiles were more reliable to fit but Voigt profiles were best for lines
which could be successfully fit.
From the spread among the individually measured line shifts we place more confidence
in the latter result, and assign uncertainties based on the $1$-$\sigma$ standard
deviation of the respective RV measurements of 4.3\,\kms\ and 3.8\,\kms.

The RV as measured by our line centering method using the \texttt{PypeIt}
reduced data for J1745$-$1640 is $36.2\pm4.4$\,\kms\ (see Table~\ref{table:radial_velocities})
which is in broad agreement with those from this independent measurement test.
The RV measured via line centre fitting as reported by~\citet{2015ApJS..220...18B} is $28\pm9$\,\kms.
Thus, we have confidence in our RV results derived from our chosen method,
which contribute to the final adopted values.

\subsubsection{Cross-correlation} \label{subsubsec:meth_val_xcorr}

To validate this second technique of measuring RVs as part of our adopted method,
and to ascertain the best way forward in its application, we used the Fourier
cross-correlation task \texttt{Fxcor} within \texttt{IRAF} to conduct tests.
Our choice of RV rest-frame models were a BT-Settl model spectrum
and custom-made synthetic atomic absorption spectra.
Our object was again the bespoke \texttt{IRAF} reduced J1745$-$1640 spectrum.

The BT-Settl spectrum used was the best fitting model with the physical parameters of
$\teff = 2000$\,K, $\logg = 5$\,dex and Fe/H$ = 0$\,dex, corresponding to ${\simeq}$L1 in spectral type.
We smooth the spectrum using a Gaussian kernel to match the dispersion and resolution of the J1745$-$1640,
and appropriate FITS header keywords added for the \texttt{Fxcor} task to recognise
the template spectrum as rest-frame.

To help highlight any potential systematic wavelength shifts introduced by the use
of the BT-Settl model, and therefore to help assess its suitability as an RV template,
we measured the line centre locations of the most reliable Rb\,\textsc{i}-b and
Cs\,\textsc{i}-a lines by fitting Voigt profiles in \texttt{Splot}.
BT-Settl is known to generate models using a different line list to those selected in this work,
where we used the NIST database.
A maximum difference compared to laboratory rest-frame line centres of $0.13\,\angstrom$ was found,
corresponding to 4.5\,\kms.
This shift is similar to the uncertainty found earlier from the fitted line profiles suggesting
that the BT-Settl model is reliable for use as a template, however, we add this uncertainty
in velocity units in quadrature to the subsequent \texttt{Fxcor} individual RV region measurements.

To facilitate the most accurate RV measurements we extracted sections of both
object and template spectra into discrete spectral regions, each covering the main
atomic absorption features as well as the FeH Wing-Ford band at ${\sim}9900\,\angstrom$,
then each region was appropriately pseudo-continuum subtracted and normalised.

During the RV measurements, we interactively adjusted the sample test wavelength range around the features of
interest to reduce noise in Fourier space domain.
Next, the width of the cross-correlation function (CCF) fit was changed to facilitate
a best-fit (Gaussian fit to the CCF was used).
The results of these changes to the CCF height, the goodness-of-fit `R-value'
and fit error were noted, until the best RV estimate was obtained.
The shape of the CCF profile was also informative to this end, it tended to be broad, with no apparent double
peaks seen.
No Fourier filtering was applied as it was not found to be beneficial.

For this test, three regions gave consistent results covering both of the rubidium lines,
the first caesium line ($\approx 8500\,\angstrom$) and the FeH Wing-Ford band.
The average of these individual results gave an RV of $21.2\pm5.2$\,\kms.

For our second test, we created a noise-free synthetic absorption spectrum
of unity continuum with line widths and depths as measured by Voigt profiles
of the neutral atomic lines in of J1745$-$1640, with no attempt to include the FeH band.
The line centres were fixed to the laboratory rest-frame wavelength values.
Results from all four regions were averaged which covered both of the rubidium lines,
the sodium doublet and both caesium lines.
Including the potassium doublet gave a similar result for that region
but gave a very large increase in uncertainty, so was not included.
We find a resulting RV of $24.6\pm1.7$\,\kms.

Our final test was conducted to ascertain the intrinsic level of uncertainty
in RV from the application of this method through the use of \texttt{Fxcor} on a representation of our spectral data.
This involved making a cross-correlation between two noise-free synthetic absorption spectra:
the same RV rest-frame template as used above in the second test,
and with the object being a wavelength shifted version of the same synthetic spectrum,
with the FITS header updated accordingly.
The shift in wavelength was set at a value corresponding to the adopted RV presented
in~\citet{2015ApJS..220...18B}, of $26.2\pm2.3$\,\kms.
We found the average combined RV of the four measured regions used to be $26.7\pm1.2$\,\kms,
indicating that 1.2\,\kms\ is our base level uncertainty in using this method.
This is, however, in addition to any uncertainty introduced from a real object spectrum (i.e.\ J1745$-$1640).

Both of these cross-correlation RV test results for J1745$-$1640 are in agreement
with the equivalent value presented in~\citet{2015ApJS..220...18B}, within their respective uncertainties.
The measured RV for J1745$-$1640 using the cross-correlation package we adopted and apply to our data set
(see Section~$\S$\ref{subsec:kinematicsanalysis}) has a value of $28.8\pm4.7$\,\kms.
Again, the results of this cross-correlation test validate our method and
provide us with confidence in the separately derived RVs as well as in our final
adopted values combined from both methods (see Section~$\S$\ref{subsubsec:final_rv}).

\subsection{Radial velocity measurement confidence} \label{subsec:rvconfidence}
We demonstrate here a worked example for our test object, J1745$-$1640, including measurement uncertainties
and our confidence metric.
J1745$-$1640 had a wavelength calibration RMS of $0.077\,\angstrom$.
The wavelength shifts and uncertainties excluding this wavelength calibration RMS,
i.e. the uncertainty corresponding to the fitted profile centre from the square root of the diagonal of the covariance matrix, are:
K\,\textsc{i}-a $0.767\pm0.397\,\angstrom$; K\,\textsc{i}-b $0.713\pm0.190\,\angstrom$;
Rb\,\textsc{i}-a $0.916\pm0.112\,\angstrom$; Rb\,\textsc{i}-b $0.542\pm0.168\,\angstrom$;
Na\,\textsc{i}-a $0.537\pm0.114\,\angstrom$; Na\,\textsc{i}-b $1.237\pm0.088\,\angstrom$;
Cs\,\textsc{i}-a $1.363\pm0.051\,\angstrom$; Cs\,\textsc{i}-b $0.330\pm0.264\,\angstrom$.
We had experimented with several different metrics such as $\chi^2$ but
found that the root mean square deviation divided by the interquartile range (RMSDIQR) gave the most robust metric, especially when
comparing across the two distinct techniques; those values were logged as follows.
J1745$-$1640, Line Centering:\\
K\,\textsc{i}-a -- Gaussian Profile with $17.4\,\angstrom \sigma$; $30.0\pm18.5$\,\kms; RMSDIQR$=$0.74.
K\,\textsc{i}-b -- Gaussian Profile with $12.2\,\angstrom \sigma$; $27.8\pm10.4$\,\kms; RMSDIQR$=$0.16.
Rb\,\textsc{i}-a -- Gaussian Profile with $2.1\,\angstrom \sigma$; $35.2\pm7.2$\,\kms; RMSDIQR$=$0.09.
Rb\,\textsc{i}-b -- Gaussian Profile with $2.2\,\angstrom \sigma$; $20.4\pm9.2$\,\kms; RMSDIQR$=$0.16.
Na\,\textsc{i}-a -- Voigt Profile with $2.4\,\angstrom \sigma$; $19.7\pm7.0$\,\kms; RMSDIQR$=$0.08.
Na\,\textsc{i}-b -- Voigt Profile with $2.8\,\angstrom \sigma$; $45.2\pm6.0$\,\kms; RMSDIQR$=$0.06.
Cs\,\textsc{i}-a -- Voigt Profile with $2.3\,\angstrom \sigma$; $47.9\pm4.5$\,\kms; RMSDIQR$=$0.04.
Cs\,\textsc{i}-b -- Gaussian Profile with $2.0\,\angstrom \sigma$; $11.1\pm11.4$\,\kms; RMSDIQR$=$0.25.
RV Line Centre ${=}36.2\pm4.4$\,\kms.
J1745$-$1640, Cross Correlation:\\
K\,\textsc{i}-a -- 2200\,K, $\logg = 5.0$\,dex ; $30.0\pm5.0$\,\kms; RMSDIQR$=$0.48.
K\,\textsc{i}-b -- 2200\,K, $\logg = 5.0$\,dex ; $20.0\pm5.0$\,\kms; RMSDIQR$=$0.20.
Rb\,\textsc{i}-a -- 2200\,K, $\logg = 5.0$\,dex ; $35.0\pm5.0$\,\kms; RMSDIQR$=$0.47.
Rb\,\textsc{i}-b -- 2000\,K, $\logg = 5.0$\,dex ; $25.0\pm5.0$\,\kms; RMSDIQR$=$1.25.
Na\,\textsc{i}-a -- 2100\,K, $\logg = 5.0$\,dex ; $25.0\pm5.0$\,\kms; RMSDIQR$=$1.33.
Na\,\textsc{i}-b -- 2000\,K, $\logg = 5.0$\,dex ; $35.0\pm5.0$\,\kms; RMSDIQR$=$0.79.
Cs\,\textsc{i}-a -- 2000\,K, $\logg = 5.0$\,dex ; $55.0\pm5.0$\,\kms; RMSDIQR$=$0.76.
Cs\,\textsc{i}-b -- 2000\,K, $\logg = 5.0$\,dex ; $5.0\pm5.0$\,\kms; RMSDIQR$=$0.89.\\
RV Cross Correlation ${=}28.8\pm$4.7\,\kms.
Adopted RV ${=}32.7\pm10.1$\,\kms.

\subsection{Spectral sequence} \label{subsec:sequencecomp}
\begin{figure*}
    \centering
    \includegraphics[width=\linewidth]{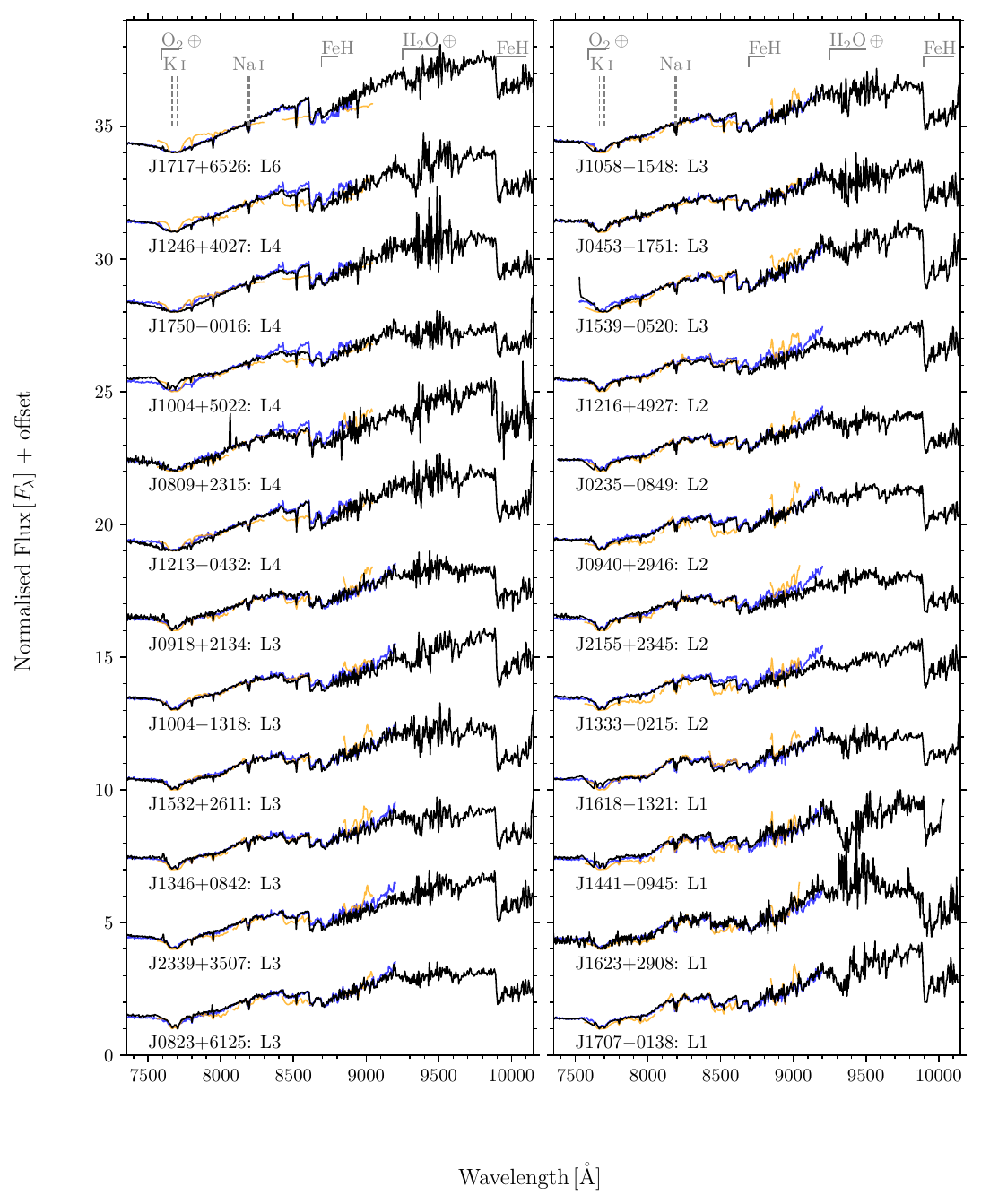}
    \caption{
        Same as Figure~\ref{fig:r2500i-full-seq1} with additional comparison spectra.
        Light blue shows the corresponding standard optical spectra whilst
        light orange is the best-fitting BT-Settl model around the relevant spectral lines.
    }
    \label{fig:r2500i-full-seq1-comp}
\end{figure*}

\begin{figure*}
    \centering
    \includegraphics[width=\linewidth]{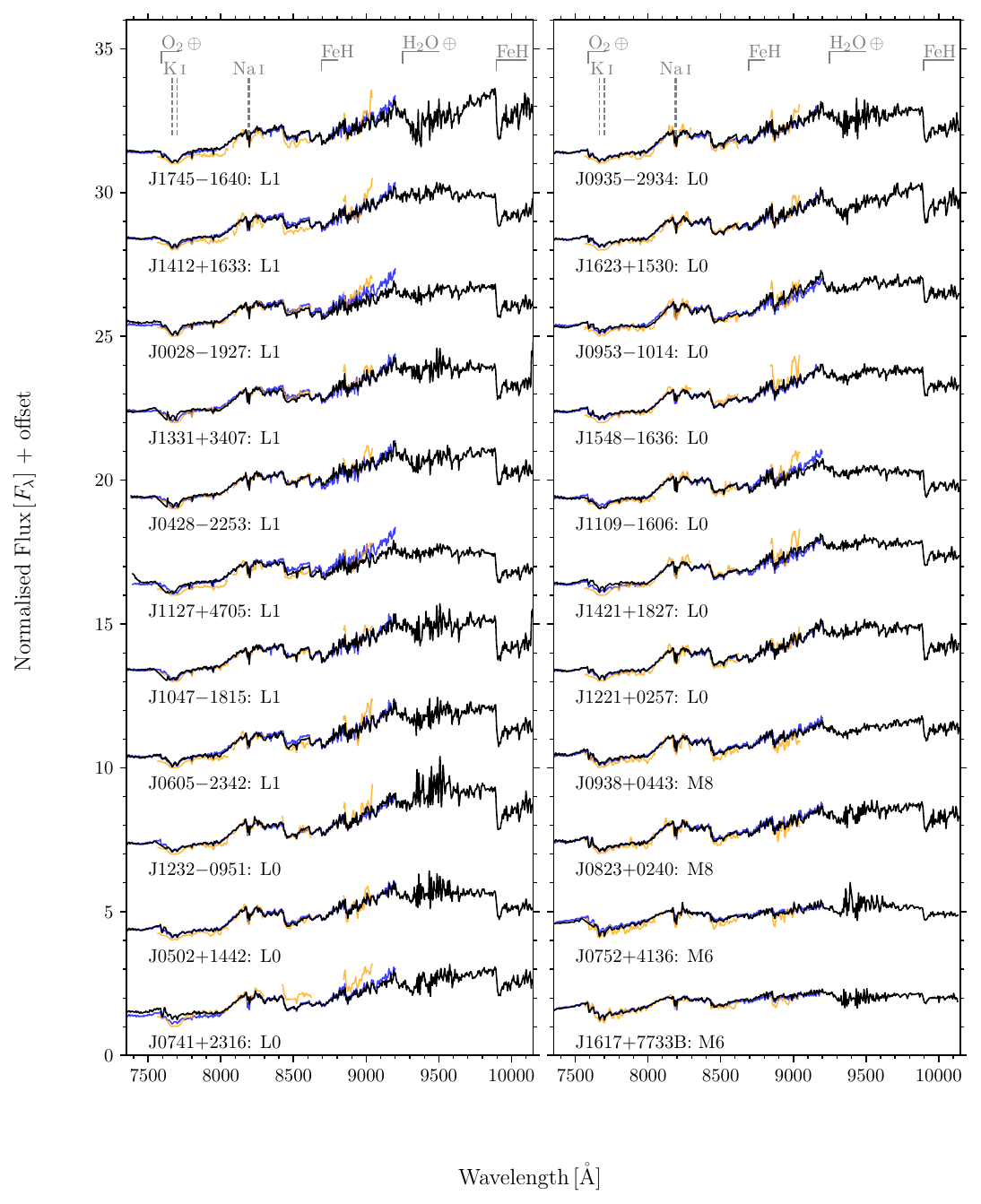}
    \caption{
        Same as Figure~\ref{fig:r2500i-full-seq1-comp} but for the second half
        of the R2500I VPHG spectral sample.
    }
    \label{fig:r2500i-full-seq2-comp}
\end{figure*}

We compare here in Figures~\ref{fig:r2500i-full-seq1-comp} and~\ref{fig:r2500i-full-seq2-comp}
the sequence of R2500I spectra, as in Figures~\ref{fig:r2500i-full-seq1} and~\ref{fig:r2500i-full-seq2},
to their appropriate standards and best-fitting BT-Settl models.
All spectra are normalised by the median flux from 8100--8200\,$\angstrom$.
The standards and BT-Settl models have been interpolated onto the wavelength grid of the spectra
from this work.
BT-Settl models have been additionally smoothed by a $2\sigma$ Gaussian kernel, so as to not ``dominate''
the plot.
These models are only plotted within $\pm 100\,\angstrom$ of each spectral line
listed in Table~\ref{table:spectrallines}.

\subsection{\texttt{PypeIt} Configuration Files} \label{sec:pypeit}
\subsubsection{Reduction}\label{subsec:reductionpypeit}
\begin{lstlisting}[label={lst:specparams}]
slitspatnum = 2:240
[calibrations]
    [[biasframe]]
        exprng = None, 1
        [[[process]]]
            apply_gain = False
            combine = median
            use_biasimage = False
            use_overscan = False
            use_pixelflat = False
            use_illumflat = False
    [[darkframe]]
        exprng = 999999, None
        [[[process]]]
            apply_gain = False
            use_biasimage = False
            use_overscan = False
            use_pixelflat = False
            use_illumflat = False
    [[arcframe]]
        [[[process]]]
            clip = False
            comb_sigrej = None
            use_overscan = False
            use_pixelflat = False
            use_illumflat = False
    [[tiltframe]]
        [[[process]]]
            comb_sigrej = None
            use_overscan = False
            use_pixelflat = False
            use_illumflat = False
    [[pixelflatframe]]
        [[[process]]]
            combine = median
            satpix = nothing
            use_overscan = False
            use_pixelflat = False
            use_illumflat = False
    [[pinholeframe]]
        exprng = 999999, None
        [[[process]]]
            use_overscan = False
    [[alignframe]]
        [[[process]]]
            satpix = nothing
            comb_sigrej = None
            use_overscan = False
            use_pixelflat = False
            use_illumflat = False
    [[traceframe]]
        [[[process]]]
            use_overscan = False
            use_pixelflat = False
            use_illumflat = False
    [[illumflatframe]]
        [[[process]]]
            satpix = nothing
            use_overscan = False
            use_pixelflat = False
            use_illumflat = False
    [[skyframe]]
        [[[process]]]
            mask_cr = True
            use_overscan = False
    [[standardframe]]
        exprng = None, 600
        [[[process]]]
            combine = median
            spat_flexure_correct = True
            mask_cr = True
            use_overscan = False
    [[wavelengths]]
        reid_arxiv = 
        method = full_template
        lamps = XeI,HgI,NeI,ArI
        fwhm_fromlines = True
        ech_fix_format = False
        n_first = 5
        n_final = 6
        match_toler = 2.
    [[slitedges]]
        sync_predict = nearest
        bound_detector = True
[scienceframe]
    exprng = 600, None
    [[process]]
        mask_cr = True
        use_overscan = False
	combine = median
	spat_flexure_correct = True
[reduce]
    [[findobj]]
        maxnumber = 2
[flexure]
    spec_method = slitcen
\end{lstlisting}
\subsubsection{Sensitivity Function}\label{subsec:sensitivity-function}
\begin{lstlisting}[label={lst:sensinit}]
[sensfunc]
    algorithm = IR
    mask_abs_lines = True
    polyorder = 5
    samp_fact = 1.0
    extrap_blu = 0.5
    extrap_red = 0.5
    [[IR]]
        objmodel = poly
        polyorder = 3
        delta_redshift = 0.
        fit_wv_min_max = [7350, 7550, 7750,
              8000, 8350, 8900, 9850, 10150]
\end{lstlisting}
\subsubsection{Flux Calibration}\label{subsec:flux-calibration}
\begin{lstlisting}[label={lst:fluxing}]
[fluxcalib]
  extinct_correct = False

flux read
 ../Science/<spec1d-standard.fits> sensfunc.fits
 ../Science/<spec1d-object.fits> sensfunc.fits
flux end

\end{lstlisting}
\subsubsection{Coadding}\label{subsec:coadding}
\begin{lstlisting}[label={lst:coaddstd}]
[coadd1d]
  coaddfile = ../Science/<standard.fits>

coadd1d read
 ../Science/<spec1d-standard.fits> 
    SPAT0240-SLIT0457-DET02
coadd1d end

\end{lstlisting}
\begin{lstlisting}[label={lst:coaddobj}]
[coadd1d]
  coaddfile = ../Science/<object.fits>

coadd1d read
 ../Science/<spec1d-object.fits> 
    SPAT0240-SLIT0457-DET02
coadd1d end

\end{lstlisting}
\subsubsection{Telluric Correction}\label{subsec:telluric-correction}
\begin{lstlisting}[label={lst:telluric}]
[telluric]
  objmodel = poly
  polyorder = 5
  fit_wv_min_max = 7350, 7550, 7750,
        8000, 8350, 8900, 9850, 10150
  maxiter = 1
  popsize = 300
  pix_shift_bounds = -10., 10.
\end{lstlisting}

\bsp	
\label{lastpage}
\end{document}